\documentclass[12pt]{article}

\AtBeginDocument{
  \addtocontents{toc}{\normalsize}
}
\pdfoutput=1 
\def\baselinestretch{1.2}

\usepackage{amsmath,amssymb,amsfonts,amsthm,amscd,mathrsfs}
\usepackage{xcolor}
\definecolor{darkblue}{rgb}{0.1,0.1,.7}
\usepackage[]{version}
\usepackage{float}
\usepackage[]{graphicx}
\usepackage[]{latexsym}
\usepackage{geometry}
\usepackage{subcaption}
\usepackage[utf8]{inputenc}
\usepackage[all,cmtip]{xy}

\usepackage{enumerate}
\usepackage{bm}

\usepackage{ifthen}
\usepackage{tikz}
\usepackage{array,setspace,yfonts,dsfont,bbm,colonequals}
\usepackage[colorlinks, linkcolor=darkblue, citecolor=darkblue, urlcolor=darkblue, linktocpage]{hyperref} 

\usepackage{cite}
\usepackage{xspace}
\usepackage{empheq}
\usepackage{xcolor}

\usepackage{amsmath}

\usepackage[export]{adjustbox}

\numberwithin{equation}{section}
\newcommand{\ba}{\begin{equation}\begin{aligned}}
\newcommand{\ea}{\end{aligned}\end{equation}}

\newcommand{\ttext}[1]{{\mbox{\tiny #1}}}
\newcommand{\hG}{\widehat{\mathcal G}}

\newcommand{\reef}[1]{(\ref{#1})}
\newcommand{\be}{\begin{equation}}
\newcommand{\ee}{\end{equation}}

\newcommand{\bea}{\begin{eqnarray}}
\newcommand{\eea}{\end{eqnarray}}

\newcommand{\ud}{\mathrm d}

\newcommand{\nn}{\nonumber \\}
\newcommand{\Df}{{\Delta_\phi}}

\newcommand{\Sp}{S_{\mbox{\tiny phys}}}
\newcommand{\af}{a_{\Delta}^{\mbox{\tiny free}}}
\newcommand{\Nh}{\hat{\mathcal N}_{\Df}}

\definecolor{darkblue}{cmyk}{0.9,0.9,0,0}
\definecolor{darkgreen}{cmyk}{0.9,0,0.9,0}
\definecolor{blueblue}{cmyk}{0.73,0.28,0,0.5}
\definecolor{lightblue}{RGB}{55,171,200}
\definecolor{grey}{gray}{0.55}
\definecolor{pink}{cmyk}{0., 0.9859943977591037, 0.3571428571428571, 0.16000000000000003}
\definecolor{lightpink}{cmyk}{0., 0.5, 0.5, 0.}
\definecolor{lightgreen}{cmyk}{0.24175824175824182, 0., 0.9615384615384616, 0.28627450980392155}

\def\({\left(}
\def\){\right)}
\def\[{\left[}
\def\]{\right]}
\def\<{\langle}
\def\>{\rangle}
\newcommand{\beq}{\begin{equation}}
	\newcommand{\eeq}{\end{equation}}
\newcommand{\beqq}{\begin{equation*}}
	\newcommand{\eeqq}{\end{equation*}}
\newcommand\beqa{\begin{eqnarray}}
	\newcommand\eeqa{\end{eqnarray}}
\newcommand{\dphi}{{\Delta_\phi}}
\newcommand{\cP}{\mathcal{P}}
\newcommand{\cF}{\mathcal{F}}
\newcommand{\cG}{\mathcal{G}}

\newcommand{\la}[1]{\label{#1}} 

\def\Xint#1{\mathchoice
	{\XXint\displaystyle\textstyle{#1}}%
	{\XXint\textstyle\scriptstyle{#1}}%
	{\XXint\scriptstyle\scriptscriptstyle{#1}}%
	{\XXint\scriptscriptstyle\scriptscriptstyle{#1}}%
	\!\int}
\def\XXint#1#2#3{{\setbox0=\hbox{$#1{#2#3}{\int}$}
		\vcenter{\hbox{$#2#3$}}\kern-.5\wd0}}

\def\dashint{\Xint-}

\begin{document}

\vspace*{-.6in} \thispagestyle{empty}
\begin{flushright}
\end{flushright}
\vspace{1cm} {\Large
\begin{center}
{\bf From conformal correlators to analytic S-matrices: CFT$_1$/QFT$_2$}\\
\end{center}}
\vspace{1cm}
\begin{center}
{Luc\'ia C\'ordova $^a$, Yifei He $^a$, Miguel F.~Paulos $^b$}\\[1cm] 
{
\small
{\em Institut de Physique Th\'eorique Philippe Meyer $^a$ \& Laboratoire de Physique $^b$ \\
de l'\'Ecole Normale Sup\'erieure PSL University, CNRS,\\
Sorbonne Universit\'es, UPMC Univ. Paris 06 24 rue Lhomond, 75231 Paris Cedex 05, France}
}\normalsize
\\
\end{center}

\begin{center}
	{\texttt{lucia.cordova, yifei.he, miguel.paulos, @ens.fr} 
	}
	\\
\end{center}

\vspace{4mm}

\begin{abstract}
We study families of one-dimensional CFTs relevant for describing gapped QFTs in AdS$_2$. Using the Polyakov bootstrap as our main tool, we explain how S-matrices emerge from the flat space limit of CFT correlators. In this limit we prove that the CFT OPE density matches that of a generalized free field, and that this implies unitarity of the S-matrix. We establish a CFT dispersion formula for the S-matrix, proving its analyticity except for singularities on the real axis which we characterize in terms of the CFT data. In particular positivity of the OPE establishes that any such S-matrix must satisfy extended unitarity conditions. We also carefully prove that for physical kinematics the S-matrix may be more directly described by a phase shift formula. Our results crucially depend on the assumption of a certain gap in the spectrum of operators. We bootstrap perturbative AdS bubble, triangle and box diagrams and find that the presence of anomalous thresholds in S-matrices are precisely signaled by an unbounded OPE arising from violating this assumption. Finally we clarify the relation between unitarity saturating S-matrices and extremal CFTs,  establish a mapping between the dual S-matrix and CFT bootstraps, and discuss how our results help understand UV completeness or lack thereof for specific S-matrices.

\end{abstract}
\vspace{2in}


\newpage

{
\setlength{\parskip}{0.05in}
\tableofcontents
\renewcommand{\baselinestretch}{1.0}\normalsize
}


\setlength{\parskip}{0.1in}
\newpage

\section{Introduction}\label{sec:introduction}

Recently, guided by conceptual and technical advances in the {\em conformal} bootstrap, the S-matrix bootstrap program \cite{Eden:1966dnq} has been brought back to life \cite{Paulos:2016fap,Paulos:2016but,Paulos:2017fhb,Homrich:2019cbt}. In this approach to describing strongly coupled quantum field theories, we make elementary assumptions on the scattering processes, such as analyticity and unitarity, and use these to constrain typically low energy observables such as couplings and scattering lengths. While this approach has already led to many interesting results\cite{Doroud:2018szp,He:2018uxa,Cordova:2018uop,Paulos:2018fym,Guerrieri:2018uew,Gabai:2019ryw,Homrich:2019cbt,EliasMiro:2019kyf,Cordova:2019lot,Bercini:2019vme,Karateev:2019ymz,Correia:2020xtr,Bose:2020shm,Guerrieri:2020kcs,Guerrieri:2020bto,Bose:2020cod,Hebbar:2020ukp,Karateev:2020axc,Kruczenski:2020ujw,Tourkine:2021fqh,Guerrieri:2021ivu,He:2021eqn,EliasMiro:2021nul,Guerrieri:2021tak,Chen:2021pgx},\footnote{See \cite{Kruczenski:2022lot} for a recent overview.} it is clear that one of the outstanding fundamental theoretical questions in this subject is how to rigorously justify and determine the precise analyticity properties of S-matrices. While hard-fought progress can be made starting from axiomatic approaches to QFT \cite{Bros:1965kbd,Martin:1965jj,Sommer:1970mr}, we still find ourselves in the embarrassing position of being far from a satisfactory understanding of even the simplest case of two-to-two equal mass scattering.

One promising approach for dealing with this problem arises in the context of `rigid' holography \cite{Paulos:2016fap}. In this setup we consider a quantum field theory which is placed in an asymptotically AdS space, where metric fluctuations are assumed to be absent or suppressed, and study it using the standard tools of holography. It does not really matter how this placement is achieved -- i.e. for which curvature couplings, choice of boundary conditions, and so on -- as long as it {\em can} be done, in some way, for some sufficiently large AdS radius. This is because what we are really after is an understanding of the physics of the QFT for scales much smaller than the AdS radius, where such choices become irrelevant. 
Any such theory defines a set of boundary observables which behave in essentially every way as ordinary conformal correlators.\footnote{The CFTs lack a stress-tensor, but global Ward identities are still satisfied by correlators on conformally flat space.} Recent work \cite{Komatsu:2020sag} (building on \cite{Dubovsky:2017cnj}) argued that for gapped QFTs the S-matrix is essentially directly determined from such correlators\footnote{See also \cite{Hijano:2019qmi}.}. Hence, we may hope to understand how the properties of S-matrices arise from those of CFT correlators, which are under better non-perturbative and analytic control \cite{Luscher:1974ez,
Kravchuk:2020scc,
Kravchuk:2021kwe}. 

The present work serves as the starting point for carrying out this program in the context of gapped 2d QFTs on AdS$_2$, which are described by families of 1d CFT correlators. This setup is technically simpler since it avoids dealing with the intricacies of the physics of spin, yet sufficiently rich to describe a wealth of interesting systems, including integrable field theories. This case also allows us to establish a detailed dictionary between the conformal and S-matrix bootstraps, via the language of extremal functionals \cite{Mazac:2016qev,Mazac:2018mdx,Mazac:2018ycv}, enlightening us on both of them.

In detail, we will perform an in depth study of how 2d S-matrices, describing two-to-two equal particle scattering, emerge from 1d conformal correlators of identical scalar operators. Our approach is purely CFT-centric: by this we mean that we do not {\em explicitly} consider QFTs in AdS$_2$, but rather we will show that {\em any} family of CFT correlators with a sufficiently large gap in the OPE {\em automatically} leads to an S-matrix satisfying all the expected properties. These properties include crossing symmetry, analyticity away from the real axis, and unitarity. We will see that the latter corresponds to the statement that CFT correlators in Euclidean kinematics approach generalized free fields in the flat space limit. We will establish the analyticity properties by deriving a dispersion formula for the S-matrix starting from one for CFT correlators. We fully characterise the singularities of such S-matrices in terms of the CFT data, and in particular we carefully prove a certain phase shift formula for the S-matrix in physical kinematics.

We test our results on concrete perturbative examples. In particular we show how the bubble, triangle and box diagrams in AdS$_2$ can be bootstrapped in the flat space limit, and find direct agreement of our formulae with Feynman amplitudes. These examples will also allow us to illustrate how our results may fail when the gap assumption is violated. In S-matrix language this gap is necessary to avoid the possibility of anomalous thresholds. In the CFT we explain that the gap is necessary to avoid unbounded OPE coefficients, arising from the existence of unitary solutions to crossing without identity. We show that for the triangle and box diagrams the appearance of anomalous thresholds precisely correlate with unboundedness of the OPE in the flat space limit.

With our non-perturbative link between scattering and conformal physics well established, we discuss the CFT description of unitarity-saturating S-matrices, such as those describing integrable QFTs. We argue such S-matrices arise from families of extremal CFT correlators, which contain a single tower of operators in the OPE with dimensions determined by the phase shift of the S-matrix, and naturally saturate various bounds on the CFT data. In particular we establish a general mapping between S-matrix and CFT bootstrap optimization problems.  We discuss the implications of our construction to the understanding of UV completeness of S-matrices.

Let us now turn to a more detailed technical description of our main results.

\subsection*{Summary and outline}

\subsubsection*{Setting up the mapping}
We will consider a family of CFT$_1$ correlators of four copies of the same field $\phi$, and parameterized by its dimension $\Df$:%
\bea
\langle \phi(x_1)\phi(x_2)\phi(x_3)\phi(x_4)\rangle=\frac{\mathcal G(z)}{|x_{13}|^{2 \Df}\,|x_{24}|^{2 \Df}}\,,\qquad z=\frac{x_{12}x_{34}}{x_{13}x_{24}}\,,\quad x_{ij}=x_i-x_j\,.
\eea
We should have in mind the prototypical example where this family of correlators arises by taking a gapped 2d QFT in AdS$_2$ and considering AdS boundary observables in such a theory.~\footnote{Conformal correlators are obtained by pushing AdS bulk insertions towards the AdS boundary \cite{Klebanov:1999tb}.} In this case each correlator can be labeled by $\Df\sim m R_{\mbox{\tiny AdS}}$, where $m$ is the mass of some stable particle in the theory. The flat space limit corresponds to taking large AdS radius keeping physical masses fixed, and since scaling dimensions behave as $\Delta\sim M R$ we must take them all to be large. In particular we must send $\Df\to \infty$.  It was argued in \cite{Komatsu:2020sag} that in this limit one may extract the S-matrix of the QFT from the family of CFT correlators, more or less directly: the correlator becomes the S-matrix after a suitable identification of kinematic quantities. However, the limit is subtle and must be done carefully. The purpose of this work is to understand this procedure in detail. 

The concrete mapping between correlators and S-matrices is defined as:
\bea
S(s)= \mathcal F[\mathcal G(z_s)]\,, \qquad z_s:=1-s/4\,, \label{eq:flatlim}
\eea
where the operator $\mathcal F$ describes taking the flat space limit.%
\footnote{In our present 1d/2d context, our conventions are such that our S-matrix would match that of a 2-to-2 scattering process of identical particles of unit mass, written as
\bea
_{\mbox{\tiny out}}\langle p_4,p_3|p_1,p_2\rangle_{\mbox{\tiny in}}= (2\pi)^2 4 E_1 E_2 \left[\delta^{(2)}(p_1-p_3)\delta^{(2)}(p_2-p_4)+\delta^{(2)}(p_1-p_4)\delta^{(2)}(p_2-p_3)\right] S(s)\,, \qquad 
\eea
where $s=-(p_1+p_2)^2$. 
}
The action of $\mathcal F$ involves two steps. Firstly, we take the limit $\Df \to \infty$ of the correlator in a kinematic region (to be found) where the limit is finite\footnote{Note that we are implicitly assuming that every CFT datum has a well defined limit, but which does not have to be finite.}. Secondly, we must analytically continue the answer to the desired kinematic point. We illustrate this procedure schematically in figure~\ref{fig:schematic}.
\begin{figure}[t]
\centering
\includegraphics[width=1\textwidth]{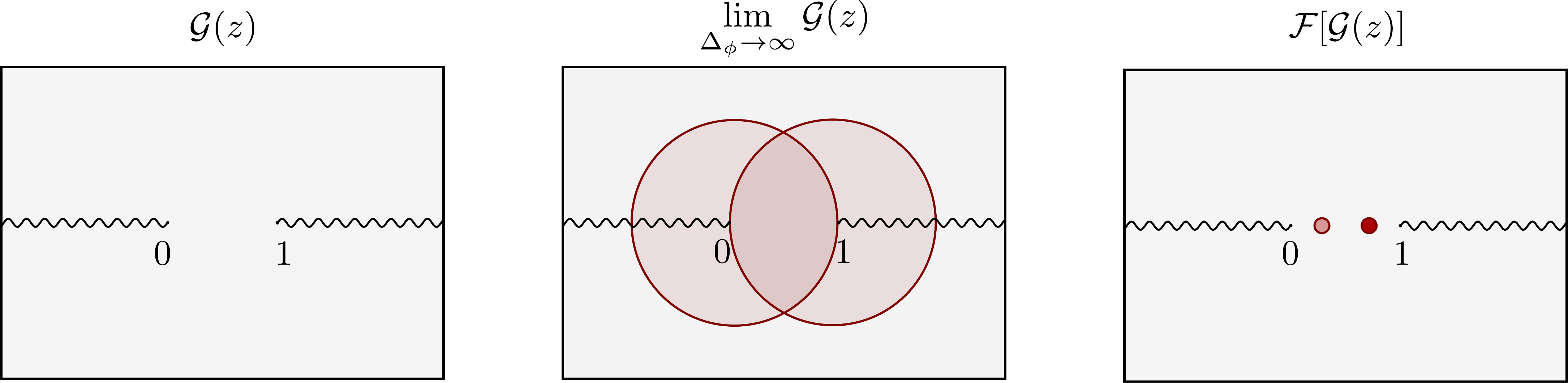}\vspace{0.2cm}
\caption{Schematic derivation of the flat space limit of the correlator. We start on the left with the conformal correlator $\mathcal G(z)$ which has branch points at $z=0,1$. We then take the limit when all dimensions are large; this limit is not well defined in certain kinematic regions as shown in the red blobs in the middle panel. Finally we consider the flat space limit where we analytically continue from the safe to the problematic regions to obtain the S-matrix; after this continuation we may encounter singularities in the S-matrix as shown in the right panel for the case of a single bound state pole.
}
\la{fig:schematic}
\end{figure}
To see how this works in a simple example, consider the two families of correlators
\bea
\mathcal G^{\pm}(z)=\pm 1+z^{-2\Df}+(1-z)^{-2\Df}\,.
\eea
These are boundary correlators for a free scalar or fermion field in AdS$_2$, or generalized free fields, with the $+/-$ sign for the Boson/Fermion. Since these are free fields we expect the corresponding $S$-matrices to be $\pm 1$. To extract this from the correlator, we act with $\mathcal F$. First we go to a region where the limit $\Df$ exists:
\ba
\lim_{\Df\to \infty} \mathcal G^{\pm}(z)&=
\begin{cases}\pm 1,& \mbox{if}\quad |z|\geq 1 \And |1-z|\geq 1\\
\infty, &\qquad \mbox{otherwise\,.}
\end{cases}
\ea
Secondly, we analytically continue to all $z$. This step is  trivial here, and we indeed have $S(s)=\mathcal F[\mathcal G]=\pm 1$. In general this second step is harder to perform, and we may encounter singularities hidden inside the ``blobs'' of figure~\ref{fig:schematic}.

Although the 2d QFT motivation described above is an important inspiration, we will want to remain agnostic about the origin of the family of correlators, and formulate everything purely in CFT language. In this sense, we can take equation \reef{eq:flatlim} as a {\em definition} of a function $S(s)$, and ask about its properties depending on our assumptions about the CFT. 
Our goals are two-fold:%
\begin{itemize}
    \item Determine the analytic properties of $S(s)$ in the complex $s$-plane.
    \item Describe $S(s)$, and in particular its singularities, purely in terms of CFT data.
\end{itemize}

\subsubsection*{The flat space limit}
To achieve these goals we need to characterize the action of $\mathcal F$. Our main tool for doing so is the Polyakov bootstrap \cite{Polyakov:1974gs,Gopakumar:2016cpb,Gopakumar:2016wkt,Gopakumar:2018xqi,Mazac:2018ycv}, which states that any CFT correlator has a representation of the form:
\bea
\mathcal G(z)=\mathcal P_0(z)+\sum_{\Delta\geq \Delta_0} a_{\Delta} \mathcal P_{\Delta}(z)\,.
\eea
This is closely related to the usual OPE expansion ($a_{\Delta}$ are the OPE coefficients squared), but where conformal blocks are replaced by Polyakov blocks, which are essentially crossing symmetric sums of Witten exchange diagrams, as we review in section \ref{sec:polyflat}. In particular, their analyticity properties are the same as those of ordinary CFT correlators. Although in general Polyakov blocks are complicated functions, we will be able to compute them explicitly in the flat space limit, allowing us to describe the action of $\mathcal F$ on each of them individually. For instance one finds for $\Delta_b<2\Df$ that
\bea
\mathcal F[a_{\Delta_b}\mathcal P_{\Delta_b}(z_s)]\propto \frac{g_b^2}{s_b-s}+\frac{g_b^2}{s+s_b-4}\,, \quad s_b=(\Delta_b/\Df)^2\,, \quad g_b^2\propto a_{\Delta_b}\,,
\eea
which illustrates that after analytic continuation one can find singularities that were not initially there. 

Next, we must argue that the $\mathcal F$ operation commutes with the infinite sum over states. This can be shown by deriving certain bounds on the OPE coefficients. To prove these bounds it will be necessary to make certain assumptions on the CFT spectrum. In practice it will be sufficient to demand that the lowest state in the OPE (above identity) should have dimension $\Delta_0>\sqrt{2}\Df$ as $\Df\to \infty$. Such a condition is expected, since below some critical gap there exist unitary solutions to crossing without identity which can always be added with an arbitrarily large coefficient to any given CFT correlator. Remarkably, in $S$-matrix language this assumption translates into demanding that particle production must begin for $s>2$, which is a sufficient condition for the absence of anomalous thresholds.  We will investigate this link in several perturbative examples.

Under these assumptions, the sum over Polyakov blocks can be performed and becomes a dispersion formula for $S(s)$:
\begin{equation}\label{CFTdispersiondensity0}
\boxed{\begin{aligned}[t]
S(s)&=1+\int_{s_0}^{4m^2} \ud s' \, \widetilde K(s,s')\,\tilde \rho(s')-\int_{4m^2}^{\infty} \ud s' \, K(s,s')\,\rho(s')\\
\rho(s)&= \lim_{\Df\to \infty}\Df^{-\alpha}\sum_{\Delta \in B_{\alpha}(s)}\left(\frac{a_{\Delta}}{\af}\right)\,4 \sin^2\left[\frac \pi 2(\Delta-2\Df)\right]\\
\tilde \rho(s)&=\lim_{\Df\to \infty}\Df^{-\alpha} \sum_{\Delta \in B_{\alpha}(s)} \left(\frac{a_{\Delta}}{\tilde a_{\Delta}^{\mbox{\tiny free}}}\right)\,.
\end{aligned}
}
\end{equation}
Here both $K,\widetilde K$ are simple Cauchy-type kernels, and $a_{\Delta}^{\ttext{free}},\tilde a_{\Delta}^{\ttext{free}}$ are both positive in the associated range of integration and determined by the OPE density of a generalized free field. The discontinuities $\rho, \tilde \rho$ are also completely specified as certain averages of the CFT data. Further details on these expressions will be given in the main text. The dispersion formula achieves our main goals: it proves that $S(s)$ is an analytic function in the complex-$s$ plane with singularities determined by the CFT data.

\subsubsection*{Phase shift formula and extremality}

It is of interest to have an expression of the S-matrix evaluated for physical kinematics:
\bea
\Sp(s):= \lim_{\epsilon\to 0^+} S(s+i \epsilon)\,, \qquad s>4m^2\,.
\eea
To compute this we must take two limits on the CFT correlator in a specific order: first large $\Df$ and then small $\epsilon$. In section \ref{physlim} we will prove that these limits can be commuted. The CFT dispersion formula for $S(s)$ already guarantees validity of the real part of the phase shift formula. To complete the proof we will derive a dispersion relation for CFT correlators involving their imaginary part (as opposed to the double discontinuity) and then taking the flat space limit. Computing the physical S-matrix with the order of limits reversed leads to a particularly nice expression for the physical S-matrix known as the ``phase shift formula'':
\bea
\Sp(s)=\lim_{\Df\to \infty}\sum_{\Delta> 2\Df} 2 \left(\frac{a_{\Delta}}{\af}\right) e^{-i\pi(\Delta-2\Df)}\, \Nh(
\Delta,s)
\eea
with $
\Nh(\Delta,s)$ approaching a delta function imposing $s=(\Delta/\Df)^2$.\footnote{This formula was first proposed in \cite{Paulos:2016fap} in higher dimensions. It was shown to follow from the OPE expansion of the correlator in \cite{Komatsu:2020sag}, but it required assuming the commuting of limits. Our main result is thus that we prove this assumption is justified for the particular case of 1d CFTs.}
Thus for these kinematics the full S-matrix is a certain average of the CFT data. Importantly, the bounds on OPE coefficients mentioned previously guarantee unitarity of the S-matrix. Schematically, we have
\bea
a_{\Delta}\sim \af \qquad \Rightarrow |\Sp(s)|\leq 1
\eea
The phase shift formula also allows us to better understand what unitarity saturation corresponds to in the CFT language: such S-matrices map onto CFT correlators whose OPE in the flat space limit is effectively (or exactly) described by a single tower of operators with dimensions described by the S-matrix phase shift. 

\subsubsection*{Outline}
The outline of this work is as follows. In section \ref{sec:polyflat} we briefly review Polyakov blocks and their computation using master functionals \cite{Paulos:2020zxx} before studying them in the flat space limit. Section \ref{sec:flatgeneral} is concerned with obtaining the action of $\mathcal F$. We first determine general bounds on the OPE density for states both above and below $2\Df$ which acts as the physical threshold in S-matrix language. Using these bounds we obtain the flat space limit of the correlator in terms of a dispersion formula for the S-matrix. In section \ref{physlim} we study the S-matrix directly in physical kinematics and prove that it satisfies a phase shift formula. Section \ref{sec:perturbative} considers various perturbative checks of our formulae. In particular we study AdS bubble, triangle and box diagrams and show that they are correctly described by our formalism. In the regime where our CFT spectrum assumptions are violated, we find that the OPE decomposition of such diagrams contains a large unbounded component coinciding with the appearance of anomalous thresholds. In section \ref{sec:nonpert} we study various consequences of our mapping between CFT correlators and S-matrices. We argue that any S-matrix which is a finite product of CDD factors may always be obtained as the flat space limit of certain families of CFT correlators. In particular we propose these families can always be chosen to be as extremal, i.e. as saturating a CFT bound. The section concludes by establishing a precise link between the functional based conformal bootstrap and the dual S-matrix bootstrap. We conclude with a discussion of the limitations and implications of our construction for understanding UV completeness of S-matrices, and future research directions. The paper is complemented by several technical appendices.

{\em Note}: while this paper was being concluded we became aware of \cite{Knop:2022} which discusses topics related to the present work.

\section{Polyakov blocks in the flat space limit}
\label{sec:polyflat}

The main goal of this section is to compute Polyakov blocks in the flat space limit. In the context of 1d CFTs, there are two simple kinds of Polyakov blocks $\mathcal P^{\pm}_\Delta$, corresponding to bosonic/fermionic boundary operators with dimension $\Df$ exchanging a scalar/pseudoscalar state bulk field with dimension $\Delta$ in the AdS bulk. Polyakov blocks are crossing symmetric combinations of Witten exchange diagrams together with carefully chosen contact terms, and in principle we could use this to compute them. Instead, we will follow a different route by using their representation in terms of master functionals \cite{Paulos:2020zxx}.

\subsection{Master functionals}\label{Mfunctional}
Consider the OPE expansion for a correlator
\bea
\mathcal G(z)=\sum_{\Delta\geq 0} a_{\Delta} G_{\Delta}(z|\Df)\,, \qquad  G_{\Delta}(z|\Df)=z^{\Delta-2\Df}\ _2F_1(\Delta,\Delta,2\Delta,z)\,,
\eea
where $a_{\Delta}:= \lambda_{\phi \phi \mathcal O_{\Delta}}^2$ are the OPE coefficients squared. 
Crossing symmetry is the statement $\mathcal G(z)=\mathcal G(1-z)$, which using the OPE becomes:
\beqa
\sum_{\Delta} a_{\Delta} F_{\Delta}(z|\Df)=0\,, \qquad 
F_{\Delta}(z|\Df)=G_{\Delta}(z|\Df)-G_{\Delta}(1-z|\Df)\,.
\eeqa
We will often drop the explicit dependence on $\Df$ below.
One way to extract information from the crossing equation is by defining suitable linear functionals \cite{Rattazzi:2008pe}. An interesting class is defined by the following ansatz: \cite{Mazac:2016qev,Mazac:2018mdx,Mazac:2018ycv}:
\bea
\Omega[F_{\Delta}]\equiv \Omega(\Delta)\equiv \frac 12\int_{\frac 12}^{\frac 12+i\infty} \ud z f(z)  F_{\Delta}(z)+\int_{\frac 12}^1 \ud z g(z) F_{\Delta}(z) \la{fnl_action}\,.
\eea
Acting with such functionals on the crossing equation leads to sum rules on the CFT data\footnote{Under certain conditions, see \cite{Rychkov:2017tpc}.} %
\bea
\Omega\left[\sum_{\Delta\geq 0} a_{\Delta}F_{\Delta}(z|\Df)\right]=\sum_{\Delta\geq 0} a_{\Delta}\Omega(\Delta)=0\,.
\eea
The bosonic/fermionic master functionals are families of functionals labeled by a cross-ratio $w$ and denoted $\Omega^+_w, \Omega^-_w$. They were introduced in \cite{Paulos:2020zxx}, where the reader can find further details on their definition and properties. They correspond to choices of kernels that will lead to particularly nice functional actions $\Omega(\Delta)$. Denoting these kernels $f^{\pm}_w, g^{\pm}_w$, they are defined by setting
\bea
g_w^{\pm}(z)=\hat g_{w}^{\pm}(z)\pm\delta(z-w)\,,\quad \hat g_{w}^{\pm}(z)=\pm (1-z)^{2\Df-2} f_w^{\pm}(\mbox{$\frac{1}{1-z}$}) \la{g_master}
\eea
together with the gluing condition\footnote{In this work we define:
\ba
\mathcal R_z f(z)\equiv \lim_{\epsilon\to 0^+} \frac{f(z+i\epsilon)+f(z-i\epsilon)}2\,,\qquad
\mathcal I_z f(z)\equiv \lim_{\epsilon\to 0^+} \frac{f(z+i\epsilon)-f(z-i\epsilon)}{2i}\,.
\ea
}
\bea
\mathcal R_z f^{\pm}_w(z)+g^{\pm}_w(z)+g^{\pm}_w(1-z)=0\,, \qquad \mbox{for}\quad z\in(0,1)\,. \la{eq:gluing}
\eea
There are also certain boundary and analyticity conditions that we must impose to ensure the solution to these constraints is unique. 

The reason we care about such  functionals here is due to their intimate connection to Polyakov blocks. As it turns out we have
\bea
\Omega^{\pm}_w(\Delta)=\mp \[ \mathcal P_{\Delta}^{\pm}(w)- G_{\Delta}(w|\Df) \]\,. \label{eq:polymaster}
\eea
The functional sum rules associated to master functionals are therefore the statement of the Polyakov bootstrap:
\bea
\sum_{\Delta} a_{\Delta} \Omega^{\pm}_w(\Delta)=0 \Leftrightarrow \mathcal G(w)=\sum_{\Delta} a_{\Delta} \mathcal P_{\Delta}^{\pm}(w)
\eea
Equivalently, the same sum rules can be re-expressed as dispersion relations for CFT correlators. These take the form \cite{Paulos:2020zxx}
\ba
\overline{\mathcal G}(w)&=-\int_0^1 \ud z \hat g_w^{+}(z) \mbox{dDisc}^+ \overline{\mathcal G}(z) \\
\underline{\mathcal G}(w)&=+\int_0^1 \ud z \hat g_w^{-}(z) \mbox{dDisc}^- \underline{\mathcal G}(z) \label{eq:disprels}
\ea
with
\bea
\overline{\mathcal G}(w):= \mathcal G(w)-\sum_{0\leq\Delta\leq 2\Df} a_{\Delta} \mathcal P^+_\Delta(w)\,, \qquad \underline{\mathcal G}(w):= \mathcal G(w)-\sum_{0\leq\Delta\leq 2\Df-1} a_{\Delta} \mathcal P^-_\Delta(w)
\eea
and
\bea
\mbox{dDisc}^{\pm} \mathcal G(z):=\mathcal G(z)\mp(1-z)^{2\Df} \mathcal R_z \mathcal G(\mbox{$\frac z{z-1}$})\,.
\eea
It follows from these results that if we know the master functional kernels we can compute the Polyakov blocks. While we do not have analytic expressions for the kernels in general, we do have them when $\Df$ becomes large:
\bea
f^{\pm}_w(z) \underset{\Df\to \infty}=\pm \frac{2}{\pi} \sqrt{\frac{w(1-w)}{z(z-1)}}\frac{z-1/2}{(z-w)(z-1+w)}\,, \qquad z>1\,. \la{f_master}
\eea
As a check, notice that this satisfies equation \reef{eq:gluing}, which becomes 
\begin{equation}
\pm\mathcal R_z f^{\pm}_w(z)\sim -\delta(w-z)-\delta(1-w-z)
\end{equation}
since $\hat g_w(z)$ is exponentially suppressed relative to $f_w(z)$ at large $\Df$.

After these preliminary remarks, we are now ready to compute Polyakov blocks in the flat space limit. But first let us point out that for the special case $\Delta=0$ there is actually no computation to perform. This is because the Polyakov blocks $\cP^\pm_0(w)$ are nothing but the generalized free boson/fermion correlators:\footnote{This is shown in detail for instance in \cite{Mazac:2018ycv}.}
\beq
\cP^\pm_0(w)=\mathcal G^\pm(z)=\pm1+z^{-2\dphi}+(1-z)^{-2\dphi}\,.
\eeq
As for the other cases,  we will be interested in the limit where both $\Delta$ and $\Df$ are large and the limit of the ratio $\Delta/\Df$ is fixed and strictly different from two. It is convenient then to split the computation according to whether $\Delta$ is larger or smaller than `threshold', which is $2\Df$. This is because the master functional actions admit different representations depending on these two cases.

\subsection{\texorpdfstring{Polyakov blocks for $
\Delta>2\Df$}{Polyakov blocks for dphi2}}\label{sec:polflat}

While the action of the master functional is generally defined by equation \reef{fnl_action}, this is not necessarily the most convenient form. Under certain conditions in $\Delta$ it is possible to deform the contours in that definition to arrive at simpler expressions for the functional actions \cite{Mazac:2018mdx}. In this way, and using \reef{eq:polymaster}, we arrive at the following expressions: 
\ba\label{eq:Pdisp}
\mathcal P^+_{\Delta}(w)&=-2 \sin^2\[\tfrac{\pi}{2}(\Delta-2\dphi)\] \int\limits_0^1 \ud z\, \hat g_w^+(z)\, G_\Delta(z|\Df)\,,&\quad \Delta&>2\Df\\
\mathcal P^-_{\Delta}(w) &=+2 \cos^2\[\tfrac{\pi}{2}(\Delta-2\dphi)\] \int\limits_0^1 \ud z\, \hat g_w^-(z)\, G_\Delta(z|\Df)\,,&\quad \Delta&>2\Df-1
\ea
In passing notice that these expressions, together with validity of the Polyakov bootstrap, imply the dispersion relations \reef{eq:disprels}. 

Let us now take the large $\Df$ limit. Since we already know the functional kernels, we just need to study that limit for the conformal block. As it turns out, the exponential suppression of the $\hat g$ kernels can be compensated by an exponential growth of the blocks. At large $\Delta$ we have \cite{Hogervorst:2013sma}:
\bea
(1-z)^{2\Df}G_{\Delta}(z|\Df)\underset{\Delta\to \infty}{=} \left(\frac{1-z}{z}\right)^{2\Df} \frac{\[4\rho(z)\]^\Delta}{\sqrt{1-\rho(z)^2}}\,, \qquad \rho(z):=\frac{1-\sqrt{1-z}}{1+\sqrt{1-z}}\,.
\label{eq:radial}
\eea
By further taking $\Df\to \infty$ one finds that for $\Delta>2\Df$ (with $\Delta/\Df$ fixed) the conformal block is exponentially suppressed except in a narrow region around a specific value of $z$. This can be expressed in the following way. Let us define:%
\bea
\mathcal N_{\Df}(z,s):=\sqrt{\frac{\Df s^3}{32 \pi(s-4)}} \exp\left[-\frac{\Df s^3}{32(s-4)}\left(z-\frac{s-4}s\right)^2\right]
\eea
which is a Gaussian in the $z$ variable centered at $z=(s-4)/s$ of width $O(\sqrt{\Df})$, tending to a delta function in $z$. Alternatively, define 
\bea
\Nh(\Delta,s)=\sqrt{\frac{2}{\pi \Df (s-4)}}\exp\left[-2\frac{(\Delta-\sqrt{s}\Df)^2}{\Df (s-4)}\right]\,, \label{nh} 
\eea
which is also Gaussian but now in the $\Delta$ variable, with width $O(\sqrt{\Df})$, and also approaching a delta function. We then have
\ba
(1-z)^{2\Df}G_{\Delta}(z|\Df)&\underset{\Delta,\Df\to \infty}{\sim} \frac{16}{\Df \, s_{\Delta}^{\frac 32}} \frac{\mathcal N_{\Df}(z, s_{\Delta})}{a_{\Delta}^{\mbox{\tiny free}}}\,, \qquad s_{\Delta}\equiv \left(\frac{\Delta}{\Df}\right)^2\,\\[12pt]
&\underset{\Delta,\Df\to \infty}{\sim}  2\,\frac{\Nh(\Delta,\mbox{$\frac{4}{1-z}$})}{a_{\Delta}^{\mbox{\tiny free}}}
\label{eq:gaussian}
\ea
where $\sim$ means up to exponentially suppressed terms and $a_{\Delta}^{\mbox{\tiny free}}$ is the OPE density for a generalized free field:\footnote{By this we mean in particular that
\bea
\mathcal G^{\pm}(w)=G_0(w|\Df)+\sum_{n=0}^\infty a_{\Delta_n^{\pm}}^\ttext{free} G_{\Delta_n^{\pm}}(w|\Df)\,,\qquad \Delta_n^{\pm}=2\Df+2n+\frac 12\mp \frac 12\,.
\eea

}
\ba\la{a_free}
	a_{\Delta}^{\mbox{\tiny free}}=\frac{2 \Gamma(\Delta)^2}{\Gamma(2\Delta-1)}\, \frac{\Gamma(\Delta+2\Df-1)}{\Gamma(2\Df)^2 \Gamma(\Delta-2\Df+1)}\,.
\ea
Since in the limit of infinite $\Df$ the Gaussian becomes a delta function the integrals \reef{eq:Pdisp} trivialize and we get:%
\ba\label{Pblockabove}
\cP_\Delta^{\pm}(w|\dphi)\underset{\Df \to \infty}{=}\mp\frac{4\sqrt{s_\Delta}}{ \pi\Df}\frac{\sin^2\[\tfrac{\pi}{2}(\Delta-2\dphi)\]}{a_{\Delta}^{\mbox{\tiny free}}} \frac{\sqrt{w(1-w)}}{\sqrt{z_\Delta(z_\Delta-1)}}\,\frac{z_\Delta-1/2}{(z_\Delta-w)(w+z_\Delta-1)}\,,
\ea
where $z_{\Delta}=1-s_{\Delta}/4$.

%

\subsection{\texorpdfstring{Polyakov blocks for $\Delta<2\Df$}{Polyakov blocks below 2dphi} \la{subsec:Polyakovbelow}}
\begin{figure}[t]
\centering
\includegraphics[width=1\textwidth]{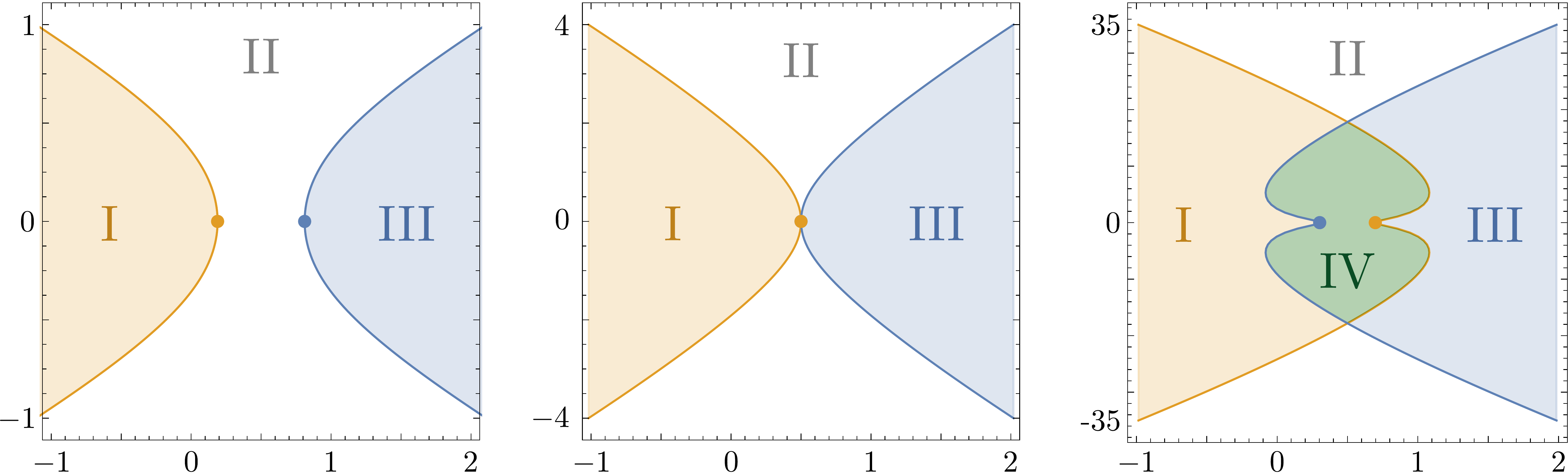}\vspace{0.2cm}
\caption{In the flat space limit Polyakov blocks become piecewise functions defined in generically four distinct regions. The boundaries of these regions are set by the conditions $\mbox{arg}\left[ G_{\Delta_b}(w|\Df)\right]=0 \cup \mbox{arg} \left[G_{\Delta_b}(1-w|\Df)\right]=0$. We show these regions in the complex $w$ plane for $\Delta_b=1.8$ (left), $\Delta_b=\sqrt2$ (center) and $\Delta_b=1.1$ (right).
}
\label{fig:regions}
\end{figure}

Now we turn to the flat space limit of Polyakov blocks with dimension $\Delta<2\dphi$. To remind us of this constraint we will sometimes denote such dimensions as $\Delta_b$ ('{\em b}' stands for bound state). The computation will be simpler to consider for general complex cross-ratio $w$ -- the special case where $w$ is real is treated in appendix \ref{app:realw} but leads to results consistent with those here. For complex $w$ the definition of the master functional action is almost exactly as before, but now the kernel $g_w^{\pm}$ no longer contains the delta function piece. However, the kernel $f_w^\pm$ still has a pair of poles at $w$ and $1-w$ with unit residue. The functional action for general $\Df$ and $\mbox{Re}[w]<\frac 12$
can be written as we have seen before in \eqref{fnl_action}:
\bea
\Omega^{\pm}_w(\Delta_b)=\frac 12\int_{\frac 12}^{\frac 12+i\infty} \ud z f_{w}^{\pm}(z)  F_{\Delta}(z)+\int_{\frac 12}^1 \ud z g^{\pm}_w(z) F_{\Delta}(z) 
\eea
Note that we can obtain the result for general values of $w$ by deforming the contour of integration to avoid the poles in $f_w^{\pm}$. In the large $\Df$ limit we can safely ignore the exponentially suppressed contribution of $g_w^{\pm}$ above if $\Delta_b<2\Df$. As for the $f_w^{\pm}$ kernel, let us write
\bea
f_w(z)=\frac{\hat f_w(z)}{\sqrt{z(z-1)}}\,.
\eea
Then the functional action can be written as
\bea
\Omega_w^{\pm}(\Delta_b)=\frac i2 \int_{\frac 12-i\infty}^{\frac 12+i\infty}\ud z\frac{\hat f_w(z)}{\sqrt{z(1-z)}}\,  G_{\Delta_b}(1-z|\Df)
\eea
Let us set $z_b=\frac{m_b^2}4$ with $m_b=\Delta_b/\Df$.
In the large $\dphi$ limit with $m_b$ fixed one finds by using the expressions of the previous subsection that the conformal block factor has a saddle point at $z=z_b$. To do the integral we can therefore deform the contour of integration to the steepest descent contour, which is defined by the condition%
\bea
\mathcal I_{z}\left\{2\sqrt{z_b} \log[4\rho(1-z)]-2\log(1-z)\right\}=0\,,
\eea
where $\rho(z)$ is defined in \eqref{eq:radial}. However, we must be careful since when doing this we may cross the poles of $f_w^{\pm}$. %
Since there are two poles at $w$ and $1-w$, this splits the complex-$w$ plane into several regions, bounded by the contour above and its image under crossing, as shown in figure \ref{fig:regions}. We therefore find
the result for the functional action is
\bea
\Omega_w^{\pm}(\Delta_b)\underset{\Delta_b,\Df\to \infty}{=}
\frac{m_b}{2\dphi} \frac{\mathcal I_{z}f^{\pm}_{w}(z_b)}{\tilde a_{\Delta_b}^\ttext{free}}+(\ldots) 
\eea
where the first term comes from the saddle point approximation, and we introduced 
\beqa
\tilde a_{\Delta_b}^\ttext{free} &=& \frac{a_{\Delta_b}^\ttext{free}}{2\sin\[\pi \dphi(2-m_b)\]}\qquad (\geq 0\quad \mbox{for}\quad \sqrt{2}\Df<\Delta_b<2\Df)%
\,.\la{a_tilde}
\eeqa
As for the pole contributions, represented as the ($\ldots$), it is easiest to write them in terms of the Polyakov block. Using the relation between the latter and the master functional actions we find:
\ba
\mathcal P_{\Delta_b}^{\pm}(w)\underset{\Delta,\Delta_b\to \infty}{=}\frac{m_b}{\pi \dphi \tilde a_{\Delta_b}^\ttext{free}} \frac{\sqrt{w(1-w)}}{\sqrt{z_b(1-z_b)}}\,\frac{z_b-1/2}{(z_b-w)(w+z_b-1)}\,+E_{\Delta_b}(w|\Df) \label{eq:pbfull}
\ea
with the crossing symmetric $E_{\Delta_b}$ satisfying:
\bea
E_{\Delta_b}(w|\Df)=\left\{
\begin{array}{l r}
G_{\Delta_b}(w|\Df) &   w\in \text{I}\\
0 & w\in \text{II}\\
G_{\Delta_b}(1-w|\Df) &   w\in \text{III}\\
G_{\Delta_b}(w|\Df)+G_{\Delta_b}(1-w|\Df) & w \in \text{IV}
\end{array}
\right.
\eea
where the regions I through IV are shown pictorially in figure \ref{fig:regions}. In particular we have for real $w$ 
\bea
\left\{\begin{array}{l c r}
w\in \text{I}\, & \Leftrightarrow &   0<w<\text{min}(z_b,1-z_b)\\
w\in \text{II} & \Leftrightarrow & 1-z_b<w<z_b\\
w \in \text{III} & \Leftrightarrow &   \text{max}(z_b,1-z_b)<w<1\\
w \in \text{IV} & \Leftrightarrow & z_b<w<1-z_b
\end{array}
\right.
\eea
This completes our calculation. As we will see in the next section, the contributions to the flat space limit arising from these Polyakov blocks are obtained by throwing out $E_{\Delta_b}(w|\Df)$.

\section{Flat-space limit: general kinematics}
\label{sec:flatgeneral}
In this section we will derive the flat space limit of the CFT correlator. For this to be possible we must have control over the behaviour of the OPE in the flat space limit. For states below $2\Df$ such bounds already exist \cite{Paulos:2016fap,Mazac:2018mdx}, under certain assumptions which we will spell out. For states above $2\Df$ we will prove new bounds, which imply that on average the OPE density of any CFT must universally match that of a generalized free field. With these results we will show that the Polyakov bootstrap translates into a dispersion formula for the S-matrix $S(s)$, establishing its analyticity properties in the complex $s$-plane.

\subsection{OPE bounds}
\subsubsection{States below threshold}
\label{sec:boundbelow} 

We would like to determine the regions where the single bound state contribution $a_{\Delta_b} \cP_{\Delta_b}(w)$ is well defined in the flat space limit. In order to do this we need to have a bound on the OPE coefficient $a_{\Delta_b}$ (recall $\Delta_b<2\dphi$). Such a bound was obtained analytically in \cite{Mazac:2018mdx} (following numerical bootstrap computations in \cite{Paulos:2016fap}, and similar results for the S-matrix bootstrap \cite{Paulos:2016but}). Let us briefly review some aspects of the derivation. 

Suppose we have some CFT correlator whose OPE has support on the identity operator and a set $\mathcal S$ of scaling dimensions. Then we can obtain bounds on OPE coefficients by constructing functionals with suitable positivity properties. Any functional $\omega$ leads to a sum rule:\footnote{We have written $\omega$ instead of $\Omega$ to emphasize that the functional giving an optimal OPE coefficient is different from the master functional.}
\bea
\omega(0)+\sum_{\Delta\in \mathcal S} a_{\Delta} \omega(\Delta)=0\,.
\eea
Suppose now that the functional action satisfies the positivity constraints
\ba
\omega(\Delta)&\geq 0\quad \mbox{for all} \quad \Delta \in \mathcal S\,.
\ea
Then the sum rule immediately implies the bounds:
\ba
\sum_{\Delta\in \mathcal S'} a_{\Delta} \omega(\Delta)&\leq -\omega(0)\quad \mbox{for any}\quad \mathcal S'\subset \mathcal S 
\qquad
\Rightarrow \qquad a_{\Delta_b}&\leq -\frac{\omega(0)}{\omega(\Delta_b)}\,, \quad \Delta_b\in \mathcal S\,.
\ea
Clearly, these bounds will in general depend on the set of states allowed on the OPE. We will find it useful to distinguish two particular sets of assumptions on this set:
\begin{itemize}
\item{\bf Weak OPE condition:} The set $\mathcal S$ does not contain any pair of states $\Delta_1,\Delta_2$ such that
\bea
s_{\Delta_1}=4-s_{\Delta_2}\,, \qquad s_{\Delta}\equiv \left(\frac{\Delta}{\Df}\right)^2\,.
\eea
\item{\bf Strong OPE condition:} The set $\mathcal S$ only contains states for which
\bea
s_{\Delta}>2\qquad \Leftrightarrow \qquad  \Delta>\sqrt{2}\Df\,.
\eea
\end{itemize}
To be more precise we have in mind imposing these conditions in the flat space limit, so that equations are meant to hold in the limit $\Df\to \infty$. Clearly the strong condition implies the weaker one. For simplicity, we will for the most part work with the stronger assumption, although essentially all our results can be straightforwardly generalized for the weaker one. Note that violation of the weak OPE condition can completely undo most results in this work: a more detailed examination is conducted in section~\ref{sec:anom}.

For CFTs satisfying the strong OPE condition, we can find an analytic functional satisfying the necessary positivity conditions in the flat space limit. It is called the sine-Gordon functional $\omega^{sG}$ and it can be obtained by choosing kernels \cite{Mazac:2018mdx}:
\ba
f^{sG}_{z_b}(z)&=\frac{2}{\pi}\frac{\sqrt{z_b(1-z_b)}}{\sqrt{z(z-1)}}\frac{z-1/2}{(z-z_b)(z-1+z_b)} \frac{1}{S_b^{sG}(s_z)}\,,\\
g^{sG}_{z_b}(z)&=(1-z)^{2\Df-2}|f^{sG}_{z_b}(\mbox{$\frac{1}{1-z}$})|\,.\label{eq:fsG}
\ea
with the sine-Gordon S-matrix:
\beq
S^\text{sG}_b(s)=\frac{\sqrt{s(4-s)}+\sqrt{m_b^2(4-m_b^2)}}{\sqrt{s(4-s)}-\sqrt{m_b^2(4-m_b^2)}}\,,\qquad (m_b^2=z_b/4)\,.
\eeq
The action of the functional $\omega^{sG}$ implies the bound
\beq
\sum\limits_{\sqrt{2}\Df<\Delta_i<2\dphi} \frac{m_i}{2\dphi} \, \(\frac{a_{\Delta_i}}{\tilde a_{\Delta_i}^\ttext{free}}\) \,  \,\mathcal I_{z_i}f_{z_b}(z_i)
\leq 1\,,\qquad z_i=\frac 14\left(\frac{\Delta_i}\Df\right)^2\,, \label{eq:boundboundstates}
\eeq
and in particular
\beq
\frac{2}{\pi} \frac{m_b}{\dphi}\(\frac{a_{\Delta_b}}{\tilde a_{\Delta_b}^\ttext{free}}\) \leq \(g^{sG}_b\)^2\,, \qquad\qquad\qquad \Delta_b=m_b\dphi<2\dphi  \,, \la{boundOPEbelow}
\eeq
where $\(g^{sG}_b\)^2= \left|2m_b^2 (4-m_b^2)(m_b^2-2)^{-1}\right|$ is the squared cubic coupling coming from the sine-Gordon breathers' S-matrix. %
This result nicely makes contact with
the flat space S-matrix bounds, since $g^{sG}_b$ is the maximum cubic coupling for a bound state of mass $m_b$ \cite{Creutz:1972ikj}. 

This result can be generalized to any spectrum $\mathcal S$ satisfying the weaker OPE condition. In this case one can obtain a suitable functional by replacing the sine-Gordon S-matrix by a carefully chosen product of CDD poles. The point we wish to emphasize is that this construction only works under the weak condition on the OPE: otherwise the functional will not be positive. Technically this happens because the functional action below $2\Df$ is essentially the value of the imaginary part of $f$ evaluated at $z=1-s/4$, and this is always antisymmetric under $s\to 4-s$.

\begin{figure}[t]
\centering
\includegraphics[width=.73\textwidth]{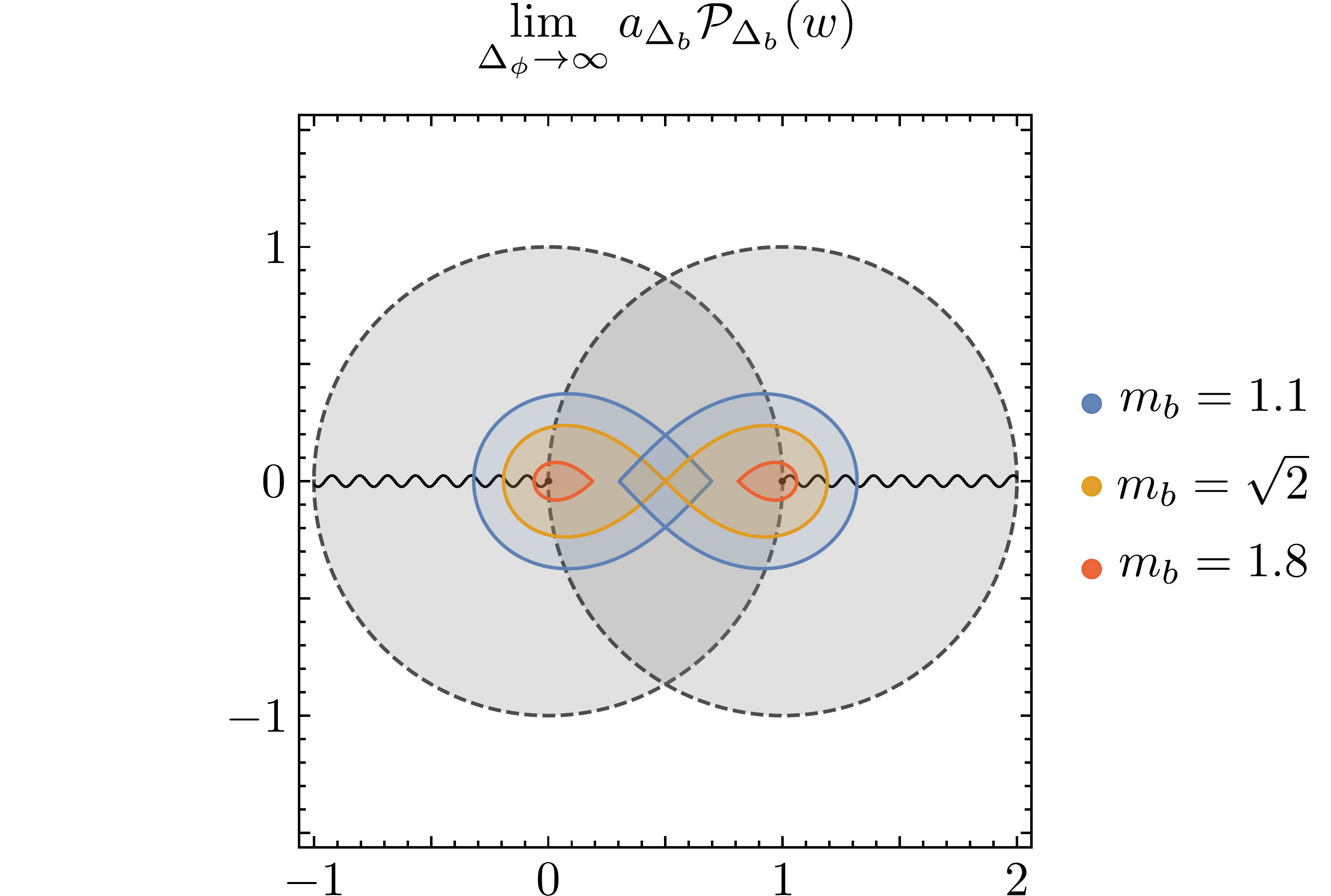}\vspace{0.2cm}
\caption{Large $\dphi$ limit of a single state below threshold $\Delta_b=m_b\dphi<2\dphi$. Each color represents different bound state masses. The limit is well defined for complex values of $w$ outside the shaded regions and diverges inside. The divergent regions (and their crossing symmetric images) start at $w=m_b^2/4$ ($w=1-m_b^2/4$) and always include a subset of the physical line $w>1$ ($w<0$); they overlap whenever $m_b<\sqrt2$. These regions lie always inside the divergent region for the identity given by the two discs depicted in gray.
}
\label{fig:bad_regions}
\end{figure}

After this review, let us assume the strong condition on the OPE so that the bounds derived above are valid. The bound \reef{boundOPEbelow} now allows us to show that the Polyakov block obtained in equation \ref{eq:pbfull} contains a finite piece in the flat space limit. We will shortly prove that this is the only relevant such piece, so that:
\beq
\boxed{\cF\[a_{\Delta_b} \cP^\pm_{\Delta_b} (w)\]= \frac{1}{4}\, \sqrt{\frac{w(1-w)}{z_b(1-z_b)}} \(\frac{g_b^2}{z_b-w} + \frac{g_b^2}{z_b-1+w}\)\,,\qquad g_b^2=\frac{2}{\pi}\frac{m_b}{\dphi} \(\frac{a_{\Delta_b}}{\tilde a^\ttext{free}_{\Delta_b}}\).}
\la{Pb}
\eeq
with the effective coupling bounded from above, $g^2_b\leq (g_b^{sG})^2$. 
To prove that this is the indeed the only contribution, we must show that the flat space limit of $a_{\Delta_b}E_{\Delta_b}(w|\Df)$ is zero. For example, in region III we have the product $a_{\Delta_b}G_{\Delta_b}(1-w|\dphi)$, which gives
\beqa
\lim_{\dphi\to \infty} a_{\Delta_b}|G_{\Delta_b}(1-w|\dphi)|&\leq& \lim_{\dphi\to \infty} \(g^{sG}_b\)^2\; 
\frac{2\dphi}{\pi m_b}\tilde a^\ttext{free}_{\Delta_b}
|G_{\Delta_b}(1-w|\dphi)|\,,\\
&\leq&\lim_{\dphi\to \infty} \(\texttt{factor}\)\times \left|\frac{(1-m_b^2/4)^2}{(1-w)^2} \frac{(2+m_b)^{m_b} (1-\sqrt w)^m}{(2-m_b)^{m_b} (1+\sqrt w)^{m}_b}\right|^\dphi \,,\nn[.5cm]
\(\texttt{factor}\)&=& \(g^{sG}_b\)^2\times 8\,\frac{\sqrt{\Df}}{\pi^{\frac 32}}\, m_b^{-1/2} (2+m_b)^{-3/2} (2-m_b)^{-1/2} \left|\frac{\sqrt w+1}{w^{1/4}}\right|\,.\nonumber
\eeqa
From the large $\dphi$ limit we get zero or infinity depending on the regions where the expression inside the absolute value in the second line is less or bigger than one. The final result in the complex plane is shown in figure~\ref{fig:bad_regions}. The ``bad'' regions where we get divergences are always contained in the ones for the identity block, and always cover a subset of the real line $1<z<2$ (which in S-matrix language is part of the physical values of the center of mass energy $4<s<8$).\footnote{These bad regions match the ones corresponding to the presence of an AdS Landau diagram  \cite{Komatsu:2020sag}.} Since outside these bad regions the large $\Df$ limit of these factors, and therefore of $a_{\Delta_b}E_{\Delta_b}$, is equal to zero, the analytic continuation is trivial and the flat space limit of a single Polyakov block with $\Delta<2\dphi$ is indeed given by \eqref{Pb}.

We thus conclude that the flat space limit of a single state below threshold \eqref{Pb} gives precisely the crossing symmetric bound state pole in our normalization.

\subsubsection{States above threshold}
\label{sec:boundabove} 

To understand the contributions of states above $2\Df$ we  have to obtain a bound on the ratio $a_{\Delta}/a_{\Delta}^{\mbox{\tiny free}}$. Although such bounds have appeared before for finite $\Df$ \cite{Mazac:2018ycv}, they were derived for large $\Delta$ and fixed $\Df$, whereas here we are interested in the limit where both are large and of the same order of magnitude.

We begin by considering the quantity:  
\ba
N_z&\equiv  (1-z)^{2\Df}\left[\mathcal G(z)-\sum_{0\leq \Delta\leq 2\Df}a_{\Delta} G_{\Delta}(z|\Df)\right]\\
&\underset{\Df\to\infty}{=} \sum_{\Delta>2\Df} 2 \left(\frac{a_\Delta}{\af}\right) \Nh(\Delta,\mbox{$\frac{4}{1-z}$})\,,
\ea
where $\Nh$ was defined in equation \reef{nh}. We see that in the flat space limit, the Euclidean correlator (with suitable subtractions) knows about the average OPE density around $\Delta=2\Df/\sqrt{1-z}$, where the average is taken in a region of width $\sqrt{\Df}$. To obtain a bound on $N_z$, we invoke the following results proven by one of us in \cite{Paulos:2020zxx}:
\bea
\mathcal G^-(z)+\sum_{0< \Delta\leq 2\Df} a_{\Delta} \mathcal P_{\Delta}^-(z) \leq\mathcal G(z)\leq \mathcal G^+(z)+\sum_{0< \Delta\leq 2\Df} a_{\Delta} \mathcal P_{\Delta}^+(z)\,,\qquad 0<z<1\,.
\eea
Importantly, these bounds hold for any $\Df$. They imply that for $0<z<1$
\ba
(1-z)^{2\Df}(\mathcal G^-(z)-G_0(z|\Df))+R^-(z)\leq N_z\leq (1-z)^{2\Df}(\mathcal G^+(z)-G_0(z|\Df))+R^+(z)\,,\label{eq:boundsNz}
\ea
where
\be
R^{\pm}(z)=(1-z)^{2\Df}\sum_{0<\Delta\leq 2\Df} a_{\Delta}\left[\mathcal P^{\pm}_{\Delta}(z)-G_{\Delta}(z|\Df)\right] \la{eq:reminder} \,.
\ee
Let us now make the same assumptions on the spectrum as for the previous subsection, i.e. the strong condition on the OPE. In this case:
\bea
R^{\pm}(z)\underset{\Df\to \infty}=0. \la{reminder_zero}
\eea
This follows from the computation of Polyakov blocks in section~\ref{subsec:Polyakovbelow} and the results from the previous subsection. The point is that even though the combination $a_\Delta \[ \cP_\Delta(z)-G_\Delta(z|\dphi)\]$ may diverge (as depicted in the bad regions of figure~\ref{fig:bad_regions}), the extra exponential decay of the factor $(1-z)^{2\dphi}$ is enough to counterbalance the divergence and get a vanishing result. 

Using the expressions for $\mathcal G^{\pm}$, we conclude from \ref{eq:boundsNz} that:
\begin{equation}
\boxed{
\lim_{\dphi\to\infty}
\sum_{\Delta>2\Df} 2 \left(\frac{a_\Delta}{\af}\right)\Nh(\Delta,s)=1\,, \qquad s\in(4,\infty)
}
\end{equation}
Physically this means that, in the flat space limit, any CFT satisfying our assumptions on the spectrum must have an OPE density above $2\Df$ which, in a precise sense, is the same as that of a free field ``on average''.\footnote{Note for a free field we have indeed $$\sum_{\Delta>2\Df} 2 \left(\frac{a_\Delta}{\af}\right) \Nh(\Delta,s)=2\sum_{n=0}^\infty \Nh(2\Df+2n,s)=\int_{2\Df}^\infty \ud \Delta\, \Nh(\Delta,s)\underset{\Df\to \infty}{=}1\,,$$ which explains the awkward factor of two.
}

As we will later see, this result will tell us that the $S$-matrix for physical kinematics $s>4m^2$ must satisfy the unitarity condition $|S(s)|\leq 1$.

\subsection{\texorpdfstring{A dispersion formula for $S$}{CFT dispersion formula for S}}\label{cftdisp}
We will now use the expressions derived in the previous subsection to determine the flat space limit of the correlator. Let us first consider the contributions coming from Polyakov blocks above threshold. We have 
\ba
\overline{\mathcal G}(z)=\sum_{\Delta>2\Df} a_{\Delta} \mathcal P_{\Delta}(z)\underset{\Df\to \infty}{=}-\sum_{\Delta>2\Df}\frac{8\sqrt{s_{\Delta}}}{\Df} K(s_z,s_{\Delta}) \left(\frac{a_{\Delta}}{\af}\right) \sin^2\left[\frac \pi 2(\Delta-2\Df)\right]\,
\ea
with
\ba
K(s,s')=\frac{2}{\pi}\frac{\sqrt{s(4-s)}}{\sqrt{s'(s'-4)}}\,\frac{s'-2}{(s'-s)(s+s'-4)}\label{eq:dispkernel}
\ea
The idea now is to take advantage of the fact that the summand splits into two factors, one of which is varying slowly with $\Delta$, another which is varying fast. Let us define 
\ba\label{rhoabove}
\rho_\alpha(s):=\lim_{\Df\to \infty}\Df^{-\alpha}\sum_{\Delta \in B_{\alpha}(s)}\left(\frac{a_{\Delta}}{\af}\right)\,4 \sin^2\left[\frac \pi 2(\Delta-2\Df)\right]
\ea
with
\ba
B_\alpha(s):=\{\Delta:\quad \sqrt{s}\Df\leq \Delta<\sqrt{s}\Df+\Df^\alpha\}
\ea
a bin in dimension space of size $\Df^{\alpha}$. Setting $\alpha<1$, a simple calculation yields
\bea
\overline{\mathcal G}(z)\underset{\Df\to \infty}{=}-\int_4^{\infty} \ud s' K(s_z,s') \rho_{\alpha}(s')\,.\label{eq:dispgb}
\eea
It is clear that the end result must be independent of the particular choice of $\alpha$.  Hence all $\rho_\alpha$, at least for $\alpha>1/2$ converge to the same density in the sense of distributions. In fact, we must have
\bea
\rho_\alpha(s)\sim \rho(s)\equiv \mathcal R_z\,[\lim_{\Df\to \infty} \overline{\mathcal G}(z_s)]<\infty
\eea
where $\sim$ means equality as distributions. In section \ref{physlim} we will show that we may safely commute the limit with $\mathcal R_z$, giving us another expression for $\rho(s)$. Crucially, the bounds on the OPE density derived in the previous subsection imply
\bea
0\leq \rho_\alpha(s) \leq 2\,, \qquad 1/2<\alpha<1
\eea
 We conclude that $\overline{\mathcal G}(z)$ has a finite flat space limit for any complex $z$, with its values determined by formula \reef{eq:dispgb}.

Moving on, let us consider the contributions of Polyakov blocks below threshold. These contributions are written as
\bea
\mathcal F\left[\sum_{0\leq \Delta\leq 2\Df} a_{\Delta} \mathcal P_\Delta(w)\right]=1+\sum_{0< \Delta\leq 2\Df} \mathcal F[a_{\Delta} \mathcal P_\Delta(w)]\label{eq:sumbstates}
\eea
On the righthand side we have separated out the contribution of the identity and made the assumption that the flat space limit commutes with the sum over states. This is justified under the assumptions we made on the OPE.
Proceeding, let us write \reef{Pb} as:
\bea
\mathcal F[a_{\Delta} \mathcal P_{\Delta}(w)]=\frac{2\sqrt{s_{\Delta}}}{\Df}\,\left(\frac{a_{\Delta}}{\tilde a_{\Delta}^{\mbox{\tiny free}}}\right) \widetilde K(s_w,s_\Delta) \label{eq:flatpolybs2}\,,\qquad \Delta<2\Df\,,
\eea
with
\bea\label{Kbelow}
\widetilde K(s,s')=\frac{2}{\pi}\frac{\sqrt{s(4-s)}}{\sqrt{s'(4-s')}}\,\frac{s'-2}{(s'-s)(s+s'-4)}\,.
\eea
We can now distinguish between two situations. In the first, a state in the OPE with dimension $\Delta_b$ sits isolated, i.e. the nearest states to lie at a distance which scales as $\Df$. In this case we can write this expression as
\bea
\mathcal F[a_{\Delta_b} \mathcal P_{\Delta_b}(w)]=\pi g_b^2 \widetilde K(s_w,s_b)\,,\qquad g_b^2\equiv \frac{2}{\pi}\frac{\sqrt{s_{\Delta_b}}}{\Df}\left(\frac{a_{\Delta_b}}{\tilde a_{\Delta_b}^{\mbox{\tiny free}}}\right)\,.
\eea
The effective coupling $g_b^2$ is guaranteed to be finite thanks to the bound reviewed in section \ref{sec:boundbelow}. In a more general situation we again introduce a density
\bea
\tilde \rho_\alpha(s)=\lim_{\Df\to \infty}\Df^{-\alpha} \sum_{\Delta \in B_{\alpha}(s)} \left(\frac{a_{\Delta}}{\tilde a_{\Delta}^{\mbox{\tiny free}}}\right)\,.
\eea
For an isolated bound state this would give a singular density which should be interpreted as a delta function. The dependence on $\alpha$ is again essentially irrelevant and we can drop it. In this way we obtain
\bea
\mathcal F\left[\sum_{0\leq \Delta\leq 2\Df} a_{\Delta} \mathcal P_\Delta(w)\right]\overset{!}{=}1+\int_{s_0}^4 \ud s' \, \widetilde K(s_w,s')\,\tilde \rho(s')\,. \label{eq:flatbelow}
\eea
Note that $\tilde \rho(s)$ is non-negative by construction, but it may become arbitrarily large, as in the delta function example above. However, this is not true for the integrated density. Indeed, the bound \reef{eq:boundboundstates} can be translated as:
\bea
\int_{s_0}^4\ud s' \frac{\widetilde K(s_b,s')}{-S_b^{sG}(s')}\tilde \rho(s')\leq 1\,.
\eea
This implies that integrating the density against any smooth function of $s$ yields a finite result. For instance, we find for real $s$:
\ba
\int_{s_0}^4\ud s' \widetilde K(s,s')\tilde \rho(s') \leq \frac{\sqrt{s_0(4-s_0)}+\sqrt{s(4-s)}}{\sqrt{(s_0(4-s_0)}-\sqrt{s(4-s)}}\,, \qquad 4-s_0<s<s_0\,.
\ea

We have now obtained expressions for the flat space limit of states both below and above threshold. Putting them together we obtain the following {\em CFT dispersion formula for $S$}:
\begin{equation}\label{CFTdispersiondensity}
\boxed{\begin{aligned}[t]
S(s)&=1+\int_{s_0}^4 \ud s' \, \widetilde K(s,s')\,\tilde \rho(s')-\int_4^{\infty} \ud s' \, K(s,s')\,\rho(s')\\
\rho(s)&= \lim_{\Df\to \infty}\Df^{-\alpha}\sum_{\Delta \in B_{\alpha}(s)}\left(\frac{a_{\Delta}}{\af}\right)\,4 \sin^2\left[\frac \pi 2(\Delta-2\Df)\right]\\
\tilde \rho(s)&=\lim_{\Df\to \infty}\Df^{-\alpha} \sum_{\Delta \in B_{\alpha}(s)} \left(\frac{a_{\Delta}}{\tilde a_{\Delta}^{\mbox{\tiny free}}}\right)\,.
\end{aligned}
}
\end{equation}
This formula provides an expression for the flat space limit of the CFT correlator anywhere on the complex plane. An equivalent way of obtaining it would be in terms of the correlator itself. Indeed, a different way of stating our results is that
\bea
\mathcal G(w)\underset{\Df\to \infty}=\frac{1}{w^{2\Df}}+\frac{1}{(1-w)^{2\Df}}+\sum_{\Delta_0\leq \Delta_b\leq 2\Df} a_{\Delta_b} E_{\Delta_b}(w|\Df) + S(s_w)
\eea
with $S(s)$ defined by \reef{CFTdispersiondensity} above. This expression makes clear that $\mathcal F[\mathcal G]=S(s)$.

The CFT dispersion formula explicitly provides the desired analytic continuation of $S(s)$, which is initially defined only for sufficiently large $s$, to the entire complex plane. In particular it establishes that $S(s)$ is analytic everywhere except on the real axis. In turn, its singularities can be obtained directly from this expression and are computable in terms of the dual CFT data. Note that defining
\bea
T(s):=2\sqrt{s(4-s)}(S(s)-1)
\eea
then we have
\ba\label{rhoIsT}
\rho(s)&=\frac{\mathcal I_s T(s)}{2\sqrt{s(s-4)}}\,, \qquad s>4\,,\\
\tilde \rho(s)&=\frac{\mathcal I_s T(s)}{2\sqrt{s(4-s)}}\,, \qquad s<4
\ea
Then the bounds $0\leq \rho(s)\leq 2$ are consistent with unitarity of the S-matrix, while $\tilde \rho(s)\geq 0$ is usually called ``extended unitarity'' \cite{Cutkosky:1960sp,Landau:1959fi,Eden:1966dnq}. To the best of our knowledge the latter has not been established non-perturbatively in QFT, but here we see it follows simply from the definition of $\tilde \rho$ in terms of the CFT OPE.

To reiterate, really what we have learned from this construction are the analyticity properties of $S$. It would of course be straightforward to arrive at a dispersion relation for $S(s)$ of the form above, interpreting $\rho$, $\tilde \rho$ as discontinuities of $S$ in the relevant cuts, if we had assumed analyticity to begin with. Here we have instead begun with the CFT correlator and its analyticity properties, which are well established, and ended up proving those of $S(s)$.

\section{Flat space limit: physical kinematics}\label{physlim}
The goal of this section is to determine and justify a simple expression for the S-matrix in physical kinematics, the phase shift formula. An expression very similar to ours was first argued to hold for general holographic QFTs in \cite{Paulos:2016fap}. The present derivation is essentially the same as in \cite{Komatsu:2020sag}, but simpler because it is specialized to $d=1$. However, as we explain, this derivation cannot be rigorously justified unless certain  limits commute, and we will therefore have to prove that this is the case.

\subsection{Phase shift formula}
We are interested in determining the S-matrix in physical kinematics. That is: 
\ba 
\Sp(s_z)&:=\lim_{\epsilon\to 0^+} \mathcal F[\mathcal G(z-i\epsilon)]\,, \qquad z<0\,.
\ea
Ignoring for the moment the issue of analytic continuation, this involves taking the two limits -- flat space, and going to physical kinematics -- in a definite order. Suppose that the limits commute, so that we can write instead:
\ba
\Sp(s_z)&\overset{!}{=}\lim_{\Df\to +\infty} \lim_{\epsilon \to 0^+}\mathcal G(z-i\epsilon), \qquad \mbox{for}\quad s_z>s_0>4\,.
\ea
Proving this is non-trivial and will be justified in the next subsection. For now let us take it as given, and see what expression we may obtain for $\Sp$. Using the OPE we have 
\bea
\Sp(s_z)=\lim_{\Df\to +\infty} \sum_{\Delta\geq 0} a_\Delta \frac{G_{\Delta}(\frac{z}{z-1})}{(-z)^{2\Df}} e^{-i\pi(\Delta-2\Df)}\,.
\eea
for $z<z_0$ which we determine as follows. Separate contributions below and above threshold, so that
\bea
	\Sp(s_z)=\lim_{\Df\to+\infty} \left[ \sum_{\Delta<2\Df}a_\Delta\frac{G_{\Delta}(\frac{z}{z-1})}{(-z)^{2\Df}}e^{-i \pi (\Delta-2\Df)}+ \sum_{\Delta>2\Df} a_\Delta\,\frac{G_{\Delta}(\frac{z}{z-1})}{(-z)^{2\Df}} e^{-i \pi (\Delta-2\Df)}\right]\,,
\eea
and let us assume the strong condition on the OPE, so that the bounds on OPE coefficients reviewed in section \ref{sec:boundbelow} are satisfied.
Then it is not hard to see that all terms in the first sum are exponentially suppressed for $z<-1$, by the same logic used in section \ref{sec:boundabove}. So $z_0=-1$ and the analytic continuation of $\Sp$ for all $z<0$ is simply
\ba
\Sp(s_z)&=\lim_{\Df\to+\infty} \sum_{\Delta>2\Df} a_\Delta\,\frac{G_{\Delta}(\frac{z}{z-1})}{(-z)^{2\Df}} e^{-i \pi (\Delta-2\Df)}\,.
\ea
Using expression \reef{eq:gaussian} we get the phase shift formula:
\begin{equation}
\label{eq:phaseshift} 
\boxed{
\Sp(s)=\lim_{\Df \to +\infty}\sum_{\Delta>2\Df} 2 \left(\frac{a_\Delta}{a_\Delta^{\ttext{free}}}\right) \Nh(\Delta,s)
\, e^{-i \pi (\Delta-2\Df)}\,.
}
\end{equation}
This result allows us to express the physical S-matrix directly in terms of the CFT data.
As promised, the bounds on the OPE density derived in section \ref{sec:boundabove} nicely translate into unitarity of the S-matrix:
\bea
|\Sp(s)|\leq \lim_{\Df \to +\infty}\sum_{\Delta>2\Df} 2 \left(\frac{a_\Delta}{a_\Delta^{\ttext{free}}}\right) \Nh(\Delta,s)=1\,.
\eea

To justify the phase shift formula, we must show that the flat space and physical kinematics limits commute, i.e.:
\bea
\lim_{\epsilon \to 0^+} S(s+i\epsilon)=\Sp(s)\,,
\eea
where the right hand side is understood to be computed by the phase shift formula. We will prove this in two steps, first by showing this relation holds when taking the real part, and then when taking the imaginary part.

\subsection{Commuting limits: real part}

We begin by recalling the dispersion relation \reef{eq:disprels} for the subtracted correlator:
\ba
\overline{\mathcal G}(w)&=-\int_0^{1} \ud z\, \hat g_w^+(z) \mbox{dDisc}^+\,\overline{\mathcal G}(z) 
\ea
which holds for any $\Df$. Taking the large $\Df$ limit we already know that the lefthand side becomes
\bea
\lim_{\Df\to \infty}  \overline{\mathcal G}(z_s)= S(s)-1+\int_{s_0}^4 \ud s' \tilde K(s,s') \tilde \rho(s')
\eea
since it is the sum of Polyakov blocks above $2\Df$. On the other hand, the righthand side can be written as
\bea
\lim_{\Df\to\infty} \int_0^{1} \ud z\, \hat g_w^+(z) \mbox{dDisc}^+\,\overline{\mathcal G}(w)= \int_4^{\infty} \ud s' K(s,s') \rho_{\mbox{\tiny phys}}(s')\,,
\eea
with
\bea
\rho_{\mbox{\tiny phys}}(s):= \lim_{\Df\to \infty} (1-z)^{2\Df}\mbox{dDisc}^+\, \overline{\mathcal G}(z)\bigg|_{z=\frac{s-4}s}\,.
\eea
This means that the dispersion relation for $\overline{\mathcal G}$ directly translates into a dispersion relation for $S(s)$. But we already had such a relation, and since they involve the same kernels $K,\tilde K$ they must agree. More precisely, this implies that as distributions we must have
\bea
\rho_\alpha(s)\sim \rho_{\mbox{\tiny phys}}(s)\,.
\eea
Now let us compute $\rho_{\mbox{\tiny phys}}$ directly. Using:
\ba
 \lim_{\Df\to \infty} (1-z)^{2\Df}\mbox{dDisc}^+\,\overline{\mathcal G}(z)&=
 \lim_{\Df\to \infty} (1-z)^{2\Df}\mbox{dDisc}^+ \left[\mathcal G(z)-\sum_{\Delta\leq 2\Df} a_{\Delta} G_{\Delta}(z)\right]\,,
 \ea
 together with the OPE and the unitarity bound, we can find
 \ba
 \rho_{\mbox{\tiny phys}}(s)
 &=\lim_{\Df \to +\infty}\sum_{\Delta>2\Df}  \left(\frac{a_\Delta}{a_\Delta^{\ttext{free}}}\right)\, 4 \sin^2\left[\frac{\pi}2(\Delta-2\Df)\right]\,\Nh(\Delta,s)
\\
&=\mbox{Re}\left[1-\Sp(s)\right]
\ea
where on the last line we use the phase shift formula for $\Sp$. We see that indeed as distributions  $\rho_\alpha(s)\sim \rho_{\mbox{\tiny phys}(s)}$ when $\alpha>1/2$, as can be seen by splitting the sum over states in the latter into bins of size $\sqrt{\Df}$. Hence we have proven:
\bea
\mathcal R_s[1-S(s)]= \mbox{Re}[1-\Sp(s)]
\eea
which establishes that the phase shift formula holds for the real part.

\subsection{Commuting limits: imaginary part}

The previous derivation makes clear that if we are to prove the phase shift formula for the imaginary part, we must write a dispersion relation for CFT correlators which will depend on the imaginary part of $\mathcal G$. Such a dispersion relation has appeared before in the literature \cite{Bissi:2019kkx}, but as we explain it suffers from some minor issues concerning subtractions which we clarify and settle here.

The idea is to use Cauchy's formula for the correlator,
\bea
\mathcal G(z)=\oint \frac{\ud z}{2\pi i} \frac{\mathcal G(z')}{z'-z}
\eea
and now deform the contour to pick up the discontinuities of $\mathcal G$. However, there are two issues. Firstly, we have to worry about contributions from $z=\infty$. Secondly, we may get divergences from the contribution of the contour close to $z=0,1$. To deal with these issues we must implement subtractions. The cleanest way to do this is to define a subtracted correlator $\widetilde{\mathcal G}(z):=\mathcal G(z)-
\sum_{\Delta\leq 2\Df+1} a_{\Delta} \mathcal P_{\Delta}(z)$. This has improved soft behaviour near $z=0,1$ as compared to $\overline {\mathcal G}(z)$ and leads to a dispersion relation 
\bea
\widetilde{\mathcal G}(w)=-\int_{1}^{\infty}\frac{\ud z}{\pi}\,\frac{w(1-w)}{z(1-z)}\, \frac{2z-1}{(w-z)(w+z-1)}\, \mathcal I_{z} \widetilde{\mathcal G}(z)\,.
\eea
which holds for any CFT.
For the purpose at hand however, it is more convenient to work with a slightly different dispersion relation, where we improve the behaviour near $z=\infty$ instead. Let us set\footnote{Note that any 1d CFT correlator is bounded by a constant at $z=\infty$, see e.g. \cite{Rychkov:2017tpc}.}
\bea
C:= \lim_{z \to \infty} \overline{\mathcal G}(z)
\eea
Then we can write the dispersion relation
\bea
\overline{\mathcal G}(w)=C-\int_{1}^{\infty}\frac{\ud z}{\pi}\,\, \frac{2z-1}{(w-z)(w+z-1)}\, \mathcal I_{z} \overline{\mathcal G}(z)\,.
\eea
This holds for arbitrary $\Df$. Using the CFT dispersion formula for $s>4$ we have:
\begin{multline}
 \mathcal I_s\left[\lim_{\Df\to \infty}\overline{\mathcal G}(z_s) \right]=\mathcal I_s\left[S(s)-1-\int_{s_0}^4 \, \ud s' \widetilde K(s,s') \tilde \rho(s')\right]\\=\mathcal I_s S(s)-\int_{s_0}^4 \ud s' \mathcal I_s \widetilde K(s,s') \tilde \rho(s')\,.
\end{multline}
Alternatively, we can compute the same quantity directly using our new dispersion relation. Matching the two we get:
\bea
\mathcal I_s S(s)= \lim_{\Df\to \infty} \,\mathcal I_s \overline{\mathcal G}(z_s)+\int_{s_0}^4 \ud s' \mathcal I_s \widetilde K(s,s') \tilde \rho(s')
\eea
Our goal will be to prove that the righthand side is the same as $\mbox{Im}\, \Sp$. Consider first the case where there are no states below threshold other than the identity. Using the OPE, a computation which by now should be familiar gives
\ba
\lim_{\Df\to \infty}\mathcal I_s \overline{\mathcal G}(z_s)&=
\sum_{\Delta\geq 2 \Df}\,2\, \left(\frac{a_{\Delta}}{\af}\right) \sin\left[\pi(\Delta-2\Df)\right] \,\Nh(\Delta,z_s)\\
&=\mbox{Im}\,\Sp(s)
\ea
as desired.  To finish the proof, we must show that states below threshold do not change the result. This will follow if we can show that:
\bea
\lim_{\Df\to \infty}\mathcal I_s \left[\sum_{0<\Delta\leq 2 \Df} a_{\Delta}\left[\mathcal P_{\Delta}(z_s)-G_{\Delta}(1-z_s)\right]\right]=-\int_{s_0}^4 \ud s' \mathcal I_s \widetilde K(s,s') \tilde \rho(s')\,, \quad s>4\,.
\eea
The meaning of this equation is that the phase shift formula for Polyakov blocks below $2\Df$ (which will appear on the left side of this equation)
should give the same answer as the flat space limit of those blocks as computed in sections \ref{subsec:Polyakovbelow} and \ref{sec:flatgeneral} (the right side of the equation). That is, we must check the limits commute by hand for these blocks. We show this by direct computation in appendix \ref{sec:phaseshiftpoly}\,, which completes the proof.

\section{Perturbative checks}
\label{sec:perturbative}
In this section we check our flat space prescription in a few perturbative examples. Our starting point is the theory of generalized free bosons $\phi_1$ with dimension $\Delta_1$. Throughout this section we will focus on the four point correlator of this field, and so we will set $\Df\equiv \Delta_1$. We will also refer to the mass of the corresponding AdS bulk field as $m_1$ but we will set this to one
, so that e.g. $\Delta_2/\Delta_1\equiv m_2/m_1=m_2$.

\subsection{General remarks}

Before any perturbation we have the generalized free boson correlator which has the following $s$-channel conformal block decomposition:
\beq
\cG_0(w)=1+\sum_n a^\ttext{free}_{[11]_n} G_{[11]_n}(w) = \cP_0(w)\,,
\eeq
with the double trace dimensions $[11]_n=2\Delta_1+2n$. We then perturb the theory by 
coupling $\phi_1$ to another generalized free boson $\phi_2$ with conformal dimension $\Delta_2$. The four point function of $\phi_1$ then admits a perturbative expansion of the form
\beqa
\cG(w)=\cG_0(w)&+ &g^2\sum_n  \(a_{[11]_n}^{(1)}G_{[11]_n}(w|\Delta_1)+a_{[11]_n}^\ttext{free}\gamma_{[11]_n}^{(1)}\partial G_{[11]_n}(w|\Delta_1)\)+ \nonumber\\
&+&g^2 \sum_n a^{(1)}_{[22]_n}G_{[22]_n}(w|\Delta_1) + \ldots\,,
\label{eq:Gperturb}
\eeqa
where $\partial G\equiv \partial_{\Delta} G$. In the first line we have the first order corrections to the dimensions and OPE coefficients of double trace operators $[\phi_1\phi_1]_n$ and in the second line the contributions coming from the introduction of $\phi_2$~\footnote{
We are assuming there is no coupling of the form $g\phi_1^2\phi_2$ so that only double traces appear in \eqref{eq:Gperturb}. One can consider particle $\phi_1$ and $\phi_2$ transforming differently under a $Z_2$ symmetry so that this cubic interaction is not allowed.}. We can write \eqref{eq:Gperturb} alternatively in terms of Polyakov blocks as follows
\beqa
\cG(w)=\cP_0(w)&+ &g^2\sum_n  \(a_{[11]_n}^{(1)}\cP_{[11]_n}(w)+a_{[11]_n}^\ttext{free}\gamma_{[11]_n}^{(1)}\partial \cP_{[11]_n}(w)\)+ \nonumber\\
&+&g^2 \sum_n a^{(1)}_{[22]_n}\cP_{[22]_n}(w) + \ldots\,.
\label{eq:Pperturb}
\eeqa
Using the Polyakov blocks obtained from the free master functionals for $\phi_1$ many of the terms above are zero
\beq
\cG(w)=\cP_0(w)+g^2\[
\sum_n a^{(1)}_{[22]_n}\cP_{[22]_n}(w)
+a_{[11]_0}^\ttext{free}\gamma_{[11]_0}^{(1)}\partial \cP_{[11]_0}(w)\]+\ldots\,.
\label{eq:Pexpansion}
\eeq
 This is because the Polyakov blocks have double zeros at the double trace dimensions $\Delta=[11]_n$. For the bosonic blocks this is true for every $n$ except for $n=0$, with the last term amounting to a four-point contact interaction in $AdS_2$, i.e. a D-function $D_{\Delta_1\Delta_1\Delta_1\Delta_1}(z)$ \cite{Freedman:1998bj,DHoker:1999mic,Hijano:2015zsa}.
 
\begin{figure}[t]
\centering
\includegraphics[width=.85\textwidth]{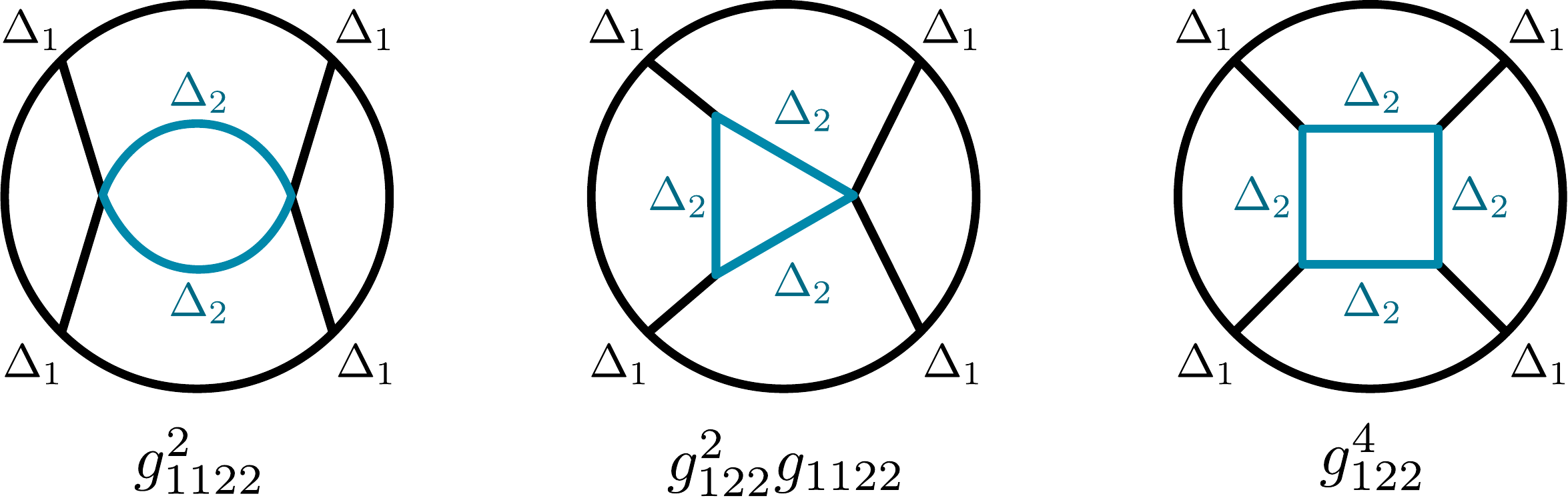}\vspace{0.2cm}
\caption{Bubble, triangle and box Witten diagrams considered in the main text.}
\label{fig:witten}
\end{figure}
 
 So far we have kept the discussion fairly general regarding the type of couplings between $\phi_1$ and $\phi_2$. We now restrict to cubic $g_{122}\phi_1\phi_2\phi_2$ and quartic $g_{1122}\phi_1\phi_1\phi_2\phi_2$ interactions, where at one loop level we encounter bubble, triangle and box diagrams each associated with different powers of $g_{122}$ and $g_{1122}$ (see figure~\ref{fig:witten})
\begin{equation}\label{Gexpand}
    \mathcal{G}(w)=\mathcal{G}_0(w)+g_{1122}^2\mathcal{G}^{\text{bubble}}(w)+g_{1122}g_{122}^2\mathcal{G}^{\text{triangle}}(w)+g_{122}^4\mathcal{G}^{\text{box}}(w)+\ldots
\end{equation}
 
 The idea is to illustrate in these specific examples how the flat space amplitude is recovered from the CFT data using our dispersion relation \eqref{CFTdispersiondensity}. In other words, we want to exhibit how the weighted sum of the Polyakov blocks for a specific diagram becomes the interacting part of the amplitude 
 \beq\la{eq:gralcomparison}
 \; \; \; \;g\sum_n a^\text{diagram}_{[22]_n}\, \cP^\text{diagram}_{[22]_n}(w) \; \xrightarrow{\;\cF\;} \;  \frac{T^\text{diagram}(s)}{2\sqrt{s(4-s)}}=g\int ds' \rho^\text{diagram}(s') K(s,s')\,,
 \eeq
 and the OPE coefficients the spectral density
 \beq\label{densitydiscT}
 \(\frac{a^\text{diagram}_{[22]_n}}{a^{\ttext{free,11}}_{[22]_n}}\)\; \xrightarrow{\;\cF\;} \; \rho^\text{diagram}(s) = \frac{\mathcal{I}_{s}T(s)}{2\sqrt{s(s-4)}}\,.
 \eeq
 as given in \eqref{rhoIsT}.

Before we proceed to the flat space limit of the one-loop diagrams, let us discuss two building blocks needed to extract $a^\text{diagram}_{[22]_n}$. Given the quartic $g_{1122}$ and cubic $g_{122}$ couplings, the building blocks for the one-loop diagrams are the contact and exchange Witten diagrams, with associated coefficients $a^\text{contact,1122}_{[22]_n}$ and $a^\text{t,exchange}_{[22]_n}$. 
\begin{figure}[t]
\centering
\includegraphics[width=0.7\textwidth]{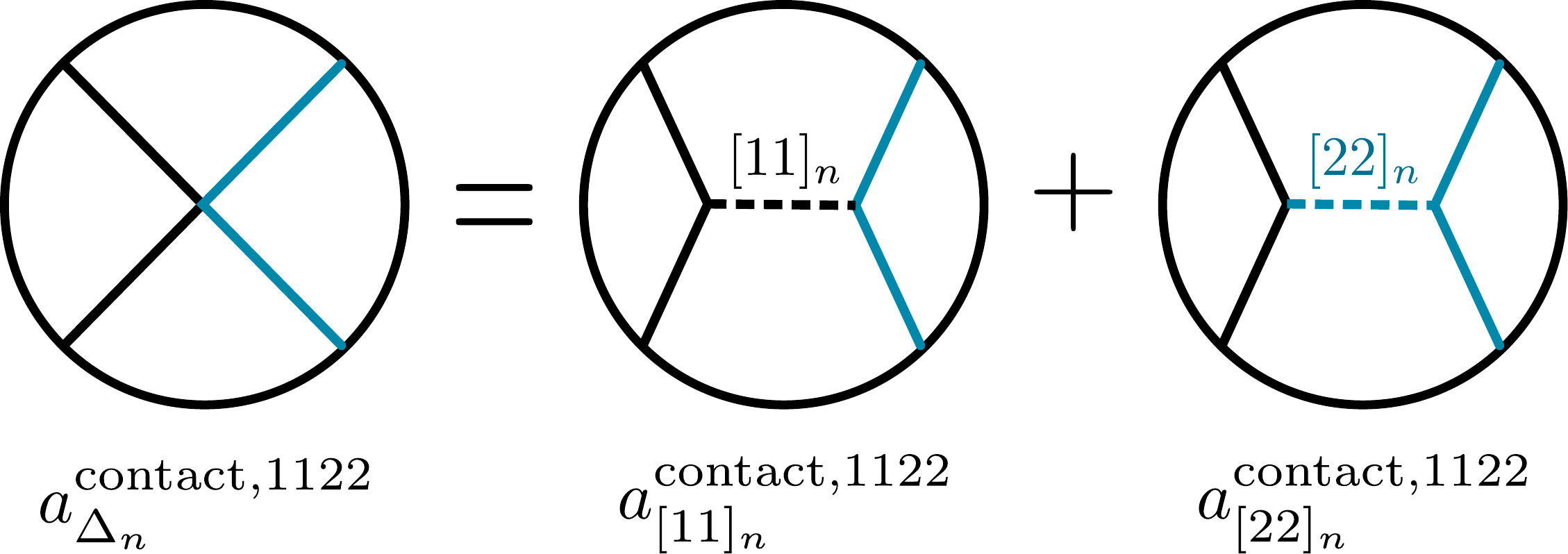}
\caption{
Pictorial representation of equation \eqref{eq:contactreln} showing the OPE coefficients of the double trace operators $[11]_n$ and $[22]_n$ appearing in the $s$-channel conformal block decomposition of the contact diagram written in \eqref{eq:contactreln}.
}
\label{fig:contactreln}
\end{figure}
These coefficients appear in the well known expansions of tree-level diagrams in terms of (s-channel) conformal blocks. For the 1122 contact diagram we have\footnote{The relevant expressions and flat space limit for the equal dimensions contact term are discussed in appendix~\ref{sec:phaseshiftpoly}.}
\begin{equation}\label{eq:contactreln}
         \cG^\text{contact,1122}(w)=  D_{\Delta_1\Delta_1\Delta_2\Delta_2}(w)=\sum_{n=0}^{\infty}a_{[11]_n}^{\text{contact},1122}\,G_{[11]_n}(w) + \sum_{n=0}^{\infty}a_{[22]_n}^{\text{contact},1122}\,G_{[22]_n}(w)\,,
\end{equation}
which we represent pictorially in figure~\ref{fig:contactreln}. Here  $G^{11,22}_{\Delta}$ are appropriate conformal blocks for a $1122$ correlator, and the coefficient  $a^\text{contact,1122}_{[22]_n}$ was computed in \cite{Hijano:2015zsa,Zhou:2018sfz} and reads up to a normalization
\begin{equation}\label{eq:acontact}
\begin{aligned}
 a^{\text{contact},1122}_{[22]_n}
 &=\,\lambda^{(1)}_{11[22]_n}\lambda^{(0)}_{22[22]_n}\\[12pt]
&\propto\begin{cases}
\frac{\sqrt{\pi } (-1)^{n} \Gamma \left(-n+\Delta _1-\Delta _2\right) \Gamma
   \left(n+\Delta _2\right){}^4 \Gamma \left(n+\Delta _1+\Delta
   _2-\frac{1}{2}\right) \Gamma \left(n+2 \Delta _2-\frac{1}{2}\right)}{2 n!
   \Gamma \left(\Delta _1\right){}^2 \Gamma \left(\Delta _2\right){}^2 \Gamma
   \left(2 \left(n+\Delta _2\right)\right) \Gamma \left(2 n+2 \Delta
   _2-\frac{1}{2}\right)},&2\Delta_2+2n<2\Delta_1\,,\\[12pt]
   \frac{\pi^{\frac{3}{2}} \Gamma
   \left(n+\Delta _2\right){}^4 \Gamma \left(n+\Delta _1+\Delta
   _2-\frac{1}{2}\right) \Gamma \left(n+2 \Delta _2-\frac{1}{2}\right)}{2 n!\sin\left[\pi(\Delta_1-\Delta_2)\right]\Gamma(1+n-\Delta_1+\Delta_2)
   \Gamma \left(\Delta _1\right){}^2 \Gamma \left(\Delta _2\right){}^2 \Gamma
   \left(2 \left(n+\Delta _2\right)\right) \Gamma \left(2 n+2 \Delta
   _2-\frac{1}{2}\right)},&2\Delta_2+2n>2\Delta_1\,.
 \end{cases}     
\end{aligned}
\end{equation}
Note that the zeroth order OPE coefficients $\lambda^{(0)}_{22[22]_{n}}$ appear in the generalized free field correlator $\cG^{2222}(z)$ where the external particles are $\phi_2$, so that we have the relation
\beq
a^{\ttext{free}}_{[22]_n}\bigg|_{\Df=\Delta_2}\equiv a^{\ttext{free},22}_{[22]_n}=\(\lambda^{(0)}_{22[22]_n}\)^2\,.
\eeq

Now we review the second building block we use in our examples: the expansion of the $t$-exchange diagram in $s$-channel blocks. It is given by
\begin{equation}\label{eq:exchange}
    \cG^\text{t-exchange}(w)=  \sum_{n=0}^{\infty} a_{[11]_{\tilde n}}^{\text{t-exchange}}\,G^{11,22}_{[11]_{\tilde n}}(w) + \sum_{n=0}^{\infty}a_{[22]_{\tilde n}}^{\text{t-exchange}}\,G^{11,22}_{[22]_{\tilde n}}(w)\,,
\end{equation}
where the double traces contain both parity even and odd states which we implement with the notation $[jj]_{\tilde n}=2\Delta_j+n$. The coefficients $a_{[jj]_{\tilde n}}^{\text{t-exchange}}$ can be obtained in terms of $a_{[jj]_{\tilde n}}^{\text{contact}}$ by means of the recursion relation of \cite{Zhou:2018sfz}. \footnote{The schematic form of the recursion relation is
\beq
\mu_{\tilde n-1} a^\text{t-exch}_{\tilde n-1} + \nu_{\tilde n} a^\text{t-exch}_{\tilde n} + \rho_{\tilde n+1} a^\text{t-exch}_{\tilde n+1} = a^\text{contact}_{n=2\tilde n}\,,
\label{eq:recursion}
\eeq
where the recursion coefficients $\mu,\nu,\rho$ are rational functions of $\Delta_1,\Delta_2,[22]_{\tilde n}$. For explicit expressions in more general scenarios see section 4.2 of \cite{Zhou:2018sfz}.}

\begin{figure}[t]
\centering
\includegraphics[width=0.95\textwidth]{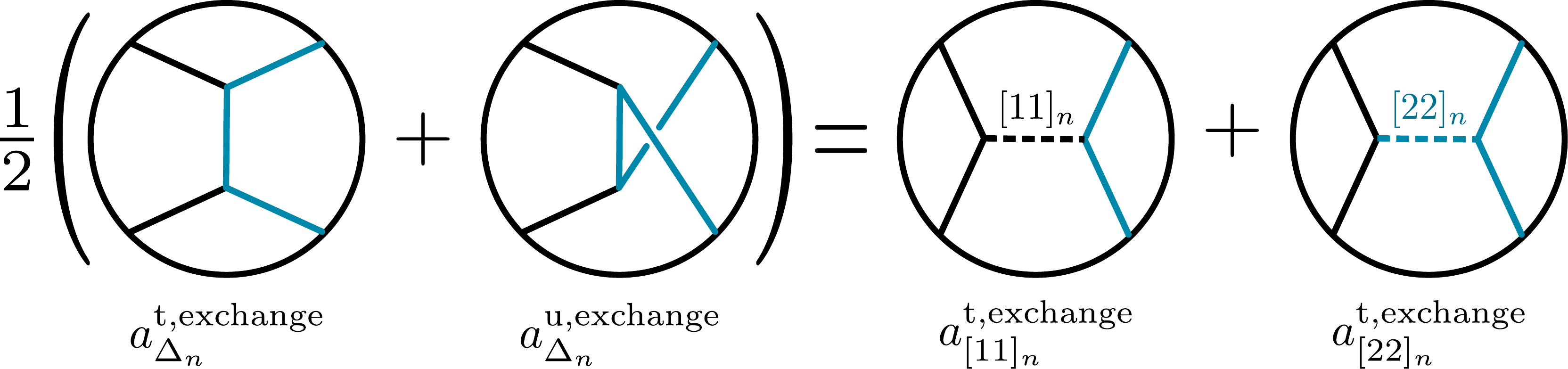}
\caption{The sum of $t$- and $u$-exchange Witten diagrams has an expansion in terms of purely parity even OPE coefficients.}
\label{fig:tuexchange}
\end{figure}

However, since we are dealing with scalar particles (as opposed to pseudo scalars), we are only interested in parity even OPE coefficients $a_{[22]_n}$. This can be simply fixed by noting the relation between $t$- and $u$-exchange OPE coefficients: $a_{[jj]_{\tilde n}}^{\text{u-exchange}}=(-1)^j a_{[jj]_{\tilde n}}^{\text{t-exchange}}$,
so that by taking the sum of $t$- and $u$-exchange Witten diagrams we have an expansion purely in terms of parity even states. The last point is displayed in figure~\ref{fig:tuexchange}.

The one-loop coefficients $a^\text{1-loop diagram}_{[22]_n}$ can be computed in terms of the tree level data discussed above $a_{[22]_{n}}^{\text{contact}}$, $a_{[22]_{n}}^{\text{t-exchange}}$. In the following sections we explain how to extract this one-loop coefficients and compute the flat space limit of each diagram.

\subsection{Bubble diagram}

\begin{figure}[t]
\centering
\includegraphics[width=0.75\textwidth]{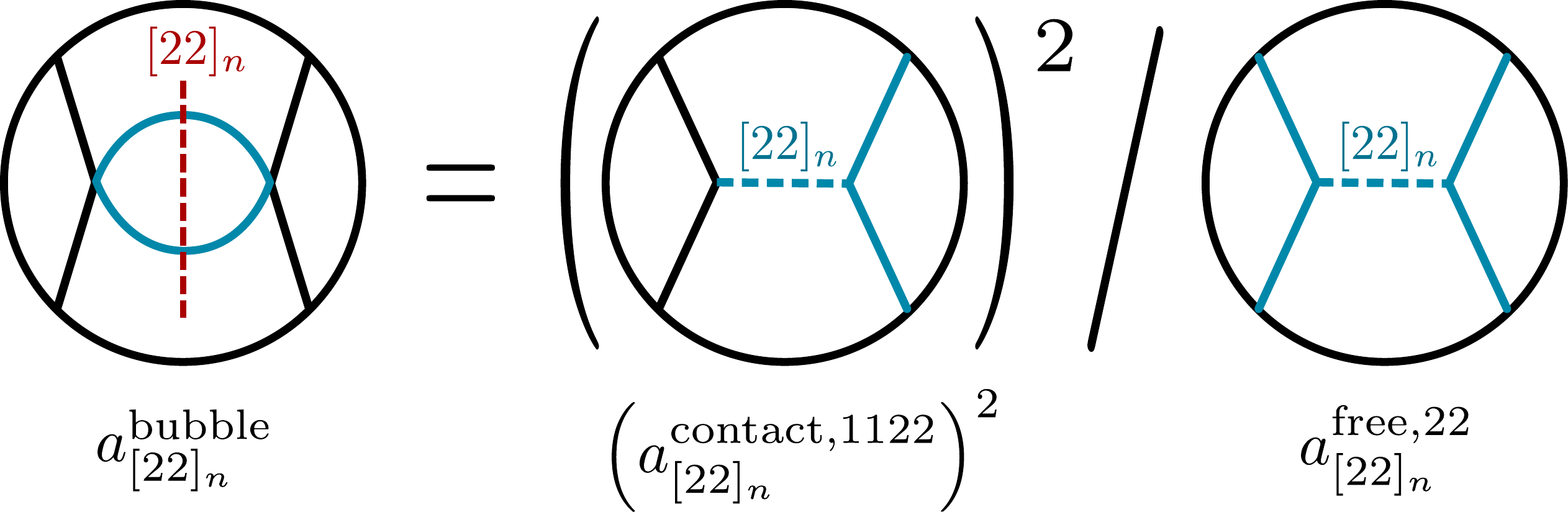}
\caption{The OPE coefficients of the double trace operators $[22]_n$ appearing in the $s$-channel bubble diagram can be extracted from the ones appearing in the conformal block expansion of the Witten contact diagram $a^\text{contact,1122}_{[22]_n}$ and $a^\text{free,22}_{[22]_n}$ as illustrated. }
\label{fig:bubblereln}
\end{figure}

Let us start with the crossing symmetric sum of bubble diagrams with external dimensions $\Delta_1$ and internal dimensions $\Delta_2$ (see first diagram in figure~\ref{fig:witten}). According to \eqref{eq:Pexpansion}, its Polyakov block expansion reads 
\begin{equation}\label{bubbleP}
    \mathcal{G}^{\text{bubble}}(w)=
    \sum_{n=0}^{\infty}a_{[22]_n}^{\text{bubble}}\mathcal{P}_{[22]_n}(w)+\text{contact}\,,
\end{equation}
where $[22]_n=2\Delta_2+2n$ indicates as usual the double trace dimensions of $\phi_2$ with even parity. All we have to do now is to understand what the coefficients $a_{[22]_n}^{\text{bubble}}$ are.  The idea is that we can get this data from lower loop level data in the same spirit as one does for amplitudes using the optical theorem. At the level of Witten diagrams, one can see this procedure as cutting the intermediate bulk-to-bulk propagators giving rise to lower loop level diagrams glued together conformally as nicely explained in \cite{Meltzer:2019nbs}. For the bubble diagram at hand we want to cut the two internal propagators and isolate the double trace contributions of $\phi_2$. 
The cutting and gluing procedure implies the following relation between the bubble and contact OPE coefficients:
\begin{equation}
    a^{\text{bubble}}_{[22]_n}=\(\lambda^{\text{contact}}_{11[22]_n}\)^2\,.
\end{equation}
To express $a^{\text{bubble}}_{[22]_n}$ in terms of $a^{\text{contact}}_{[22]_n}$ appearing in \eqref{eq:acontact} we need to divide by $\(\lambda^{(0)}_{22[22]_n}\)^2=a^{\ttext{free},22}_{[22]_n}$
\begin{equation}
    a^{\text{bubble}}_{[22]_n}=\(a^{\text{contact}}_{[22]_n}\)^2/a^{\text{free},22}_{[22]_n}\,.
\end{equation}
This relation is depicted in figure~\ref{fig:bubblereln}.

We would like to check that in the flat space limit the OPE density matches the spectral density of the flat space bubble amplitude. We will consider the less trivial case where the internal dimensions are smaller than the external ones $\Delta_2<\Delta_1$ (or $m_2<m_1$ in flat space amplitude), so that there is the two-particle cut for $m_2$ starting before the one for $m_1$.
Normalizing the external mass to one $m_1=1$, we are instructed to set
\begin{equation}
    \frac{\Delta_2}{\Delta_1}=m_2,\;\;\frac{[22]_n}{\Delta_1}=\sqrt{s}\,,
\end{equation}
and compute the CFT dispersion densities as in \eqref{CFTdispersiondensity}. In this case the averaging procedure is trivial and we get
\ba
\rho^{\text{bubble}}_\text{CFT}(s)=\lim_{\Delta_1\rightarrow\infty} \(\frac{a^\text{bubble}_{[22]_n}}{a^\text{free,11}_{[22]_n}}\)=\frac{1}{4\sqrt{s(s-4)} \sqrt{s(s-4m_2^2)}}\,,\qquad (s>4)\,,\\
\tilde\rho^{\text{bubble}}_\text{CFT}(s)=\lim_{\Delta_1\rightarrow\infty} \(\frac{a^\text{bubble}_{[22]_n}}{\tilde a^\text{free,11}_{[22]_n}}\)=\frac{1}{4\sqrt{s(4-s)} \sqrt{s(s-4m_2^2)}}\,,\qquad (s<4)\,,
\label{bubbledensADS}
\ea
where we have chosen a particular convenient overall normalization in \eqref{eq:acontact}.~\footnote{This was chosen to match with with the exact density computer for the amplitude further below. We could of course also match the overall normalization factor by carefully choosing the normalization of AdS propagators and couplings appropriately \cite{Paulos:2016fap}.}

\bigskip

Now we would like to compare our CFT density \eqref{bubbledensADS} with the spectral density of the flat space amplitude. From \eqref{eq:gralcomparison} we need the discontinuity of the amplitude. We could compute the full amplitude first and then take the discontinuity, but we will take a shortcut and get the density directly from Cutkosky rules \cite{Cutkosky:1960sp}.\footnote{See for example section 6.3.4 of \cite{itzykson2012quantum} for details.} (This method is especially useful when considering more complicated diagrams like the massive box below, where computing the exact analytic amplitude is difficult.) 
In principle we should consider all possible bubble diagrams where we take all permutations of the external legs, but the discontinuity (or imaginary part) for $s>4m_2^2$ is non-zero only for the diagram below. By replacing the internal propagators with momenta $q$ and $q_2=q-p_1-p_2$  with delta functions that put the particles on shell we find the discontinuity of the amplitude
\beqa
&&\mathcal{I}_{s}T^{\text{bubble}}(s)=\includegraphics[height=3cm,valign=c]{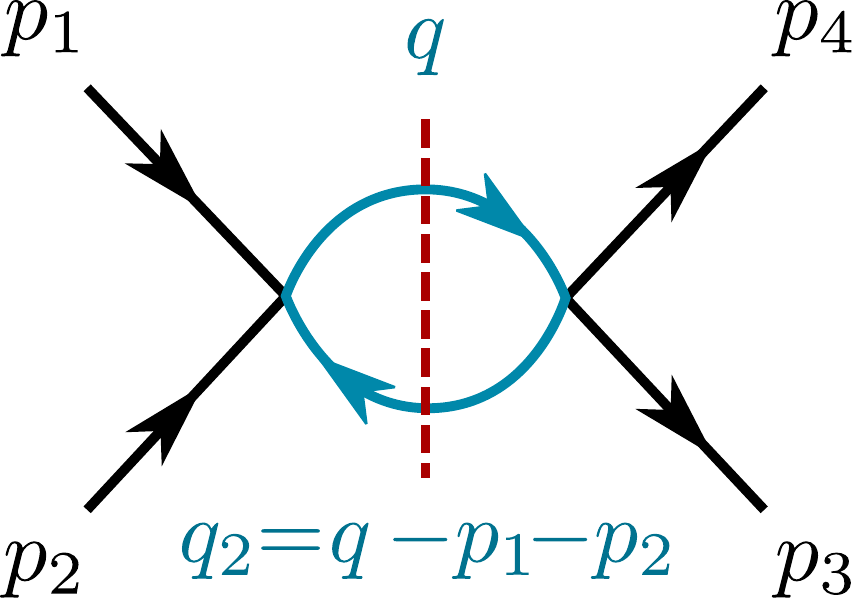}\nonumber\\
&&=\frac{(-i)^0}{2}\int\frac{d^2q}{(2\pi)^2}(2\pi)^2 \, \theta(q^0)\theta(-q_2^0) \, \delta(q^2-m_2^2)\delta(q_2^2-m_2^2)\nonumber\\
&&=\frac{1}{2}\int_{\pm}\frac{d|\vec{q}|}{8E|\vec{q}|}\delta\Big(|\vec{q}|-\sqrt{E^2-m_2^2}\Big)
=\frac{1}{2\sqrt{s(s-4m_2^2)}}
\eeqa
where we have chosen the center of mass frame to do the computation so $q=(E,\vec{q})$ where $2E=\sqrt{s}$  is the total energy. The $\pm$ indicates we need to sum over the two possible signs for the spatial component of the internal momenta.

From here we see that indeed we have a match between the flat space CFT density and the spectral density in the amplitude below and above threshold
\beqa
\rho^\text{bubble}(s)= \frac{\mathcal{I}_{s}T^{\text{bubble}}(s)}{2\sqrt{s(s-4)}} =\frac{1}{4\sqrt{s(s-4)} \sqrt{s(s-4m_2^2)}}= \rho^\text{bubble}_\text{CFT}(s)\,,\\
\tilde\rho^\text{bubble}(s)= \frac{\mathcal{I}_{s}T^{\text{bubble}}(s)}{2\sqrt{s(4-s)}} =\frac{1}{4\sqrt{s(4-s)} \sqrt{s(s-4m_2^2)}} = \tilde\rho^\text{bubble}_\text{CFT}(s)\,.
\eeqa

Finally, the constant term ambiguity in \eqref{bubbleP} can be fixed by comparing to the full $s,t,u$ crossing symmetric flat space amplitude which includes a constant term coming from the $s\rightarrow u=0$ term.

\subsection{Triangle diagram}

Let us now consider the term associated with $g_{122}^2g_{1122}$ in the expansion \eqref{Gexpand} which is the crossing symmetric sum of the triangle Witten diagrams.

\begin{figure}[t]
\centering
\includegraphics[width=0.77\textwidth]{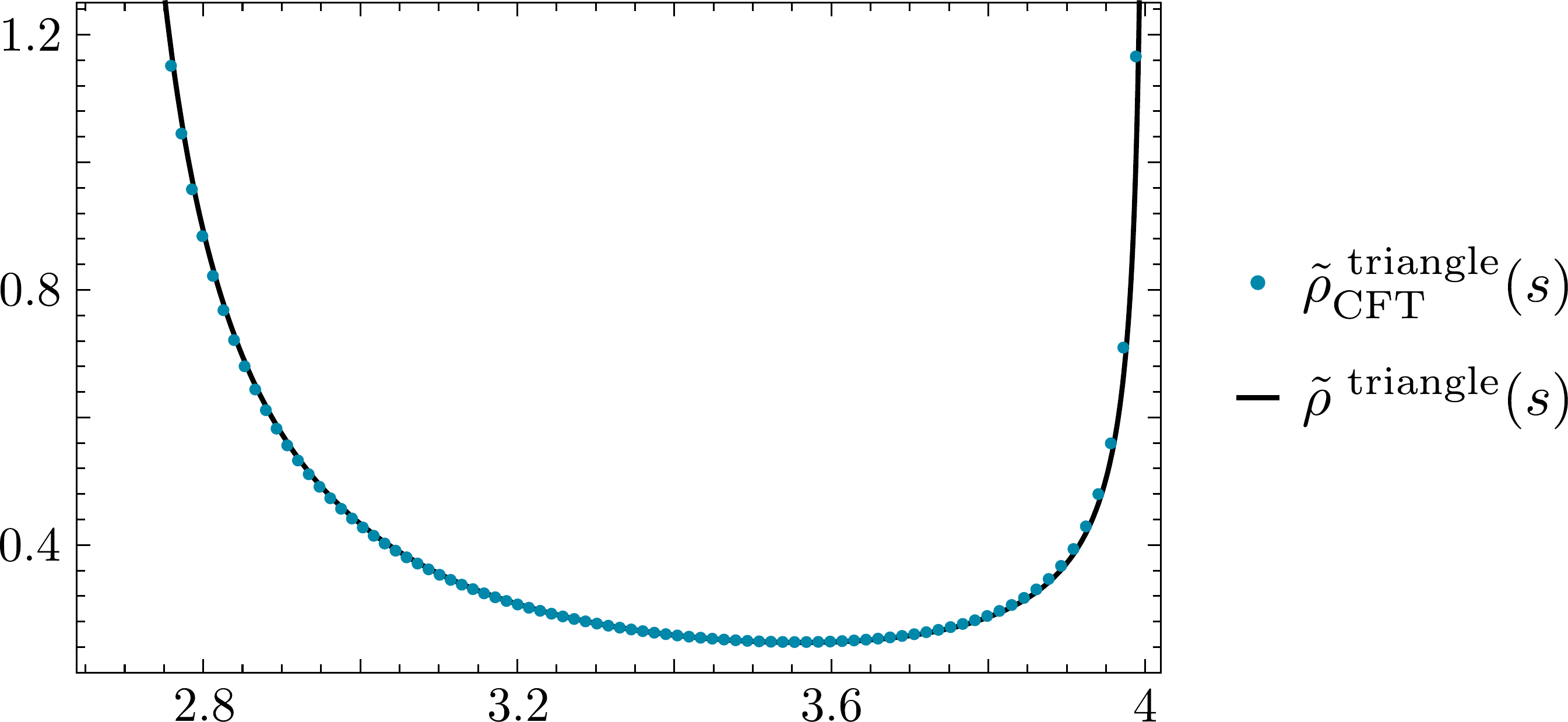}
\caption{Comparison between the amplitude's spectral density below threshold $\tilde\rho^\text{ triangle}(s)$ and large dimension ($\Delta_1=500$) numerical CFT density $\tilde\rho^\text{ triangle}_\text{CFT}(s)$ for $m_2^2=2/3$. }
\label{fig:triangle_plot}
\end{figure}

Taking again the Polyakov block expansion \eqref{eq:Pexpansion} we have 
\begin{equation}
    \mathcal{G}^{\text{triangle}}(w)=\sum_{n=0}^{\infty}a^{\text{triangle}}_{[22]_n}\mathcal{P}_{[22]_n}(w)+\text{contact}\,,
\end{equation}
where the sum is over the double trace operators of $\phi_2$ with even parity. Same as for the bubble diagram, we can extract the coefficients $a^{\text{triangle}}_{[22]_n}$ from the tree level diagrams we generate when cutting internal propagators in a way such that we have two external legs on the right and two on the left of the cut. In this case we are left with a $t$-exchange and (1122) contact diagrams. The relation between OPE coefficients is then: 
\begin{equation}
    a^{\text{triangle}}_{[22]_n}=\lambda^{\text{t,exchange}}_{11[22]_n}\times\lambda^{\text{contact}}_{11[22]_n}=\(a^\text{t,exchange}_{[22]_n}\times a^\text{contact}_{[22]_n}\)\bigg{/}a^\text{free,22}_{[22]_n}\,.
\end{equation}
This time one of the OPE coefficients $a^\text{t,exchange}_{[22]_n}$ is not available in closed form, but it can be computed via the recursion relations of \cite{Zhou:2018sfz} sketched in \eqref{eq:recursion}. This fact prevents us from computing the CFT density
\beq
\tilde\rho^{\text{bubble}}_\text{CFT}(s)=\lim_{\Delta_1\rightarrow\infty} \(\frac{a^\text{triangle}_{[22]_n}}{\tilde a^\text{free,11}_{[22]_n}}\)\,,\qquad s=\left(\frac{[22]_n}{\Delta_1}\right)^2<4\,
\eeq
analytically, so we will resort to a numerical comparison by solving the recursion relations for large enough dimensions. The result for $\Delta_1=500$ and $\Delta_2^2=2/3\Delta_1^2$ is shown in blue dots in figure~\ref{fig:triangle_plot}.

\bigskip

To compare with the flat space amplitude, let us extract the spectral density of the triangle diagram using the Cutkosky rule. Taking the $s$-channel discontinuity we find
\beqa
&&\mathcal{I}_{s}T^{\text{triangle}}(s)
=\includegraphics[height=3.2cm,valign=c]{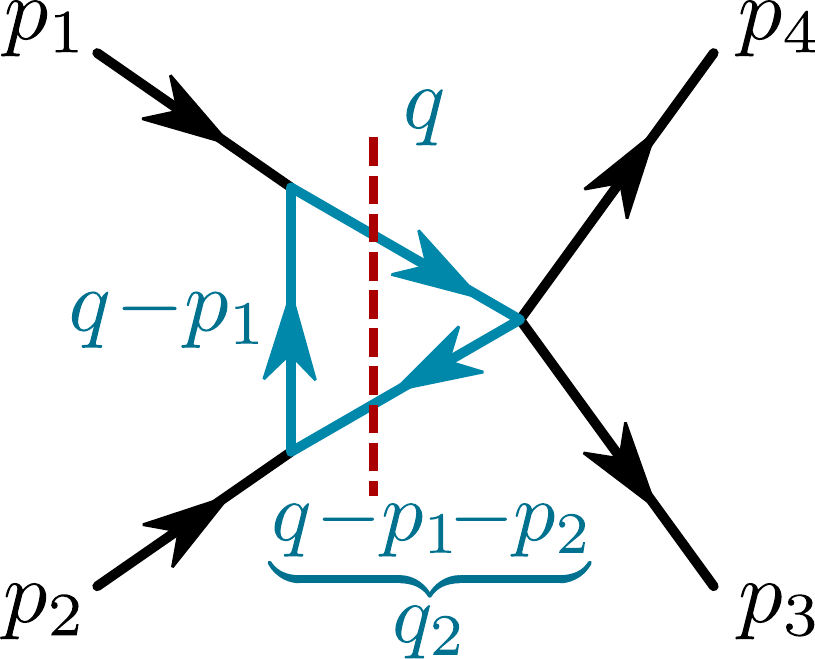}\nonumber\\
&&=\frac{(-i)}{2}\int\frac{d^2q}{(2\pi)^2}\frac{i}{(q-p_1)^2-m_2^2+i\epsilon}(2\pi)^2\theta(q^0)\theta(-q_2^0)\delta(q^2-m_2^2)\delta(q_2^2-m_2^2)\nonumber\\
&&=\frac{1}{2}\int_{\pm}\frac{d|\vec{q}|}{8E|\vec{q}|}\frac{\delta\Big(|\vec{q}|-\sqrt{E^2-m_2^2}\Big)}{\big((q-p_1)^2-m_2^2+i\epsilon\big)}=\frac{s-2}{2(1+m_2^2(s-4)) \sqrt{s(s-4m_2^2)}}\,.
\eeqa
where the convention is the same as the bubble case. Using \eqref{densitydiscT}, we then get the exact density below threshold
\beq\label{densitytriangle}
\tilde\rho^\text{ triangle}(s)=\frac{s-2}{4(1+m_2^2(s-4)) \sqrt{s(4-s)}\sqrt{s(s-4m_2^2)}}\,.
\eeq
In figure~\ref{fig:triangle_plot} we plot both the CFT and flat space densities and see that we get a perfect match. This is the case also for the density above threshold as well as for any value of $m_2$ such that $m_2^2>1/2$. We will see in section~\ref{sec:anom} what happens for internal masses below this value.

\subsection{Box diagram}

Finally, we have the crossing symmetric sum of box diagram from the term with $g_{122}^4$ in \eqref{eq:Pexpansion}. The procedure for the comparison is the same as for the triangle diagram. The Polyakov expansion reads
\begin{equation}
    \mathcal{G}^{\text{box}}(w)=\sum_{n=0}^{\infty}a^{\text{box}}_{[22]n}\mathcal{P}_{[22]n}(w)+\text{contact}\,.
\end{equation}
By cutting the box in two, we see that the coefficients $a^{\text{box}}_{[22]n}$ are given by the square of the (parity even) $t$-exchange diagram OPE coefficients
\begin{equation}
    a^{\text{box}}_{[22]n}=\big(\lambda^{\text{t,exchange}}_{11[22]_n}\big)^2=\(a^\text{t,exchange}_{[22]_n}\)^2\bigg{/}a^\text{free,22}_{[22]_n}\,.
\end{equation}
As for the triangle, these coefficients are not available in closed form, so once again we perform a numerical comparison with the flat space density.

\begin{figure}[t]
\centering
\includegraphics[width=0.75\textwidth]{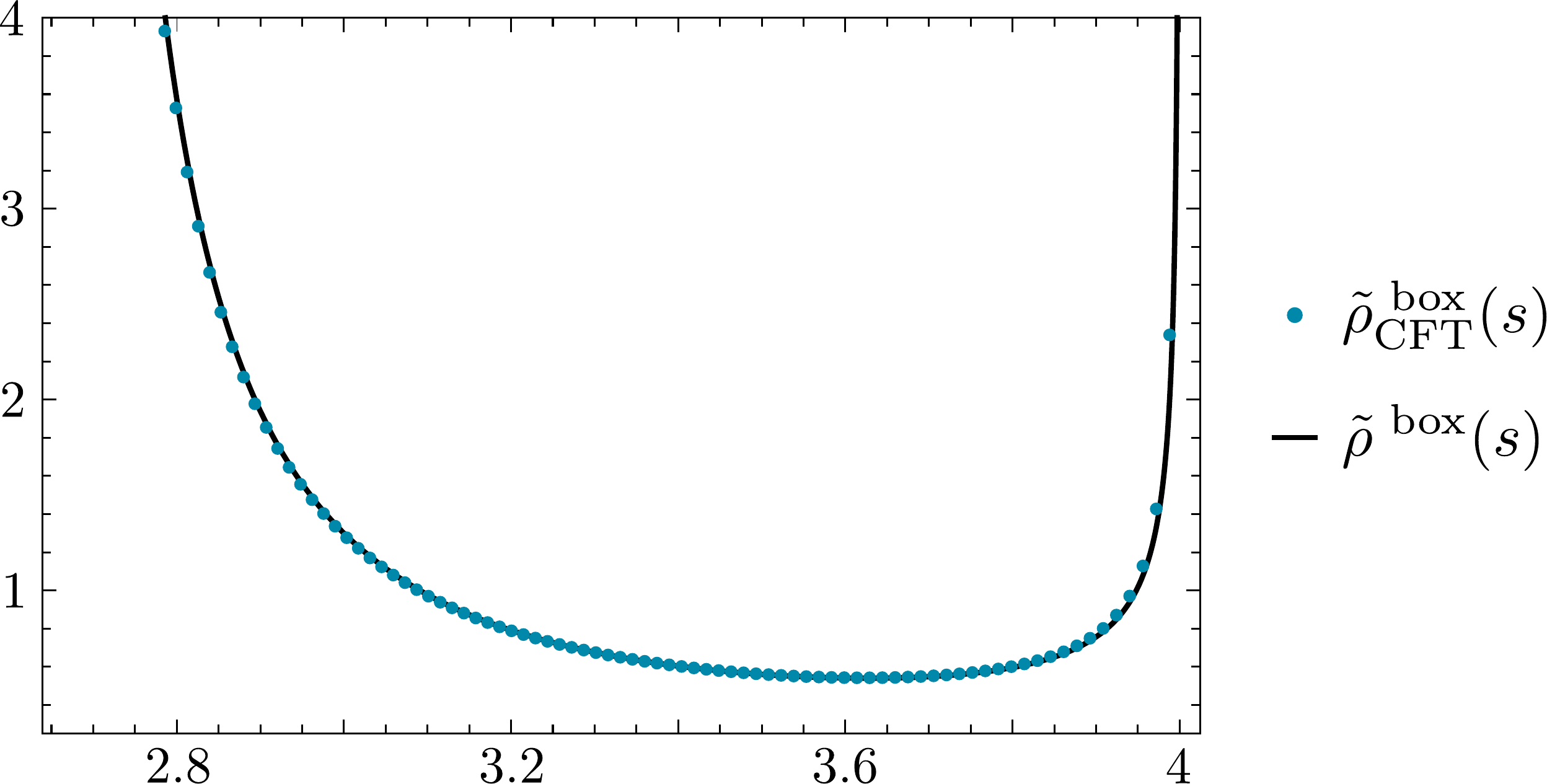}
\caption{Comparison between the amplitude's spectral density below threshold $\tilde\rho^\text{ box}(s)$ and large dimension ($\Delta_1=500$) numerical CFT density $\tilde\rho^\text{ box}_\text{CFT}(s)$ for $m_2^2=2/3$. }
\label{fig:box_plot}
\end{figure}

To compare with the flat space computation, we take the imaginary part or discontinuity of the amplitude for $s>4m_2^2$. As in the previous examples, we should consider all possible box diagrams but only the box and twisted box below have non zero discontinuity in this region. One can also make sense of this particular combination recalling that the even parity coefficients $a^\text{t,exchange}_{[22]_n}$ are extracted from the sum of $t$- and $u$-exchange Witten diagrams as in figure~\ref{fig:tuexchange}. This means that with the cutting and gluing procedure we have precisely the sum of box and twisted box diagrams as below. Using Cutkosky rules we arrive at
\begin{equation}
    \begin{aligned}
     &\mathcal{I}_{s}\[T^{\text{box}}(s)+T^{\text{twisted box}}(s)\]=\;\includegraphics[height=3.2cm,valign=c]{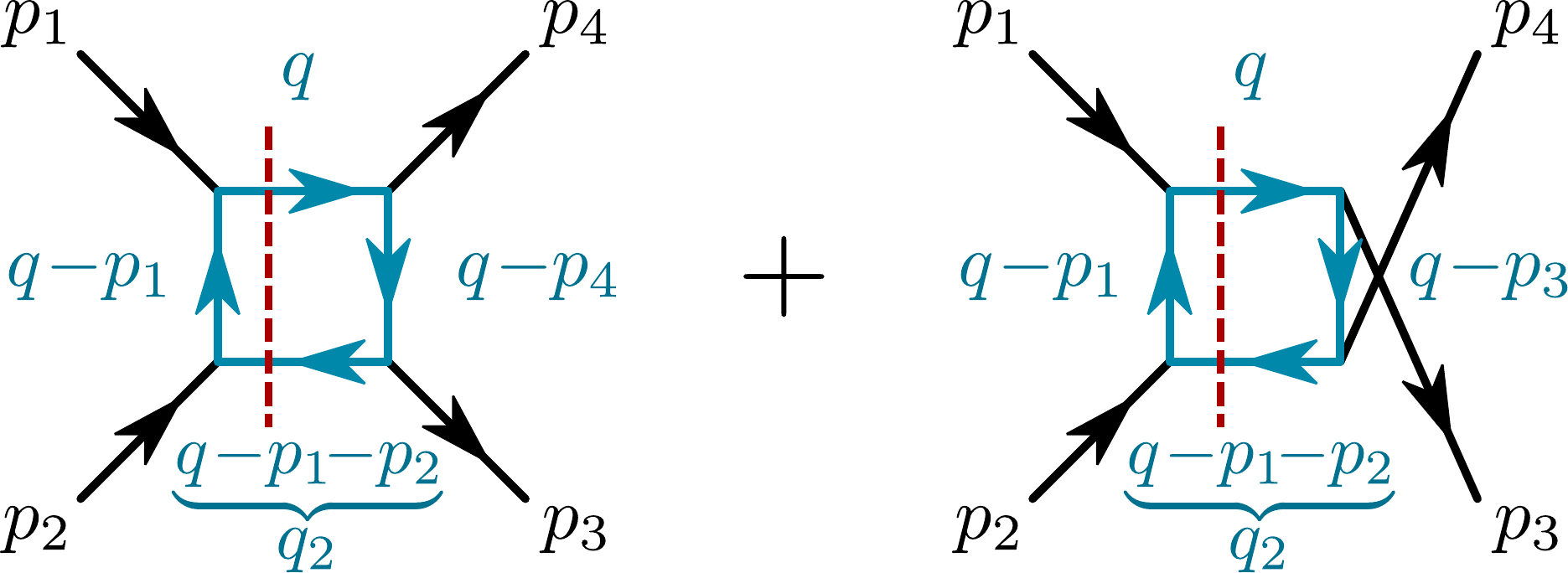}\\
&=\frac{(-i)^0}{2}\!\!\!\int\!\!\frac{d^2q}{(2\pi)^2}\frac{i}{(q\!-\!p_1)^2\!-\!m_2^2\!+\!i\epsilon}\frac{-i}{(q\!-\!p_4)^2\!-\!m_2^2\!-\!i\epsilon}(2\pi)^2\theta(q^0)\theta(-q_2^0)\delta(q^2\!-\!m_2^2)\delta(q_2^2\!-\!m_2^2)\\
&\quad+(p_4\leftrightarrow p_3)\\
&=\frac{1}{2}\int_{\pm}\frac{d|\vec{q}|}{8E|\vec{q}|}\frac{\delta\Big(|\vec{q}|-\sqrt{E^2-m_2^2}\Big)}{\big((q-p_1)^2-m_2^2+i\epsilon\big)\big((q-p_4)^2-m_2^2-i\epsilon\big)}+(p_4\leftrightarrow p_3)\\
&=\frac{(s-2)^2}{2(1+m_2^2(s-4))^2 \sqrt{s(s-4m_2^2)}}\,
    \end{aligned}
\end{equation}
and the corresponding spectral density is:
\beq
\tilde\rho^\text{ box}(s)=\frac{(s-2)^2}{4(1+m_2^2(s-4))^2 \sqrt{s(4-s)}\sqrt{s(s-4m_2^2)}}\,.
\eeq
In figure~\ref{fig:box_plot} we show the match between the expression above and the CFT density obtained numerically for $\Delta_1=500$ and $m_2^2=2/3$. As for the triangle diagram, this match persists for any $m_2^2>1/2$. For $m_2^2<1/2$ we encounter anomalous thresholds which are the topic of the next section.

\subsection{Anomalous thresholds}\label{sec:anom}

So far we have considered ranges of masses such that we avoid anomalous thresholds. The latter are singularities in the amplitude which do not have a direct interpretation in terms of intermediate physical states. One can obtain them by computing the associated Landau diagrams where internal particles are on-shell (see e.g. \cite{Eden:1966dnq}). In higher spacetime dimensions, these singularities are typically branch points whereas in two dimensions we encounter poles (a famous example are the Coleman-Thun double poles in the sine-Gordon model \cite{Coleman:1978kk}). For the diagrams discussed above we have anomalous thresholds for the triangle and box when the internal mass has values $\frac{m_1^2}{4}<m_2^2<\frac{m_1^2}{2}$.\footnote{We shall not consider $m_2^2<\frac{m_1^2}{4}$, as particle with mass $m_1$ would become unstable ($m_1>2m_2$) in this region. For the sake of clarity we write all $m_1$ factors in this discussion.}
As can be checked -- e.g. from direct computation of the Feynman integral or Landau equations -- the leading singularity of the triangle diagram is a simple pole
\beq
T^\text{triangle}_\text{anom}(s)\sim \frac{\pi m_1}{m_2^2\sqrt{4m_2^2-m_1^2}}\;\frac{1}{s-\(4m_1^2-\dfrac{m_1^4}{m_2^2}\)}\,,\qquad \frac{m_1^2}{4}<m_2^2<\frac{m_1^2}{2}\,.
\label{eq:Ttriangleanom}
\eeq
For the box diagram we have that the leading singularity is a double pole at the crossing symmetric point, occurring only at specific value $m_2^2=m_1^2/2$ 
\beq
T^\text{box, leading}_\text{anom}(s)\sim \frac{4\pi}{(s-2m_1^2)^2}\,,\qquad\qquad m_2^2=\frac{m_1^2}{2}\,,
\eeq
and the sub-leading singularity when reducing\footnote{When writing the amplitude in terms of Feynman parameters, the reduced diagrams come from some of these Feynman parameters vanishing, effectively setting the the length of the associated internal leg to zero.} the box diagram to the triangle above
\beq
T^\text{box, subleading}_\text{anom}(s)=T^\text{triangle}_\text{anom}(s)\,,\qquad\qquad \frac{m_1^2}{4}<m_2^2<\frac{m_1^2}{2}\,.
\label{eq:Tboxanom}
\eeq

In all these examples, anomalous thresholds are present when $m_2^2<1/2$, which implies that the two-particle production cut begins at $s_0<2$.
Equivalently, in CFT language,
the OPE contains a tower of states with dimensions $[22]_n$ beginning below $\sqrt{2}\Df$. But this violates our assumptions on the CFT spectrum spelled out in \ref{sec:boundbelow}, where they were required in order to ensure boundedness of the OPE. It is tempting therefore to conjecture that this unboundedness is related with anomalous behaviour in the S-matrix.  The amplitudes above provide then an excellent opportunity for understanding this.

Our first comment is that we do not believe that this unboundedness is merely an artifact of the flat space limit. That is, we believe the OPE is genuinely unbounded (and not just parametrically larger than our bounds) whenever our gap assumption fails. A rigorous proof that this is the case when $\Delta_0\leq\frac 43 \Df$ follows from considering the following family of functions:
\bea
\mathcal B_\alpha(z)=\frac{1}{[z(1-z)]^{(2-\alpha) \Df}}
\eea
This function is clearly crossing symmetric, and it admits a conformal block decomposition with a leading operator of dimension $\Delta_0=\alpha \Df$ (i.e. without identity). But furthermore the OPE is {\em positive} whenever $\Delta_0\leq \frac 43 \Df$. For instance \cite{Hogervorst:2017sfd}:\footnote{See also appendix B of \cite{Antunes:2021abs} for a related discussion.}
\beqa
\mathcal B_{\frac{4}3}(z)&=&\sum_{n=0}^\infty b_{n}\, G_{\frac 43 \Df+2n}(z|\Df)\,,\\
\text{with  }\quad b_n&=&\frac{\left(\frac{4 \Delta _{\phi }}{3}\right)_{2 n}^2 \, _3F_2\left(-2 n,\frac{2 \Delta _{\phi
   }}{3},2 n+\frac{8 \Delta _{\phi }}{3}-1;\frac{4 \Delta _{\phi }}{3},\frac{4 \Delta _{\phi }}{3};1\right)}{(2 n)!
   \left(2 n+\frac{8 \Delta _{\phi }}{3}-1\right)_{2 n}}>0\,.\nonumber
\eeqa
This means that given a general unitary CFT correlator $\mathcal G$ with $\Delta_0\leq \frac 43 \Df$, we can obtain a new one by doing $\mathcal G\to \mathcal G+\lambda\, \mathcal B_{\frac 43}$ with arbitrarily large positive $\lambda$. This establishes there exist CFT correlators for which quantities like $\tilde \rho$ and $\rho$ are not in general finite, at least when the gap is below $\frac 43\Df$. This simple example explains the general mechanism establishing that it is not possible to bound the OPE whenever the gap is below some critical value: the existence of unitary correlators without identity whose overall coefficient may therefore become arbitrarily large.
\footnote{A general argument that this must be the case is as follows: if there is no bound on a particular OPE coefficient, then it must be possible to construct unitary families of correlators $\mathcal G_{\lambda}$ where that OPE coefficient is some large number $\lambda$. Then $\partial_{\lambda} \mathcal G_{\lambda}|_{\lambda=\infty}$ is a unitarity solution to crossing without identity.}
Can we improve our proof for any gap below $\sqrt{2}\Df$? Unfortunately, an exploration of generalized free field correlators of composite operators (where the above function arises) does not seem to lead to solutions with a gap higher than $\frac{4}{3} \Df$. Perhaps we did not try hard enough, or perhaps such solutions must necessarily be interacting. Let us proceed assuming such solutions do exist, to avoid an (in our view) artificial separation between gaps below $\frac{4}{3}\Df$ and above it.

Let us test our conjecture linking OPE unboundedness and anomalous behaviour on the example of the box diagram, in the region $ \frac{m_1^2}{4}<m_2^2<\frac{m_1^2}{2}$. In this case, the correct scattering amplitude is obtained by taking 
\beqa
\tilde\rho^\text{box}(s) &\rightarrow& \tilde\rho^\text{box}(s) + \tilde\rho^\text{box}_\text{anom}(s)\,, \qquad\qquad \frac{m_1^2}{4}<m_2^2<\frac{m_1^2}{2}\,,\\
\tilde\rho^{\text{box}}_{\text{anom}}(s)&=&-\frac{i\pi s}{4(s-4)}\delta\(s-\frac{1}{m_2^2}\)\,.
\eeqa
That is, the naive analytic continuation of the density to $m_2^2<\frac{m_1^2}{2}$ fails, as we must add a term with a delta function reproducing the simple pole in \eqref{eq:Tboxanom} and \eqref{eq:Ttriangleanom}. 
We should point out that, in fact, the density is not uniquely defined in the region $(s_0,4-s_0)$. This is because
\bea
\int_{s_0}^{4} \ud s' \widetilde K(s,s') \delta \tilde \rho(s')=0\,,\qquad \delta\tilde \rho(s)=\delta\tilde \rho(4-s)\,,\quad \delta\tilde \rho(s>4-s_0)=0
\eea
owing to the antisymmetry property $\widetilde K(s,s')=-\widetilde K(s,4-s')$.

\begin{figure}[t]
\centering
\includegraphics[width=.82\textwidth]{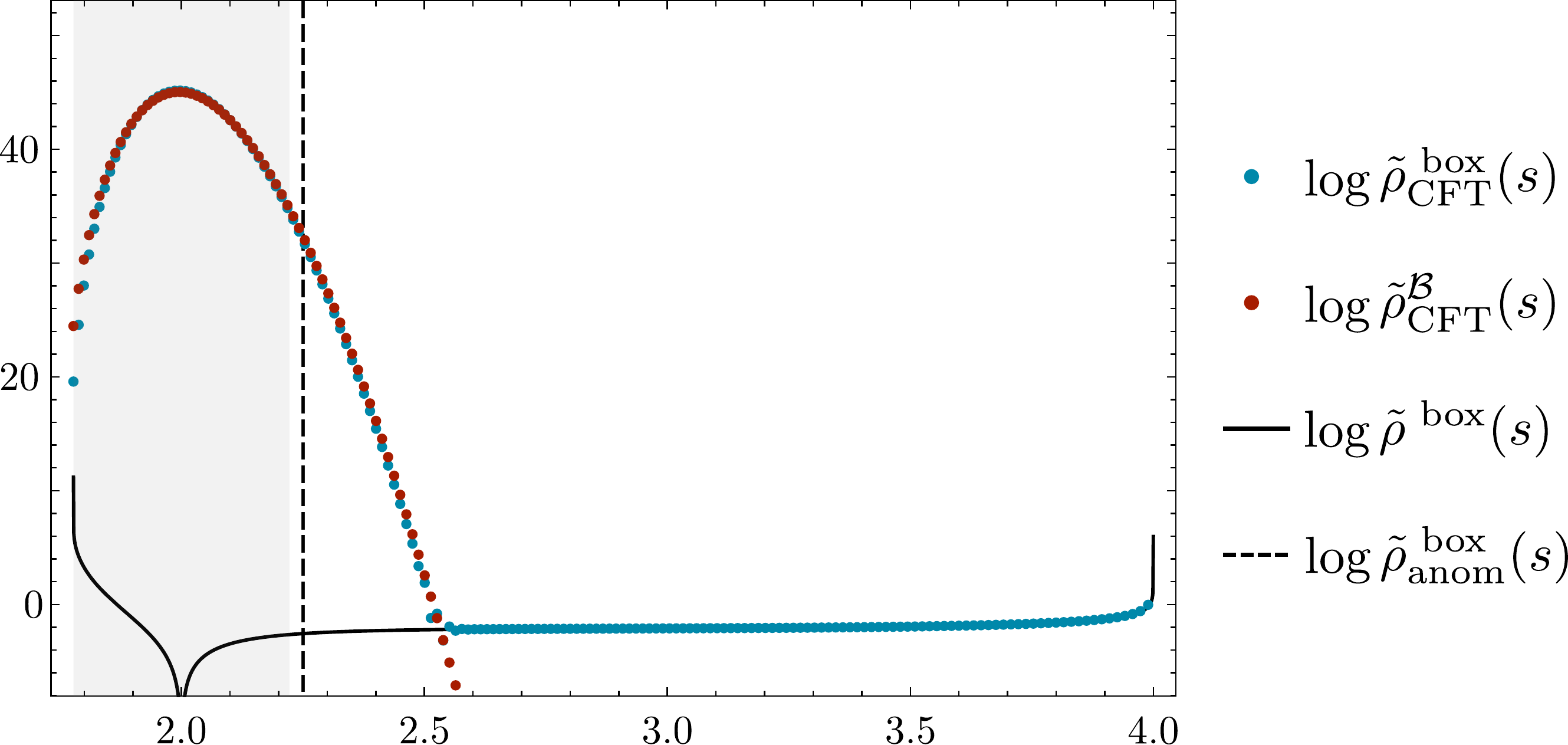}\vspace{0.2cm}
\caption{Densities $\tilde \rho$ in the presence of anomalous thresholds, in the case $2\Delta_2=\frac 43 \Df \Leftrightarrow m_2^2=\frac 49$. The densities $\tilde \rho^{\mathcal B}_{\mbox{\tiny CFT}},\tilde \rho_{\mbox{\tiny CFT}}^{\mbox{\tiny box}}$ are computed with $\Df=500$. Above a critical $s\approx 2.5$ the CFT and exact densities agree, but below this value the mismatch is exponential (getting worse with increasing $\Df$). The anomalous pole contribution is shown as a dashed line. The ambiguity in $\tilde \rho$ is only present in the shaded band. 
}
\label{fig:anom}
\end{figure}

With this caveat in mind, we can compare the expected density with the one computed from the OPE data in the CFT. This is shown in figure \ref{fig:anom}.
We have chosen to set $\Delta_0=\frac 43 \Df$ for simplicity, but the figure looks similar for other choices. We see the density computed from the CFT only matches the one obtained from the amplitude above some critical value of $s$. In our explorations we find that it varies very little with $\Df$, but it depends strongly on $\Delta_0$. Also, this value is always above $4-s_0$, so that the ambiguity in the definition of $\tilde \rho$ cannot help us cure the mismatch. 
 Furthermore, we find the anomalous pole always lies in the region where there is a mismatch. More strikingly, it seems that the anomalously large piece of the CFT density closely resembles the density computed from function $\mathcal B$, that is
 \bea
 \tilde \rho^{\mathcal B}_{\mbox{\tiny CFT}}(s)=\lim_{\Df\to \infty}\frac{b_n}{\tilde a^{\ttext{free}}_{\frac 43 \Df+2n}}\,, \qquad s=\left(\frac{\frac 43 \Df+2n}{\Df}\right)^2\,.
 \eea
 These results suggest that anomalous thresholds are indeed directly correlated with unboundedness of the OPE. They also suggest that this unboundedness could follow from the appearance of a large component in the CFT OPE which satisfies crossing symmetry by itself. Therefore, it seems that to obtain the correct flat space limit of the full correlator this piece must be first subtracted and its limit handled separately. Presumably this will lead to the appearance of new, anomalous, contributions to the CFT density. Finally, note that these results are consistent with the picture in \cite{Komatsu:2020sag}: the unbounded piece of the OPE should map into the ``wrong'' saddles corresponding to AdS Landau diagrams.


\section{Extremality}
\label{sec:nonpert}
We have established a mapping from certain families of CFT correlators and $S$-matrices. In this section we will explore general properties of this mapping and how they lead to a detailed link between extremal S-matrices, extremal CFTs, and bootstrap problems in both contexts.

\subsection{Extremality and the flat space limit}
\subsubsection*{Information loss}
Our basic formula states that
\bea
S(s)=\mathcal F[\mathcal G(z_s)]\,.
\eea
The $S$-matrix may be computed for physical kinematics via the phase shift formula \reef{eq:phaseshift}, and elsewhere on the complex plane from the dispersion relation \reef{CFTdispersiondensity}. Our definition of the operation $\mathcal F$ is linear, so that given two distinct CFT families $\mathcal G_1, \mathcal G_2$ matching with $S$-matrices $S_1$ and $S_2$, we can build an entire family of new solutions: 
\begin{multline}
S_{1+2}(s)=x S_1(s)+(1-x) S_2(s)=x\,\mathcal F[\mathcal G_1(z_s)]+(1-x)\,\mathcal F[\mathcal G_2(z_s)]\\
=\mathcal F[x\, \mathcal G_1(z)+(1-x)\mathcal G_2(z)]\,,\qquad x\in [0,1]
\end{multline}
The constraint $x\in[0,1]$ simultaneously guarantees that the identity operator appears in the sum of CFT correlators with unit coefficient, and that the unitarity condition $|S_{1+2}|\leq 1$ holds from those of $S_1, S_2$. More interestingly, starting from any two $S$-matrices, their product is also an $S$-matrix. This is ensured by
\bea
\mathcal F[\mathcal G_1(z_s) \mathcal G_2(z_s)]=\mathcal F[\mathcal G_1(z_s)] \mathcal F[\mathcal G_2(z_s)] \label{eq:prodprop}
\eea
which follows from elementary properties of the limit and of analytic continuation. This property is a bit surprising from the way we have effectively constructed the $\mathcal F$ operation, which requires the OPE, and we clarify why it is true in appendix \ref{app:products}.

These properties imply that the $\mathcal F$ operation loses information. For instance, we can take any CFT correlator and multiply it by the generalized free field correlator, the flat space limit will unaffected since $\mathcal F[\mathcal G^+]=1$. Or we could start with a correlator and add to it the sum of generalized free fermion and boson correlators, whose flat space limits ($S=\pm 1$) add up to zero. In both cases the OPE structure of the CFT correlator before and after multiplication/addition is very different, but in the flat space limit these differences are subleading and get washed out. Thus, distinct families of CFT correlators can nevertheless lead to the same $S$-matrix. This is of course in line with our general expectations: placing a QFT in AdS for specific choices of boundary conditions and curvature couplings will lead to specific families of CFTs, but any such choice must nevertheless have the same flat space limit.

\subsubsection*{Extremality}

A particularly interesting class of 2d $S$-matrices 
are those which saturate unitarity for physical kinematics, i.e. $|S(s)|=1$ for $s>4$, which we will call {\em extremal}. This is generally found to be the case for $S$-matrices saturating bootstrap bounds (see e.g. \cite{Cordova:2019lot}), and it is also true for integrable models \cite{Dorey:1996gd,Bombardelli:2016scq}. Extremal S-matrices can be expressed as products of CDD factors, which describe zeros or poles, and take the form:
\bea
S^{\mbox{\tiny pole}}_{s_b}(s)=\frac{\sqrt{s(4-s)}+\sqrt{s_b(4-s_b)}}{\sqrt{s(4-s)}-\sqrt{s_b(4-s_b)}}\,, \qquad S^{\mbox{\tiny zero}}_{s_r}(s)=\frac{\sqrt{s(4-s)}-\sqrt{s_r(4-s_r)}}{\sqrt{s(4-s)}+\sqrt{s_r(4-s_r)}}\,.
\eea
A general extremal $S$-matrix is thus written as
\bea
|S|=1 \quad \Rightarrow \qquad S(s)=\prod_i S^{\mbox{\tiny pole}}_{s_i}(s) \prod_j S^{\mbox{\tiny zero}}_{s_j}(s)\,. \label{eq:prodcdd}
\eea
Let us examine what the condition $|S|=1$ means for the CFT data. Using the phase shift formula we have
\bea
S(s)=\lim_{\Df\to \infty} \sum_{\Delta>2\Df} 2\left(\frac{a_{\Delta}}{\af}\right) \Nh(\Delta,s) e^{-i \pi(\Delta-2\Df)}\label{eq:phasesh2}
\eea
with
\bea
\lim_{\Df\to \infty} \sum_{\Delta>2\Df} 2\left(\frac{a_{\Delta}}{\af}\right) \Nh(\Delta,s)=1\,.
\eea
Recall that the Gaussian $\mathcal N_{\Df}$ implies that in these sums the only states which contribute lie inside a narrow window of width $\sqrt{\Df}$ centered around $\Delta=\sqrt{s}\Df$. The only way in which the unitarity condition can be saturated is if in \reef{eq:phasesh2} the phases remain coherent in any such window. More precisely, this should be true of the phases of the states which contribute predominantly to the OPE (see apendix \ref{app:products}). Such states should be described by a single tower of operators with dimensions $\Delta_n=2\Df+2n+\gamma_n(\Df)$ satisfying: 
\bea
\gamma_n(\Df)\underset{\substack{n,\Df\to \infty\\ n/\Df~\mbox{\tiny fixed}}}{\to}\gamma(s)\,, \qquad s:=\lim_{\substack{n,\Df\to \infty\\ n/\Df~\mbox{\tiny fixed}}} \left(\frac{2\Df+2n}{\Df}\right)^2
\eea
That is, the anomalous dimensions can be promoted to a slowly varying function of $s$ in the large $\Df$ limit. In this case not only is unitarity saturated but we can explicitly describe the S-matrix in terms of the spectrum of the CFT:
\bea
S(s)=\lim_{\Df\to \infty} \sum_{\Delta>2\Df}2 \left(\frac{a_{\Delta}}{\af}\right) \Nh(\Delta,s) e^{-i \pi\,\gamma_n(\Df)}=e^{-i\pi \gamma(s)}\,, \qquad \mbox{if}\quad |S|=1\,. \label{eq:anomdimformula}
\eea

At this point, we would like to pose two questions. The first is: can we construct families of CFT correlators which in the flat space limit describe any S-matrix of the form \eqref{eq:prodcdd}? In the following two subsections we will formulate and solve bootstrap optimization problems whose optimal solutions are CFT correlators which map to single CDD pole and CDD zero factors. Thanks to the product property \reef{eq:prodprop}, the answer to this question is therefore affirmative.

The second question is, can such families be chosen to be extremal: that is, that the OPE in the corresponding correlators contains a single tower of operators not effectively but exactly, even away from the strict $\Df\to \infty$ limit. Tensor products and sums of correlators do not satisfy this property, but extremal correlators do arise as optimal solutions of bootstrap problems \cite{Mazac:2018ycv,Paulos:2019fkw}. Therefore, here we also believe the answer is affirmative: in section \ref{sec:bootstraps} we will show that there is a direct mapping between large classes of S-matrix and CFT bootstrap problems. This means that the associated optimal solutions are also mapped into each other, thus establishing a link between extremal S-matrices and families of extremal correlators. 

\subsection{Bootstrapping the CDD pole}\label{CDDpole}

We will begin by describing an extremal correlator which in the flat space limit describes the CDD pole. Although it is already known that this correlator can be obtained by maximizing the OPE coefficient of a state corresponding to the pole \cite{Paulos:2016fap,Mazac:2018mdx}, here we will follow a different route based on master functionals, by maximizing the correlator.\footnote{The correlator maximization problem with a gap $\Delta_0$ is expected to be extremized by the same correlator which maximizes the OPE coefficient at $\Delta_0$ \cite{Paulos:2020zxx}. }

Consider then the problem of maximizing the value of a CFT correlator, $\mathcal G(w)$, whose spectrum starts at some dimension $\Delta_0>
\sqrt{2}\Df$, i.e. it satisfies the strong OPE condition. A bound may be obtained by constructing a functional $\Omega_w^{\ttext{int}}$ satisfying the properties
\bea
\Omega_w^{\ttext{int}}(\Delta)\geq G_{\Delta}(w|\Df)\,, \qquad \mbox{for all}\quad \Delta\geq \Delta_0\,,
\eea
since acting with such a functional on the crossing equation is easily seen to lead to
\bea
\mathcal G(w)\leq \mathcal P_{0}^{\ttext{int}}(w|\Df)\,, \qquad \mathcal P_{\Delta}^{\ttext{int}}(w):=G_{\Delta}(w|\Df)-\Omega_w^{\ttext{int}}(\Delta)\,.
\eea
If $\mathcal P_0^{\ttext{int}}(w)$ matches a physical CFT correlator $\mathcal P_0^{\ttext{int}}(w)=\mathcal G^{\ttext{int}}(w)$ then the bound is optimal and saturated by that correlator. Note that we should think of $\mathcal P_{\Delta}^{\ttext{int}}(w)$ as a kind of interacting version of ordinary Polyakov blocks, and the superscript serves to remind us of this fact.

Such a functional can be defined in the same way as the master functionals $\Omega^\pm_w$ of section \ref{Mfunctional}, but with the important difference that for general $\Df$ the kernels $g_w^{\ttext{int}}$ and $f_w^{\ttext{int}}$ aren't simply related. However, for $\Df\to \infty$ we can choose:
\bea
g^{\ttext{int}}_w(z)=\hat g_w^{\ttext{int}}(z)+\delta(w-z)\,, \qquad \hat g_w^{\ttext{int}}(z) &\underset{\Df\to \infty}{\sim} (1-z)^{2\Df-2} |f_w^\ttext{int}(\mbox{$\frac{1}{1-z}$})|\,,\label{eq:largedphikernels}
\eea
In the limit of large $\Df$ the remaining constraint on $f_w^{\ttext{int}}$ becomes
\bea
\mathcal R_z f^{\ttext{int}}_w(z)\sim -\delta(z-w)-\delta(1-z-w) \qquad \mbox{for}\quad z\in(0,1) \label{eq:deltafuncsinglargedphi}
\eea
which can be solved as
\bea\label{fstar}
f^{\ttext{int}}_w(z)=4 K(s_w,s_z)\,\frac{S^{\ttext{int}}(s_w)}{S^{\ttext{int}}(s_z)}+\ldots
\eea
where we assumed $S^{\ttext{int}}(s)$ that has the analyticity properties of an S-matrix, but that it does not have zeros for any complex $s$. The corrections shown as $\ldots$ depend on the details of $S^{\ttext{int}}$, and would allow us to relax the absence of zeros, but we will not need to write them out explicitly for the time being. Note that with $S^\ttext{int}=\pm 1$ we recover the master functionals $f_w^{\pm}$.

The sum rule for the functional $\Omega^{\ttext{int}}_w$ can be stated as validity of the interacting Polyakov bootstrap:
\bea
\Omega_w^\ttext{int}(0)+\sum_{\Delta\geq \Delta_0} a_\Delta \Omega_w^\ttext{int}(\Delta)=0\quad \Leftrightarrow \quad \mathcal G(w)=\mathcal P_0^{\ttext{int}}(w)+\sum_{\Delta\geq \Delta_0} a_{\Delta} \mathcal P_{\Delta}^{\ttext{int}}(w)
\eea
The computation of these interacting Polyakov blocks is essentially identical to the free case, and we will give the result below. This equation translates into a dispersion relation for the CFT, which we will write directly in the flat space limit in terms of the S-matrix: 
\bea\label{Sdual}
S(s)=S^{\ttext{int}}(s)-S^{\ttext{int}}(s)\int_{s_0}^\infty \ud s' \mathcal I_{s'}\left[ \widetilde K(s,s')\left(1-\frac{S(s')}{S^{\ttext{int}}(s')}\right)\right]
\eea

The bound can now be obtained as follows. The positivity conditions on $\Omega^{\ttext{int}}$ are the statement that
\bea
\mathcal P^{\ttext{int}}_{\Delta}(w)\leq 0\,, \qquad \Delta\geq \Delta_0\,.
\eea
In the dispersion relation this is the constraint that the integral is positive. Indeed if that's the case the dispersion relation immediately implies the bound $S(s_w)\leq S^{\ttext{int}}(s_w)$. We split the integral into two pieces:
\beq
S^{\ttext{int}}(s_w)
\int_{s_0}^4 \ud s' \widetilde K(s_w,s')\mathcal I_{s'}\left[-\frac{S(s')}{S^{\ttext{int}}(s')}\right]
+
S^{\ttext{int}}(s_w)\int_{4}^\infty \ud s'  K(s_w,s')\mathcal R_{s'}\left[1-\frac{S(s')}{S^{\ttext{int}}(s')}\right]
\eeq
The second term on the right represents the contributions above threshold. Setting
\ba
|S^{\ttext{int}}(s)|=1\,, \qquad s>4
\ea
and using $|S(s)|\leq 1$,
then the integral will be manifestly non-negative as long as $S^{\ttext{int}}(s_w)\geq 0$. As for the first term, notice that $\mathcal I_s S(s)=\tilde \rho(s)\geq 0$, and hence we can make those contributions positive as well by demanding that $S^{\ttext{int}}(s)$ is real and negative for $s_0<s<4$.

We are nearly done. The constraints on $S^{\ttext{int}}$ imply that it can be written as a product of CDD poles, with pole positions $s_p\leq s_0$ and chosen such that both $S^{\ttext{int}}(s_w)\geq 0$ and $S^{\ttext{int}}(s_0\leq s\leq 4)\leq 0$. Any  $S^{\ttext{int}}$ of this form gives a valid upper bound, but the optimal such bound is obtained by choosing an isolated CDD pole at precisely $s_0$, since in this case it can be saturated an S-matrix satisfying our assumptions. We conclude
\ba
S(s)\leq S^{\ttext{pole}}_{s_0}(s)\,.
\ea

To wrap up, let us check that the CFT extremal correlator $\mathcal P_0^{\ttext{int}}$ which saturates our bound is related to the extremal S-matrix in the way we expect, namely $\mathcal F[\mathcal P_0^{\ttext{int}}]=S^{\ttext{int}}$. Although below strictly speaking we have in mind the case $S^{\ttext{int}}=S^\ttext{pole}_{s_0}$, we will stick to the 'int' notation since most steps in our computation hold more generally. We will therefore compute:
\bea
\mathcal P_0^{\ttext{int}}(w)=G_0(w|\Df)-\Omega_w^{\ttext{int}}(0)
\eea
To obtain $\Omega^{\ttext{int}}(0)$ we use the fact that it is a valid functional, and so the corresponding sum rule must be satisfied by any CFT correlator. In particular it must be satisfied by a generalized free field. Therefore
\bea
\Omega_w^{\ttext{int}}(0)=-\sum_{n=0}^\infty a_{\Delta_n}^\ttext{free} \Omega_w^{\ttext{int}}(\Delta_n)=1+\frac{1}{(1-w)^{2\Df}}+\sum_{n=0}^\infty a_{\Delta_n}^\ttext{free} \mathcal P_{\Delta_n}^{\ttext{int}}(w)
\eea
with $\Delta_n=2\Df+2n$. To compute the interacting Polyakov blocks, we use that in this case the full interacting master functional is written
\bea
f^{\ttext{int}}_w(z)=4 K(s_w,s_z)\,\frac{S^{\ttext{int}}(s_w)}{S^{\ttext{int}}(s_z)}+f_{z_0}^{sG}(z)\, E_{\Delta_0}(w|\Df)
\eea
with $s_0=4 z_0$ and $f^{sG}$ the sine-Gordon functional \reef{eq:fsG}. The last term is chosen to insure that $\mathcal P^{\ttext{int}}_{\Delta_0}(z)=0$. The computation of the interacting Polyakov blocks is now almost exactly the same as the one we did for the free case in sections \ref{sec:polyflat} and \ref{sec:flatgeneral}. In particular we find
\bea
\sum_{n=0}^\infty a_{\Delta_n}^\ttext{free} \mathcal P_{\Delta_n}^{\ttext{int}}(w)=S^{\ttext{int}}(s)\int_4^{\infty} \ud s' K(s,s')\mathcal R_{s'}\left[1-\frac{1}{S^{\ttext{int}}(s')}\right]
+a_{\Delta_0}^{sG}\, E_{\Delta_0}(w|\Df)
\eea
where we used $\sum_{n=0}^{\infty} a_{\Delta_n}^\ttext{free} \omega^{sG}(\Delta_n)=-\omega^{sG}(0)=a_{\Delta_0}^{sG}$. Evaluating the first line and putting everything together we find
\bea
\mathcal G^{\ttext{int}}(w)\equiv \mathcal P_0^{\ttext{int}}(w)=\frac{1}{w^{2\Df}}+\frac{1}{(1-w)^{2\Df}}+a_{\Delta_0}^{sG}\, E_{\Delta_0}(w|\Df)+S^{\ttext{int}}(s_w)\,.
\eea
In this way we see that the flat space limit of the extremal correlator for the conformal bootstrap problem is indeed the extremal S-matrix for the corresponding S-matrix bootstrap problem.

\subsection{Bootstrapping the CDD zero}
\label{sec:bootzero}
We will now show that there is an extremal CFT that leads to the CDD zero S-matrix in the flat space limit. 

Let us introduce a functional $\Omega^{\partial}_{w}$ whose kernels are given by
\ba
f_w^{\partial}(z)&=4 K(s_w,s_z) \frac{d_w}{S^{\ttext{zero}}_{s_w}(s_z)}\,.
\ea
with
\bea
d_w:= \frac{\ud}{\ud z} S^{\ttext{zero}}_{s_w}(s_z)\bigg|_{z=w}=\frac{2w-1}{4w(1-w)}\,.
\eea
To satisfy the gluing condition we  need to set
\ba
g_w^{\partial}(z)&=(1-z)^{2\Df-2} |f_w^\partial(\mbox{$\frac{1}{1-z}$})|+\delta'(w-z)\,.
\ea
Let us set
\bea
\mathcal P_{\Delta}^{\partial}(w)\equiv \partial_w G_{\Delta}(w|\Df)-\Omega^{\partial}_w(\Delta)\,.
\eea
Note that the $f^\partial_w,g^\partial_w$ kernels above are {\em not} derivatives of the master functional kernels $f^+_w,g^+_w$, and accordingly the $\mathcal P^\partial_\Delta$ are {\em not} derivatives of ordinary Polyakov blocks.
The sum rule of $\Omega_w^\partial$ now yields
\bea
\mathcal G'(w)=\mathcal P^{\partial}_0(w)+\sum_{\Delta_0\leq \Delta_b\leq 2\Df} a_{\Delta} \mathcal P^{\partial}_{\Delta_b}(w)\,-d_w\int_4^\infty \ud s K(s_w,s)\,\mathcal R_s\left[1-\frac{S(s)}{S^{\ttext{zero}}(s)}\right]
\eea
It is easy to compute
\ba
\mathcal P_0^\partial(w)&=\partial_w\left[\frac{1}{w^{2\Df}}+\frac{1}{(1-w)^{2\Df}}\right]+d_w\\
a_{\Delta_b}\mathcal P_{\Delta_b}^\partial(w)&=\partial_w E_{\Delta_b}(w)+2\frac{m_b}{\Df} \widetilde K(s_w,m_b^2) \frac{d_w}{S_{s_w}(m_b^2)}
\ea
Positivity now implies
\bea
\mathcal G'(w)\leq \partial_w\left[\frac{1}{w^{2\Df}}+\frac{1}{(1-w)^{2\Df}}\right]+d_w\,, \qquad \frac 12< w<\frac 14 \left(\frac{\Delta_0}{\Df}\right)^2\label{eq:boundder}
\eea
with optimality achieved when $S(s)=S^{\ttext{zero}}_{s_w}(s)$. We conclude that the CDD zero S-matrix arises from the flat space limit of the family of correlators which saturates an upper bound on the {\em derivative} of the correlator at a point. This nicely ties in with the S-matrix derivation in \cite{Doroud:2018szp}.

An apparent puzzle is how to characterize such CFTs from the point of view of the OPE, since they have no bound states. In appendix \ref{app:cddzero} we argue that these CFTs arise as a deformation of the generalized free boson where the leading scalar dimension $\Delta_0$ is pushed parametrically close to the maximal gap $2\Df+1$.

\subsection{Dual S-matrix and conformal bootstraps}
\label{sec:bootstraps}
The goal of this section is to explain how the S-matrix bootstrap in its dual formulation \cite{Cordova:2019lot,Guerrieri:2020kcs,He:2021eqn} is related to the conformal bootstrap written in the language of functionals.

\subsubsection*{S-matrix bootstrap}
We begin in the S-matrix picture. We want to solve the optimization problem:
\begin{equation}
    \text{max }\mathcal{F}_{P}^{\ttext{S-mat}}\,, \qquad \mathcal F_{P}^{\ttext{S-mat}}:=\int_{s_0}^{4}\ud s\,\tilde{c}(s)\tilde{\rho}(s)+\int_{4}^{\infty}\ud s\, c(s)\rho(s)\label{fprimal}
\end{equation}
with
\begin{equation}\label{CFTdispS}
    S(s)=1+\int_{s_0}^4 \ud s' \, \widetilde K(s,s')\,\tilde \rho(s')-\int_4^{\infty} \ud s' \, K(s,s')\,\rho(s')\;,\qquad |S(s^+)|\leq 1 
\end{equation}
The primal variables are $\rho,\tilde \rho \geq 0$ and $c,\tilde c$ are some chosen cost functions.
To get an upper bound we will introduce dual variables to get a quantity larger than $\mathcal F_P$\,:
\bea
\mathcal F_P^\ttext{S-mat}\leq\mathcal F_P^{\ttext{S-mat}}+\int_4^\infty \ud s\left[|k(s)|-\mathcal R_s\left(k(s) S(s)\right)\right]+\int_{s_0}^4 \ud s \tilde k(s) \tilde \rho(s)
\eea
where by assumption:
\bea
\tilde k(s)\geq 0 \qquad \mbox{for}\quad s\in (s_0,4)\,.\label{eq:posktilde}
\eea
Note that positivity of the first term follows from unitarity of the S-matrix. To get a bound valid for any $S$-matrix, we first write $S(s)$ in terms of the $\rho,\tilde \rho$ using \reef{CFTdispS}. We then impose conditions on $k(s)$, $\tilde k(s)$ to eliminate all dependence on the primal variables from the right hand side of the above. (An example will be given further below {\em cf.} \reef{2ks}). The result is:
\bea\label{Fd}
\mathcal F_D^{\ttext{S-mat}}:=\int_4^\infty\ud s\left[|k(s)|-\mathcal R_s k(s)\right]\,,
\eea
and by construction we have
\bea
\mbox{max}\, \mathcal F_{P}^{\ttext{S-mat}}\leq \mbox{min}\, \mathcal F_D^{\ttext{S-mat}}\,.
\eea
Optimality is achieved if \footnote{Note that in the region $s>4$, optimality can also be attained through $k(s)=0$ as numerically observed in \cite{Cordova:2019lot}. Here we do not consider this possibility but focus on unitarity saturating S-matrices.}
\begin{equation}\label{optimality}
\begin{cases}
     \tilde k(s)=0\; \text{or}\; \tilde{\rho}(s)=0,& s_0<s<4\\
     S_{\text{ext}}(s)=|k(s)|/k(s),& s>4
\end{cases}
\end{equation}
where $S_{\text{ext}}$ stands for the optimal (extremal) S-matrix.

\subsubsection*{CFT bootstrap}
Now consider the following CFT bootstrap optimization problem:
\bea
\mbox{max}\,\mathcal F_P^{\ttext{CFT}}\,, \qquad \mathcal F_{P}^{\ttext{CFT}}=\sum_{\Delta\geq \Delta_0} a_{\Delta} \mu_\Delta
\eea
with
\bea
\mathcal G(z)=G_0(z|\Df)+\sum_{\Delta\geq \Delta_0} a_{\Delta} G_{\Delta}(z|\Df)\,, \qquad \Delta_0\geq \sqrt{2}\Df\,.
\eea
The primal variables are now the OPE coefficients $a_{\Delta}\geq 0$, and $\mu_\Delta$ are again some chosen cost functions. For example, OPE maximization would correspond to $\mu_{\Delta}=\delta_{\Delta,\Delta_b}$, and correlator minimization to $\mu_{\Delta}=-G_{\Delta}(w)$. The gap assumption may be replaced by the weak OPE condition spelled out in section~\ref{sec:boundbelow}, and as we already know, it plays the role of unitarity for the S-matrix problem since it implies the OPE is bounded.

To obtain a bound we will again add a positive quantity to the primal objective. Let us introduce a functional satisfying:
\bea
\Omega(0)+\sum_{\Delta\geq \Delta_0} a_{\Delta} \Omega(\Delta)=0\,, \qquad\mbox{and}\qquad \Omega(\Delta)\geq \mu_{\Omega}(\Delta) \quad \mbox{for}\quad \Delta\geq \Delta_0\,. \label{funcconds}
\eea
To get an upper bound we consider
\bea
\mathcal F_P^{\ttext{CFT}}\leq\mathcal F_P^{\ttext{CFT}}+\sum_{\Delta\geq \Delta_0} a_{\Delta}[\Omega(\Delta)-\mu_{\Omega}(\Delta)]\,.
\eea       
We now constrain our functional such that $\mu_{\Omega}=\mu$. Using \reef{funcconds} we define
\bea
\mathcal F_D^{\ttext{CFT}}:=-\Omega(0)\,,
\eea
from which we obtain an upper bound
\bea
\mbox{max}\,\mathcal F_P^{\ttext{CFT}}\leq \mbox{min}\,\mathcal F_D^{\ttext{CFT}}\,.
\eea
Optimality is achieved if
\bea
a_{\Delta}=0\quad \mbox{or}\quad \Omega(\Delta)=\mu(\Delta)\,, \qquad \Delta\geq \Delta_0\,.
\eea

\subsubsection*{Mapping the problems}
We have presented these optimization problems in a suggestive way which makes clear that they are closely related. Let us now make the link more precise starting from the CFT problem in the flat space limit.  

First, it is clear that since $\rho,\tilde \rho$ are directly related to OPE coefficients, the CFT correlator may also be expressed in terms of those  variables. It is also clear that it is possible to choose CFT cost functions $\mu_\Delta$ which will give $\mathcal F_P^{\ttext{S-mat}}=\mathcal F_P^{\ttext{CFT}}$.\footnote{In detail, this is achieved by setting
\ba
\mu_{\Delta}&=\frac{2\sqrt{s_{\Delta}}}{\Df \tilde a^\ttext{free}_{\Delta}}\tilde c(s_{\Delta})\,,& \qquad \Delta_0\leq \Delta&<2\Df\\
\mu_{\Delta}&=\frac{2\sqrt{s_{\Delta}}}{\Df \af}\,4\sin^2\left[\frac{\pi}2(\Delta-2\Df)\right] c(s_{\Delta})\,,& \qquad \Delta&\geq 2\Df\,.
\ea
}
Let us set
\bea
\hG(z)=\mathcal G(z)-\sum_{\Delta\leq 2\Df} a_{\Delta} G_{\Delta}(z|\Df)\,.
\eea
Then the phase shift formula and the bounds of section \ref{sec:boundabove} can be written: 
\ba
\hG(z)&\underset{\Df\to \infty}{=}S(s_z)\,,& \quad \quad z&<0\,\\ 
(1-z)^{2\Df} \hG(z)&\underset{\Df\to \infty}=1\,,& \qquad z&\in(0,1)
\ea
We now parameterize our functional in terms of kernels $f,g$ as usual. Inspired by previous examples, we will set 
\bea
g(z)=(1-z)^{2\Df-2}|f(\mbox{$\frac{1}{1-z}$})|+\delta g(z)\,, \qquad z\in(0,1)\,,
\eea
with $\delta g(z)$ introducing extra freedom and assumed to not scale with $\Df$,
so that
\bea\label{Rfdg}
\mathcal R_z f(z)\sim -\delta g(z)-\delta g(1-z)
\eea
Using the definition of the functional action we can find
\ba
\sum_{\Delta\geq 2\Df} a_{\Delta} [\Omega(\Delta)-\mu_\Omega(\Delta)]&=\int_{-\infty}^0 \ud z\left\{ |f(z)|-\mathcal R_z[f(z) \hG(z)]\right\}\\
\sum_{\Delta_0\leq \Delta< 2\Df} a_{\Delta} \left[\Omega(\Delta)-\mu_{\Omega}(\Delta)\right]&=\int_{z_0}^1 \ud z \left[\mathcal I_z f(z) \tilde \rho(4-s_z)\right]
\ea
where the $\mu_{\Omega}(\Delta)$ are in general non-zero, since computing just $\Omega(\Delta)$ leads to more contributions other than those shown on the RHS. For instance on the first line we should have
\bea
\sum_{\Delta\geq 2\Df} a_{\Delta}\mu_\Omega(\Delta)=\int_0^1 \ud z \,\delta g(z) \hG(z)
\eea
It is now clear we should identify:
\bea\label{kfidentify}
k(s)=4f(z_s)\,, \qquad \tilde k(s)=4 \mathcal I_z f(z_s)
\eea
In particular, this leads to
\ba
\mathcal F_D^\ttext{S-mat}=\int_4^\infty\ud s\bigg(|k(s)|-\mathcal R_s k(s)\bigg)
=\int_{-\infty}^0\ud z\bigg(|f(z)|-\mathcal R_z f(s)\bigg)=\mathcal F_D^{\ttext{CFT}}\,.
\ea
which completes our mapping between the two bootstrap problems.

For completeness, and to conclude, let us point out that it is straightforward to generalize these constructions to enforce specific constraints on S-matrices/correlators, such as presence of states with definite couplings. In this case we should set
the costs of such states to be zero, and since they do not need to be eliminated we do not to constrain the associated dual variables. For instance, in CFT bootstrap language, this leads to a modified dual objective of the form
\bea
\mathcal F_D^{\ttext{CFT}}=-\Omega(0)-\sum_{i} a_{\Delta_i} \Omega(\Delta_i)=\int_{-\infty}^0\ud z\left[|f(z)|-\mathcal R_z f(s)\right]+\sum_i \mathcal I_z f(z_i) \tilde \rho_i\,.
\eea
where the $\tilde \rho_i$ are given and fixed.

\subsubsection*{Example: correlator maximization }

Let us now illustrate the above map between the CFT functionals and the dual S-matrix bootstrap problem through the example of the correlator maximization problem considered in section \ref{CDDpole}. In the flat space limit, the corresponding S-matrix bootstrap problem becomes bounding the amplitude evaluated at a point $S(s_w)$ with $s_w\in (4-s_0,s_0)$. In particular, in this case the primal objective \eqref{fprimal} is given by the cost functions
\begin{equation}
F_P^\ttext{S-mat}=S(s_w)-1\quad \Rightarrow\quad \begin{cases}
       \tilde{c}(s')=\tilde{K}(s_w,s')\ge 0,&s'\in (s_0,4)\\
       c(s')=-K(s_w,s')\le 0,& s'>4
\end{cases}
\end{equation}
as can be seen by directly evaluating \eqref{CFTdispS} at $s=s_w$.
The standard dual formulation gives the minimization problem with the dual functional \eqref{Fd} where the dual variables $k,\tilde{k}$ are subject to the constraints:
\begin{equation}\label{2ks}
\begin{aligned}
 \mathcal{R}_sk(s)&=-c(s)-\frac{2}{\pi}\dashint_4^{\infty}\!\!\!ds'\frac{\sqrt{s'(s'-4)}}{\sqrt{s(s-4)}}\frac{s-2}{(s-s')(s'+s-4)}\mathcal{I}_sk(s'),\;\; s>4\;,\\
  -\tilde{k}(s)&=\tilde{c}(s)-\frac{2}{\pi}\int_4^{\infty}\!\!\!ds'\frac{\sqrt{s'(s'-4)}}{\sqrt{s(4-s)}}\frac{s-2}{(s-s')(s'+s-4)}\mathcal{I}_sk(s'),\;\; s_0<s<4\;.
\end{aligned}
\end{equation}
which arise by demanding decoupling of the primal variables. Plugging in the cost functions, we can solve these constraints  by setting
\ba
    k(s)&=K(s_w,s)-\frac{2}{\pi}\int_{4}^{\infty}\!\!\!ds'\frac{\sqrt{s'(s'-4)}}{\sqrt{s(s-4)}}\frac{s-2}{(s-s')(s'+s-4)}\text{Im}k(s')\,,\\
    \tilde{k}(s)&=\mathcal I_s k(s)\,.
    \label{eq:dispk}
\ea
Notice this is consistent with \eqref{kfidentify}. We also have
\ba
\mathcal{R}_sk(s)&=-\delta(s-s_w)-\delta(4-s-s_w)\;.
\ea
which is the same as equation \reef{eq:deltafuncsinglargedphi} satisfied by $f^\ttext{int}_w(z)$, again in agreement with \eqref{kfidentify}. It follows that we can also solve for $k$ in the same way, by setting
\bea
k(s)=K(s_w,s)\frac{S^\ttext{int}(s_w)}{S^\ttext{int}(s)}\,.
\eea
The analyticity properties of $k(s)$ imply that the function $S^\ttext{int}(s)$ should be crossing-symmetric and meromorphic with branch points at $s=0$ and $s=4$. Recall \eqref{optimality}, i.e., optimality implies $S_{\text{ext}}(s)=S^{\ttext{int}}(s)$ with $|S^{\ttext{int}}(s^+)|=1$ and $S^{\ttext{int}}(s)$ can then be chosen to be a product of CDD poles. Now it is the positivity constraint \reef{eq:posktilde} which forces these poles to be located below $s_0$, and choosing without loss of generality $S^{\ttext{int}}(s_w)>0$, we can compute the dual objective \reef{Fd} as:
\bea
F_D^\ttext{S-mat}=S^\ttext{int}(s_w)\int_4^\infty \ud s K(s_w,s)\left(1-\mathcal R_s\left[\frac{1}{S^\ttext{int}(s)}\right]\right)=S^\ttext{int}(s_w)-1\,.
\eea
Optimality is achieved when $F_D^{\ttext{S-mat}}=F_P^{\ttext{S-mat}}$, i.e. when $S(s_w)=S^{\ttext{int}}(s_w)$, which happens if both are given by a single CDD pole at $s=s_0$, as before in section \ref{CDDpole}.

\section{Discussion and Outlook}
In this work we have studied the detailed relation between families of 1d CFT correlators and 2d S-matrices. Our main results are a derivation of the analyticity properties of such S-matrices under certain gap assumptions, and a characterization of their singularity structure in terms of the CFT data.

It would be important to clarify for which S-matrices do our results apply. The starting point of our construction is that the S-matrix arises from a gapped QFT. Can any such QFTs be placed in an AdS space of sufficiently large radius? This is certainly the case for any Lagrangian QFT, but more generally it seems hard to imagine an obstruction given that correlators in any such theory decay exponentially at scale separations larger than the gap. For instance, it seems we could define the correlators of the theory in the following manner: starting from local QFT correlation functions, we push fields away from each other distances much larger than the inverse mass gap while keeping the effects of curvature negligible. At this scale correlators factorize into products of two point functions given by free massive propagators. We can then ``complete'' each such propagator to an AdS bulk-to-bulk propagator by adding a small correction. At this point we can move around operators to arbitrarily large distances, thus leading to fully well defined correlators in AdS.

Conversely, we can hope to use our results to show that some S-matrices {\em cannot} arise from a gapped QFT. Indeed, many if not most  2d  S-matrices consistent with crossing, analyticity and unitarity have a non-trivial UV behaviour, which can be understood in the context of generalized $T\bar T$ deformations \cite{Smirnov:2016lqw,Cavaglia:2016oda,Camilo:2021gro,Cordova:2021fnr}, and it would be great to understand if we can use our results to investigate if these theories can arise from gapped QFTs in AdS. One idea would be to show that such S-matrices do not arise from QFTs with local observables: in practice one would need to show that it would not be possible to construct local operators from such a theory's S-matrix, i.e. by ``inverting'' LSZ. In CFT language we would have to show that the bulk reconstruction problem \cite{Kabat:2016zzr} would not be solvable given the CFT data implied by the S-matrix. 

A different kind of argument relies on the fact that our construction implicitly acts as an UV completion for S-matrices. Consider an S-matrix of the form:
\bea
S_{T\bar T}(s)=e^{-\ell^2 \sqrt{s(4-s)}}
\eea
which arises in the context of $T\bar T$ deformations. One of the peculiarities of this S-matrix is that the associated density oscillates indefinitely with $s$:
\bea
\rho_{T\bar T}(s)=2\sin^2\left(\ell^2\sqrt{s(s-4)}\right)\,.
\eea
But now recall that in CFT language the density $\rho$ is given by the double discontinuity of the correlator,
\bea
\rho(s)=\lim_{\Df\to \infty}\left[(1-z)^{2\Df}\mbox{dDisc}^+ \overline{ \mathcal G}(z)\right]\bigg|_{z=\frac{s-4}s}\,.
\eea
Generally we would expect therefore that $\ud \rho(s)/\ud s$ should instead decay with some power related to the Regge behaviour of the correlator \cite{Paulos:2020zxx}. To make this expectation precise, we can use the bounds on the correlator and its derivative derived in sections \ref{CDDpole} and \ref{sec:bootzero} to find
\bea
\frac{\ud }{\ud s} \rho(s)\leq \frac{4\Df}{s-4}\qquad \mbox{for}\quad s>8\,,
\eea
where the bound holds up to exponentially small corrections in $\Df$. We should understand this expression as telling us that $\Df$ acts as a hidden UV cut-off, beyond which we must have $\rho(s)$ decay to a constant. S-matrices arising from a UV complete QFT must therefore satisfy this constraint above some scale no larger than $\Df$. This makes sense: $\Df=m R_{\ttext{AdS}}$ can also be thought of as an IR cutoff, and hence there should be no reason why high energy processes should know about it. As an example, notice that any S-matrix expressed as an arbitrary large (but finite) product of CDD factors satisfies this property. \footnote{It is important to point out however that there should be extra consistency conditions in order for the S-matrix to describe a UV complete theory. For the example of CDD factors at hand, one such condition goes along the lines of not having more resonances than bound states in the theory (see \cite{Camilo:2021gro,Cordova:2021fnr}).} In contrast, for general S-matrices with an essential singularity at infinity such as $S_{T\bar T}$, the indefinite oscillations in the density imply that there have to be corrections at a UV scale $s\sim \Df/\ell^2$ if they are to arise from a CFT$_1$/QFT$_2$ system.
Physically, one way of understanding how an S-matrix can possibly fail this UV constraint is if it arises from a gravitational theory. In this case there is indeed UV/IR mixing: we cannot scatter particles with arbitrarily large energies while ignoring the arbitrarily large but finite size of the AdS box, since eventually we can create black holes whose size grows with the center of mass energy. Thus our argument suggests that essential singularities signal the presence of gravitational physics, in agreement with the logic of \cite{Dubovsky:2017cnj}.

There are several open questions and directions of research to pursue in the future. The most important is to clarify the singularity structure in the presence of anomalous thresholds. In the CFT these appear when the the gap in the OPE is sufficiently small, causing a catastrophic loss of control in the OPE data. We have conjectured that a resolution could lie in an understanding of unitary solutions to crossing without identity and in particular their flat space limit. It would be nice to understand what this means in practice, and possible links to the observations based on the Mellin amplitude prescription made in~\cite{Komatsu:2020sag}, for which the 1d Mellin amplitudes of \cite{Bianchi:2021piu} might prove useful. 

An exciting direction to explore is  scattering processes of higher numbers of particles, with the most interesting being three-to-three. Firstly because analyticity properties in this case are poorly understood, and secondly because of our map between extremal CFTs and extremal S-matrices. Integrability can be formulated in terms of the Yang-Baxter equation which expresses three-to-three scattering processes in terms of two-to-two, and it would be nice to  understand what these conditions mean for the CFT data. 

It would also be interesting to generalize our study to CFT correlators of distinct fields, which map onto S-matrices describing distinguishable particles. A special and rich set of examples corresponds to S-matrices with global symmetry. In this case not only have S-matrix bootstrap studies been performed, but also CFT ones. Some of the S-matrix components now have ``left'' cuts which are constrained by unitarity, and it would be interesting to understand how these are determined by the CFT data. There are also  S-matrices saturating bounds with intricate analytic structure. It would be nice to understand their CFT interpretation.

\vspace{1 cm}
\section*{Acknowledgments}
We would like to thank Alessandro Georgoudis, Kausik Ghosh, Shota Komatsu, Martin Kruczenski, Balt van Rees,  and Zechuan Zheng for useful discussions. We also thank the Simons Collaboration on the Nonperturbative Bootstrap for leading to opportunities for discussion and collaboration. LC and YH were supported in part by the National Science Foundation under Grant No. NSF PHY-1748958. MFP is supported in part by the Agence National de Recherche, Tremplin-ERC grant ANR-21-ERCC-0006 'FUNBOOTS'.
\pagebreak

\appendix

\section{\texorpdfstring{Polyakov blocks for $\Delta<2\Df$: real argument }{Polyakov blocks below 2dphi} \la{app:realw}}

In this subsection we perform the computation of Polyakov blocks with dimension $\Delta<2\dphi$ for the special case where the cross-ratio $w$ takes real values. The computation follows in the footsteps of a similar one in \cite{Mazac:2018mdx}. Again, we set $\Delta=\Delta_b$. After a change of variables we can rewrite a general functional action in terms of a single conformal block
\beq
\Omega(\Delta_b)=-\[\int\limits_{\frac{1}{2}}^{\frac 12+i\infty} dz\,\frac{f(z)}{2} -\int\limits_{\frac{1}{2}-i\infty}^{\frac{1}{2}} dz\,\frac{f(1-z)}{2} +\int\limits_{\frac{1}{2}}^{1} dz\, g(z) -\int\limits_{0}^{\frac{1}{2}} dz\, g(1-z) \] G_{\Delta_b}(1-z|\dphi)\,. \la{action_below1}
\eeq
In the large $\dphi$ limit with $m_b$ fixed the conformal block factor has a saddle point at $z=z_b\equiv m_b^2/4$ with steepest descent direction now running along the imaginary axis. We will therefore modify shift our contour, making use of the gluing condition \eqref{eq:gluing} to obtain
\beq
\Omega(\Delta_b)=-\[\int\limits_{z_b}^{z_b+i\infty} dz\,\frac{f(z)}{2} -\int\limits_{z_b-i\infty}^{z_b} dz\,\frac{f(1-z)}{2} +\int\limits_{z_b}^{1} dz\,  g(z) -\int\limits_{0}^{z_b} dz\,  g(1-z) \] G_{\Delta_b}(1-z|\dphi)\,. \la{action_below2}
\eeq
Now we restrict to master functionals $\Omega^{\pm}_w(\Delta_b)$. The first two integrals in the above equation can be performed directly by the steepest descent method. Since the poles of $f_w^{\pm}$ lie on the real axis we don't have to worry about the steepest descent contour crossing them. Instead those contributions now effectively appear through the last two terms above, since  $g_w^\pm(z)=\hat g_w^\pm(z)\pm\delta(w-z)$, where the kernel $\hat g_w(z)$ is exponentially suppressed with respect to $f_w(z)$. That leads us to the expression
\begin{multline}
\Omega_w^{\pm}(\Delta_b)\underset{\Delta_b,\Df\to \infty}{=}
\frac{m_b}{\dphi} \frac{\mathcal I_{z}f^{\pm}_{w}(z_b)}{\tilde a_{\Delta_b}^\ttext{free}}\mp\[ \int\limits_{z_b}^1 dz\, \delta(z-w)- \int\limits_0^{z_b} dz\, \delta(z-1+w) \] G_{\Delta_b}(1-z|\dphi)\,, \la{action_deltas}
\end{multline}
where
\beqa
\tilde a_{\Delta_b}^\ttext{free} &=& \frac{a_{\Delta_b}^\ttext{free}}{2\sin\[\pi \dphi(2-m_b)\]}\qquad (\geq 0\quad \mbox{for}\quad \sqrt{2}\Df<\Delta_b<2\Df)%
\,.
\eeqa
The last two terms with delta functions in \eqref{action_deltas} evaluate to different
combinations of individual conformal blocks, depending on the possible configurations between the saddle point $z_b$ and the master functional parameter $w$. Using the relation between the Polyakov blocks and the master functional actions we find:
\ba
\mathcal P_{\Delta_b}^{\pm}(w)\underset{\Delta,\Delta_b\to \infty}{=}\frac{m_b}{\pi \dphi \tilde a_{\Delta_b}^\ttext{free}} \frac{\sqrt{w(1-w)}}{\sqrt{z_b(4-z_b)}}\,\frac{z_b-1/2}{(z_b-w)(w+z_b-1)}\,+E_{\Delta_b}(w|\Df)
\ea
with the crossing symmetric $E_{\Delta_b}$ satisfying:
\bea
E_{\Delta_b}(w|\Df)=\left\{
\begin{array}{l r}
G_{\Delta_b}(w|\Df) &   0<w<\text{min}(z_b,1-z_b)\\
0 & 1-z_b<w<z_b\\
G_{\Delta_b}(1-w|\Df) &   \text{max}(z_b,1-z_b)<w<1\\
G_{\Delta_b}(w|\Df)+G_{\Delta_b}(1-w|\Df) & z_b<w<1-z_b
\end{array}
\right.
\eea
This is in agreement with the results in the main text.

\section{The phase shift formula for Polyakov blocks}
\label{sec:phaseshiftpoly}
The goal of this section is to show that the phase shift formula applied to Polyakov blocks leads to the same result as their flat space limit computed in section \ref{sec:polflat}. Consider then
\bea
\left[ \mathcal P_{\Delta}(z)-G_{\Delta}(z)\right]=-\sum_{n=0}^{\infty} \left[ \alpha_n(\Delta) G_{\Delta_n}(z_s)+\beta_n(\Delta)\, \partial_{\Delta}G_{\Delta_n}(z_s)\right]
\eea
We want to prove that
\bea
\lim_{\Df \to \infty} \lim_{\epsilon\to 0}\left[ \mathcal P_{\Delta}(z_s)-G_{\Delta}(z_s)\right]= i g^2\, \frac{\sqrt{s(s-4)}}{\sqrt{s_{\Delta}(4-s_{\Delta})}}\,\frac{2s_{\Delta}-4}{(s-s_{\Delta})(s-4+s_{\Delta})}
\eea

The sum over states localizes on those $\Delta_n$ satisfying $\Delta_n\sim \Df \sqrt{1-z}$. This means that to compute the above we need to determine the functional actions in the limit where $n,\Df\to \infty$ with fixed ratio. We will do this relying on fact that we can write the Polyakov block as a sum of Witten exchange diagrams:
\bea
\mathcal P_{\Delta}(z)=W_{\Delta}^{(S)}(z)+W_{\Delta}^{(T)}(z)+W_{\Delta}^{(U)}(z)+n(\Delta) \mathcal C(z)
\eea
with some suitably chosen $n(\Delta)$ and $\mathcal C(z)$ the $\Psi^4$ contact term in AdS$_2$. We have
\bea
\mathcal C(z)=\sum_{n=0}^{\infty}\left[ a_n G_{\Delta_n}(z)+b_n \partial_{\Delta} G_{\Delta_n}(z)\right]
\eea
with
\bea
b_n=a_{\Delta_n}^{\mbox{\tiny free}}\frac{(2 n)! (\Df)_n^4 (4 \Df-1)_{2 n}}{2 (n!)^2 (2 \Df)_n^2 (2 \Df)_{2 n}^2}\,, \qquad a_n=\frac 12 \partial_n b_n
\eea
We begin by applying the phase shift formula to the contact term. We begin by rearranging
\bea
\mathcal C(z)=\sum_{n=0}^\infty\left\{ \left(\frac{a_n}{a_{\Delta_n}^{\mbox{\tiny free}}}-\frac{\partial_n a_{\Delta_n}^{\mbox{\tiny free}}}{2 a_{\Delta_n}^{\mbox{\tiny free}}}\right)a_{\Delta_n}^{\mbox{\tiny free}} G_{\Delta_n}(z)+ \frac{b_n}{a_{\Delta_n}^{\mbox{\tiny free}}}\, \partial_{\Delta}\left[a_{\Delta_n}^{\mbox{\tiny free}}G_{\Delta_n}(z)\right]\right\}
\eea
In the flat space limit the second set of terms dominate. Using the asymptotic form of conformal blocks \reef{eq:gaussian} we find 
\bea
\mathcal C(z_s)\sim \frac{-i\pi}{\sqrt{s(s-4)}}\,,
\eea
with $s>4$, up to an irrelevant overall constant. The $s$-channel exchange diagram is simply related to the contact diagram by the action of the Casimir operator (see e.g \cite{Zhou:2018sfz}). We have
\bea
W^{(S)}_{\Delta}(z_s)\sim \sum_{n=0}^\infty \frac{\beta_n^{(S)}(\Delta)}{a_{\Delta_n}^{\mbox{\tiny free}}}\, \partial_{\Delta}\left[a_{\Delta_n}^{\mbox{\tiny free}}G_{\Delta_n}(z)\right]
\eea
with $\beta_n^{(S)}\sim b_n/[\Delta(\Delta-1)-\Delta_n (\Delta_n-1)]$. Including the precise normalisation and taking the flat space limit one finds
\bea
W^{(S)}_{\Delta}(z_s)=-i \frac{2 \sqrt{s_{\Delta}}}{\pi \Df \tilde a_{\Delta}^{\mbox{\tiny free}}} \frac{\sqrt{s_{\Delta}(4-s_\Delta)}}{ \sqrt{s(s-4)}} \frac{1}{ s-s_{\Delta}}\,.
\eea
Doing the same computation for the other channels is not trivial, since the OPE coefficients are not known in closed form for general $n, \Df$. However, we do know that whatever the phase shift formula gives has to be a crossing symmetric expression, so we can simply sum this result over images. This leaves the overall contact term to be fixed. Its coefficient is determined by demanding that in the OPE expansion of $\mathcal P_{\Delta}$ we have $\beta_0(\Delta)=0$. In the flat space limit this means that we must tune the contact term such that $\mathcal P_{\Delta}(z_s)$ is suppressed at threshold, $s\to 4$. Doing this leads to the result:
\bea
\tilde a_{\Delta}^{\mbox{\tiny free}}\left(\mathcal P_{\Delta}(z_s)-G_{\Delta}(z_s)\right)=-i \, \frac{2 \sqrt{s_{\Delta}}}{\Df}\,\frac{2}{\pi} \frac{\sqrt{s(s-4)}}{\sqrt{s_{\Delta}(4-s_{\Delta})}}\, \frac{s_{\Delta}-2}{(s-s_{\Delta})(4-s-s_{\Delta})}\,.
\eea
This agrees on the nose with the limit of expression \reef{eq:flatpolybs2}.

\section{Products of CFT correlators and S-matrices}
\label{app:products} 
Let us adress an apparent puzzle. Consider two extremal CFT correlators leading to two $S$-matrices saturating unitarity. The product property \reef{eq:prodprop} then guarantees that the product CFT correlator will also lead to an extremal S-matrix simply given by the product of the previous two. But this is surprising, since the product CFT correlator is not ``extremal'', in the sense that it will not only the two towers of operators in $\mathcal G_1$ and $\mathcal G_2$, but also new operators arising from their tensor product. So naively these three towers of states, each of which have different anomalous dimensions, should interfere with each other and lead to $|S|<1$. The solution as we is that there is a single subset of these operators which dominates the OPE in the flat space limit. 

First note that if we say that $\mathcal G=\mathcal G_1 \mathcal G_2$ corresponds to a given $\Df$, then we must assign $x \Df$ to $\mathcal G_1$ and $(1-x) \Df$ to $\mathcal G_2$. For simplicity let us set $x=1/2$. There are then three towers of (non-identity) states with dimensions
\ba
\Delta_{1,n}&=\Df+2n+\gamma_{1}(n)\\ \Delta_{2,m}&=\Df+2m+\gamma_{2}(m)\\ \Delta_{nm,p}&=\Delta_{1,n}+\Delta_{2,m}+2p
\ea
Then the statement is that the first two towers actually give subleading contributions to the OPE in the flat space limit, while the third tower is then equivalent to a single tower of states with dimensions $\Delta_{12}=2\Df+2p+\gamma_1(s)+\gamma_2(s)$, leading to the expected result $S(s)=e^{-i \pi[\gamma_{1}(s)+\gamma_{2}(s)]}$.

Let us see how the OPE coefficients of each tower compare to $\af$, introducing the explicit dependence on $\Df$. Then at a given $\Delta$ the first two towers give contributions of the form $\af(\Df/2)/\af(\Df)$ which is exponentially suppressed. The last tower has OPE coefficients
\bea
a_{\Delta_1,n}^{\mbox{\tiny free}}(\Df/2)a_{\Delta_2,m}^{\mbox{\tiny free}}(\Df/2)\times \lambda(\Delta_1,\Delta_2,p)
\eea
with $n+m+p\sim (\Delta-2\Df)/2$ and the $\lambda$ coefficients appear in
\bea
G_{\Delta_1}(z)G_{\Delta_2}(z)=\sum_{p=0}^\infty \lambda(\Delta_1,\Delta_2,p) G_{\Delta_1+\Delta_2+2p,z}(z)\,,
\eea
and are given explicitly by
\bea
\lambda(\Delta_1,\Delta_2,p)=\frac{2^{-4 p} \Gamma \left(p+\frac{1}{2}\right) \left(\Delta _1\right)_p \left(\Delta _2\right)_p \left(\Delta
   _1+\Delta _2+\frac{1}{2}\right)_{p-1} \left(\Delta _1+\Delta _2+1\right)_{2 p-1}}{\sqrt{\pi } \Gamma (p+1)
   \left(\Delta _1+\frac{1}{2}\right)_p \left(\Delta _2+\frac{1}{2}\right)_p \left(\Delta _1+\Delta
   _2+\frac{1}{2}\right)_{2 p-1} \left(\Delta _1+\Delta _2+1\right)_{p-1}}
\eea
We begin by noting that $\lambda$ decreases exponentially with $p$ even for $p,\Delta_1,\Delta_2$ large, so the dominant contribution to the sum comes from $p\sim 0$.  We are left with a sum over $n,m$, but it is now easy to show
\bea
\lim_{\Df\to \infty} \frac{a_{\Delta_1,n}^{\mbox{\tiny free}}(\Df/2)a_{\Delta-\Delta_{1,n}}^{\mbox{\tiny free}}(\Df/2)}{\af}=\left\{\begin{array}{cc}
1& \Delta_{1,n}=\Delta/2\\
0 & \mbox{otherwise}
\end{array}
\right.
\eea
and so the dominant contributions come from states with $n\sim m$ and $p=0$, i.e. with
\bea
\Delta=2\Df+2n+\gamma_{1}(s)+\gamma_{2}(s)\,.
\eea

\section{Extremal CFTs for the CDD zero}
\label{app:cddzero}
There is a simple one parameter family of extremal CFT correlators obtained as a deformation of the generalized free boson.\footnote{See \cite{Mazac:2018ycv} and \cite{Paulos:2019fkw} for analytic and numerical studies.} These correlators saturate an  upper bound on the OPE coefficient of an operator sitting at the gap $\Delta_0=2\Df+g$, where $g< 1$, so that each correlator is labeled by both $g$ and $\Df$\,, i.e. $\mathcal G(z)\equiv\mathcal G_g(z|\Df)$. One finds both numerically and analytically in perturbation theory that the spectrum of the correlator $\mathcal G_g$ varies smoothly as we dial $g$ away from zero, where the correlator matches the generalized free boson. By this we mean that not only no new states appear beyond those already contained in the generalized free solution, but furthermore that the dimensions of the states vary continuously with $g$. The deformation is relevant, in the sense that for any fixed $g<1$, anomalous dimensions of operators eventually decay to zero at high energies, but as $g\to 1$ the spectrum approaches that of a generalized free fermion. Another way of putting it is that the same family can in principle be obtained as an irrelevant deformation of the latter solution. 

We will argue that in the limit $\Df\to \infty$ this family of extremal CFTs maps onto the family of S-matrices described by a single CDD factor. This factor will be a zero or a pole depending on $g$. For definiteness we will focus on the former case, and explain the relationship to the latter in due course. That is, we claim:
\ba
S^{\mbox{\tiny CFT}}(s)=S_{s_0}^{\mbox{\tiny zero/pole}}(s)\,, \qquad S^{\mbox{\tiny CFT}}(s)\equiv \mathcal F[\mathcal G_g(z_s)]\,,
\ea
where $g$ is related to $s_0$ and must be taken to scale with $\Df$ in a precise way. The picture is that for $g/\Df<0$ and fixed for large $\Df$ we recover CDD pole, where as for $0<g<1$, we have a CDD zero, where in particular a finite position of the zero requires $1-g\sim 1/\sqrt{\Df}$. 

We can check this mapping perturbatively when $g\ll 1$. In this limit the CFT is a small deformation of the free boson solution, and is described by a free boson in AdS$_2$ with a small quartic contact interaction. The interaction shifts the dimensions of double trace operators in the correlator, but does not introduce any new states to the theory. The anomalous dimensions of operators are given by \cite{Mazac:2018ycv}
\bea
\gamma_n(\Df):=\Delta_n-2\Delta_n^B=2g\,\frac{(2 n)! (\Df)_n^4 (4 \Df-1)_{2 n}}{2 (n!)^2 (2 \Df)_n^2 (2 \Df)_{2 n}^2}+O(g^2) \label{eq:gammas}
\eea
This leads to
\bea
\gamma(s)=\frac{4g \sqrt{2}}{\sqrt{\pi\Df}\sqrt{s(s-4)}}+O(g^2)
\eea
from which we can determine the S-matrix $S^{\mbox{\tiny CFT}}$. Expanding it we find a perfect match with $S_{s_0}^{\mbox{\tiny zero}}$ if we equate 
\bea
\sqrt{s_0(4-s_0)}=\sqrt{\frac{2\pi}{\Df}}\, g\,. \label{eq:idcdd}
\eea
which implies that in this regime $s_0$ is parametrically close to threshold. Negative $g$ requires continuing $s_0$ around the cut beginning at $s=4$, which turns the CDD zero into a CDD pole. Hence, in a certain sense, the second sheet of the S-matrix is the region $2\Df<\Delta<2\Df+1$. 
The function $\gamma(s)$ has an interesting form. The CFT computation which leads to \reef{eq:gammas} requires that all anomalous dimensions are small. But this is true even when $g$ is not parametrically small, as long as we are not too close to threshold $s\sim 4$. Hence, for finite $g$ we should expect that
\bea
\gamma(s)=\frac{c(g)}{\sqrt{\Df s(s-4)}}\,,\qquad \sqrt{s_0(4-s_0)}=\frac{c(g)}{\sqrt{\Df}}
\eea
The function $c(g)$ captures the non-perturbative dynamics which are only relevant for $\Delta-2\Df=O(1)$ in the flat space limit. Unfortunately we do not have information about this function beyond leading order\footnote{In reference \cite{Mazac:2018ycv} anomalous dimensions were computed up to cubic order in $g$ but only for fixed $\Df=1$.}. However, we expect that for any fixed $g<1$, $c(g)$ remains finite, and hence the CFT family maps onto a single CDD zero parametrically close to threshold. 

Let us now assume the correspondence between $S^{\mbox{\tiny CFT}}$ and $S_{s_0}^{\mbox{\tiny zero}}$ is true to see what can be learned about the mapping between $s_0$ and $g$. Since $g$ is the anomalous dimension of operators close to threshold we impose the condition
\bea
S_{s_0}^{\mbox{\tiny zero}}(s=4+\epsilon)=g
\eea
where $\epsilon=\epsilon(g)$ should be parametrically small. In fact, we should expect $\epsilon=O(\Df^{-1})$. A simple calculation yields
\bea
\sqrt{s_0(4-s_0)}=\sqrt{\epsilon(g)} \tan\left(\frac{\pi g}2\right) \label{eq:matchingcddzero}
\eea
Consistency with our results at finite $g<1$ determines
\bea
\sqrt{\Df\epsilon(g)}\tan\left(\frac{\pi g}2\right) =c(g)\,.
\eea
For small $g$ this yields $\epsilon(g)=8\pi/\Df$,  i.e. $\Df \epsilon(g)$ is order one as expected. Assuming this remains true for any $g$ we see that to have a finite $s_0$ requires taking $g$ parametrically close to unity, $1-g=O(\Df^{-1/2})$. In this regime $c(g)$ becomes large and perturbation theory breaks down.

\small
\parskip=-10pt

\newpage

\bibliographystyle{utphys}
\bibliography{functionalreferences}

\end{document}